\documentclass[12pt,a4paper]{article}

\usepackage{amsmath,amssymb,amsfonts,amsthm,mathtools,dsfont}

\usepackage{algorithm}
\usepackage{algpseudocode}

\usepackage{graphicx}
\usepackage{float}
\usepackage{placeins}
\usepackage{array,multirow,makecell,cellspace,rotating,diagbox,booktabs,adjustbox,longtable}
\usepackage{pdflscape}
\usepackage[table]{xcolor}

\usepackage{enumitem}
\usepackage{url}
\usepackage[T1]{fontenc}
\usepackage{lmodern}
\usepackage[margin=0.85in]{geometry}
\usepackage[authoryear,round]{natbib}
\usepackage[colorlinks=true,linkcolor=blue,citecolor=blue,urlcolor=blue]{hyperref}
\usepackage{xurl}
\usepackage[nameinlink,noabbrev]{cleveref}

\newtheorem{theorem}{Theorem}
\newtheorem{definition}{Definition}
\newtheorem{remark}{Remark}
\newtheorem{lemma}{Lemma}
\newtheorem{assumption}{Assumption}
\newtheorem{prop}{Proposition}

\numberwithin{equation}{section}
\allowdisplaybreaks
\title{Sparse Convoluted Smooth Hybrid Quantile-Expectile Regression in High Dimensions}
\author{Abdellah Atanane$^{1,2}$\thanks{Corresponding author: \texttt{atanane.abdellah@courrier.uqam.ca}}\quad Abdallah Mkhadri$^{2,\dagger}$\quad Karim Oualkacha$^{1,\dagger}$\\[6pt]
\small $^1$Department of Mathematics, Université du Québec à Montréal, Canada\\
\small 201 avenue Président-Kennedy, Montreal, H2X 3Y7, Quebec, Canada\\
\small $^2$Department of Mathematics, Cadi Ayyad University, Morocco\\
\small Abdelkrim Khattabi Avenue, Marrakesh, 40000, Marrakesh-Safi, Morocco\\[4pt]
\small $^\dagger$These authors contributed equally to this work.}
\date{}
\begin{document}
\maketitle
\begin{abstract}
We recently proposed hybrid quantile--expectile regression (HQER), which combines quantile regression and expectile regression through a convex combination of their loss functions. Although HQER provides a flexible framework for asymmetric regression, its objective function is not differentiable and does not incorporate variable selection, which limits its use in high-dimensional settings. To address these issues, we introduce smooth hybrid quantile--expectile regression (SHQER), based on convolution-type kernel smoothing of the quantile loss, and combine it with Lasso-type regularization. We also develop a unified penalized SHQER algorithm that accommodates a broad class of convex penalties, including the Lasso and Elastic Net. The algorithm is based on a local adaptive majorize--minimization principle and allows arbitrary smooth kernels. We establish theoretical properties of the L1-penalized SHQER estimator, including existence, uniqueness, consistency, and non-asymptotic high-probability error bounds. These bounds match the classical rates obtained for L1-penalized quantile regression under comparable assumptions. Numerical experiments show that SHQER is competitive with quantile, expectile, and related penalized approaches in estimation accuracy and support recovery across a range of sparse high-dimensional settings. An application to gene expression data illustrates its practical usefulness.
\end{abstract}


\section{Introduction}

\noindent In modern statistical learning, sparse linear modeling has become a central tool for analyzing data sets in which the number of covariates may be large, possibly much larger than the sample size. Classical penalized least-squares procedures are primarily designed to recover the conditional mean structure of the response while enforcing sparsity through suitable regularization. Although this paradigm has proved highly effective in many applications, the conditional mean may provide only a partial description of the response distribution. In several practical contexts, including economics, finance, biomedical studies, and genomic data analysis, covariate effects may change substantially across the conditional distribution, and the error distribution may be asymmetric, heteroscedastic, or heavy-tailed. For example, in the Job Training Partnership Act study analyzed by \citet{Abadie2002}, the impact of the program was more visible in the lower part of the earnings distribution than at the average level. Such examples motivate regression methods that go beyond the conditional mean and target distributional features such as conditional quantiles and conditional expectiles.

\noindent Quantile regression (QR), introduced by \citet{koenker_regression_1978} and further developed in \citet{Koenker2005} and \citet{Koenker2017}, provides a direct way to study conditional quantile functions and is particularly attractive when the interest lies in heterogeneous effects across the response distribution. Expectile regression (ER), introduced by \citet{newey_asymmetric_1987}, offers a related distributional regression framework based on asymmetric least squares; see also \citet{GuZou2016}. While quantiles are naturally interpreted through the inverse cumulative distribution function, expectiles are defined through asymmetric quadratic loss functions and can be viewed as alternative location-type summaries of the conditional distribution. As shown by \citet{Jones1994}, expectiles can also be related to quantiles of a suitably transformed distribution. From a computational and inferential point of view, ER has the advantage of relying on a smooth loss function, which simplifies numerical optimization and facilitates covariance estimation without requiring direct nonparametric estimation of conditional densities.

\noindent Expectile-based methods have attracted increasing attention in several applied areas. They have been used in risk management and financial econometrics \citep{taylor_estimating_2008,kuan_assessing_2009,Xie2014,BelliniBernardino2017,daouia2018estimation}, in the analysis of inflation dynamics \citep{Busetti2021}, and in studies involving life expectancy and economic production \citep{schnabel2009}. In financial risk analysis, expectiles are closely connected to the notion of expectile value-at-risk and provide a coherent way to quantify tail risk under suitable conditions \citep{BelliniBernardino2017,GuZou2019}. Their interpretation as risk measures has also been studied in \citet{kuan_assessing_2009}, \citet{BelliniBignozzi2015}, and \citet{Ziegel2016}. Nevertheless, the quadratic nature of the expectile loss may reduce robustness in the presence of outliers or heavy-tailed errors. In contrast, quantile regression is typically more robust, but its non-smooth loss may create computational and theoretical difficulties, especially in high-dimensional penalized settings.

\noindent To combine the robustness of QR with the efficiency and smoothness advantages of ER, we recently introduced in \citet{atanane2025efficient} a Hybrid Quantile--Expectile Regression (HQER) framework. The HQER loss is constructed as a convex combination of the quantile and expectile losses, governed by a mixing parameter \( \gamma \in [0,1] \). When \( \gamma=0 \), the model reduces to QR; when \( \gamma=1 \), it coincides with ER; and for intermediate values of \( \gamma \), it produces a continuum of hybrid estimators. This formulation allows the practitioner to adjust the balance between robustness and efficiency according to the degree of asymmetry, tail heaviness, and noise contamination in the data. The initial HQER framework in \citet{atanane2025efficient} focused mainly on estimation and theoretical properties in low-dimensional or moderate-dimensional settings.

\noindent Related attempts to bridge QR and ER include the use of $k$th power asymmetric loss functions \citep{efron1991regression}, which provide a smooth transition between quantile-type behavior when \(k=1\) and expectile-type behavior when \(k=2\). The theoretical properties of such $k$th power expectile models have been studied by \citet{jiang2021k} and \citet{lin2022k}, while high-dimensional penalized versions have been investigated by \citet{zha2021}. More generally, hybrid loss constructions have been used to combine the advantages of different convex criteria and to improve stability under asymmetric or heavy-tailed designs. Examples include penalized composite quasi-likelihood methods \citep{bradic2011penalized} and systemic risk modeling procedures \citep{fan2018single}. These developments show that carefully designed hybrid losses can provide both modeling flexibility and improved statistical performance.

\noindent Despite its flexibility, the original HQER criterion still inherits the non-differentiability of the quantile component. This non-smoothness can complicate both the optimization procedure and the derivation of sharp high-dimensional theory. To overcome this difficulty, we introduce in this paper a smoothed version of HQER, called Smoothed Hybrid Quantile--Expectile Regression (SHQER). The proposed SHQER loss is obtained by applying a convolution-type smoothing mechanism to the hybrid loss. This construction yields a differentiable surrogate objective while preserving the distributional interpretation and robustness-efficiency trade-off of the original HQER model.

\noindent Smoothing ideas in quantile regression have a long history, going back to \citet{Horowitz1998}. More recently, \citet{fernandes2021smoothing} developed a convex and twice-differentiable framework for smooth quantile regression. Further advances by \citet{He2021} and \citet{Tan2022} demonstrated that convolution smoothing can lead to computationally efficient and theoretically tractable estimators in both low-dimensional and high-dimensional regimes. Building on these contributions, our work extends the smoothing principle from the quantile loss to the hybrid quantile--expectile loss. This extension leads to a differentiable SHQER objective that is suitable for sparse estimation with modern first-order optimization methods.

\noindent We further incorporate a Lasso-type penalty into the SHQER objective in order to perform variable selection in high-dimensional models. For computation, we develop a unified penalized SHQER algorithm based on the Local Adaptive Majorize--Minimization (LAMM) principle \citep{Fan2018}. At each iteration, the algorithm replaces the smoothed objective by a local quadratic majorant and then solves a regularized surrogate problem. This strategy accommodates several convex penalties, including the Lasso \citep{Tibshirani1996} and the elastic net \citep{ZouHastie2005}. The resulting procedure can be viewed as a generalization of the Iterative Shrinkage--Thresholding Algorithm (ISTA) \citep{BeckTeboulle2009}, while being well adapted to the smoothed hybrid loss considered here. Because each LAMM iteration reduces to a proximal thresholding update with adaptive backtracking, the resulting procedure provides a simple first-order implementation for high-dimensional penalized estimation. We do not claim solver-level superiority over interior-point, coordinate-descent, or ADMM implementations; a systematic runtime benchmark across software platforms is beyond the scope of the present paper.

\noindent On the theoretical side, we establish rigorous guarantees for the proposed $L_{1}$-penalized SHQER estimator. In particular, we prove existence and uniqueness of the estimator under suitable conditions, derive consistency results, and obtain non-asymptotic estimation error bounds. We also provide high-probability bounds showing that the proposed penalized SHQER estimator attains the same order of estimation error as the classical $L_{1}$-penalized quantile regression estimator under comparable assumptions. These results show that smoothing the hybrid loss does not degrade the statistical rate under the stated assumptions, while making gradient-based optimization directly applicable.

\noindent The main contributions of this paper are summarized as follows.  
(i) We introduce SHQER, a smoothed version of the Hybrid Quantile--Expectile Regression framework, which provides a differentiable loss function for high-dimensional distributional regression.  
(ii) We formulate a sparse SHQER estimator by adding an $L_{1}$-type regularization term to the smoothed hybrid objective, thereby enabling variable selection in high-dimensional settings.  
(iii) We develop a unified optimization algorithm based on the Local Adaptive Majorize--Minimization (LAMM) principle. The proposed algorithm uses local quadratic majorization and supports several convex penalties, including the Lasso and the elastic net.  
(iv) We establish theoretical guarantees for the proposed estimator, including existence, uniqueness, consistency, and non-asymptotic error bounds. We also prove high-probability estimation bounds matching the order obtained for classical $L_{1}$-penalized quantile regression under analogous assumptions.  
(v) We assess the finite-sample performance of SHQER through extensive simulation studies and illustrate its usefulness on a real gene expression data application.  
(vi) We provide an open-source implementation of the proposed methodology in the R package \texttt{shqer}, available at \url{https://github.com/abdellah-atanane/shqer}, to facilitate reproducibility and practical use.

\noindent The remainder of the paper is organized as follows. Section \ref{section 2.1} introduces the HQER and SHQER frameworks, together with the convolution-type smoothing methodology. Section \ref{section 2.2} presents the LAMM algorithm for computing sparse SHQER estimators. Section \ref{section 3} establishes the theoretical properties of the proposed estimators in high-dimensional settings. Section \ref{section 4} reports the simulation results and the real-data application. Section \ref{section conclusion} concludes the paper. All technical details and proofs are collected in the Appendix.

\noindent\textbf{Notations:} Throughout this paper, bold uppercase letters denote matrices and bold lowercase letters denote vectors. For $p \ge 2$, $\mathbb{I}_p$ denotes the $p \times p$ identity matrix and $\mathbb{S}^{p-1}=\{\boldsymbol{u}\in\mathbb{R}^p:\|\boldsymbol{u}\|_2=1\}$ denotes the unit sphere in $\mathbb{R}^p$. For any symmetric positive semidefinite matrix $\boldsymbol{A}\in\mathbb{R}^{p\times p}$, let $\mu(\boldsymbol{A})=(\mu_1(\boldsymbol{A}),\ldots,\mu_p(\boldsymbol{A}))$ be the vector of its eigenvalues arranged in decreasing order, so that $\|\boldsymbol{A}\|_2=\mu_1(\boldsymbol{A})$. For any $\boldsymbol{u}\in\mathbb{R}^p$, we write $\|\boldsymbol{u}\|_{\boldsymbol{A}}=\|\boldsymbol{A}^{1/2}\boldsymbol{u}\|_2$. For $s,t\in\mathbb{R}$, we use $s\vee t=\max(s,t)$ and $s\wedge t=\min(s,t)$. For two sequences of nonnegative numbers $\{a_n\}_{n\ge1}$ and $\{b_n\}_{n\ge1}$, the notation $a_n\gtrsim b_n$ means that there exists a constant $C>0$, independent of $n$, such that $a_n\ge Cb_n$. Similarly, $a_n\lesssim b_n$ means that there exists a constant $C'>0$, independent of $n$, such that $a_n\le C'b_n$, and $a_n\asymp b_n$ means that both inequalities hold.
\section{Methodology}\label{section 2}
\noindent In Section \ref{section 2.1}, we review HQER and introduce its smoothed version (SHQER) via convolution smoothing of the quantile loss. We also present a sparse SHQER formulation for high-dimensional settings. Section \ref{section 2.2} then develops a LAMM algorithm for computing the resulting penalized estimators.

\subsection{Smooth HQER via Convolution Smoothing}\label{section 2.1}

\noindent Let $y$ denote a real-valued response and let $\boldsymbol{x}$ denote a vector of $p$ covariates, with $\boldsymbol{x}\in\mathbb{R}^p$. For a fixed quantile level $\tau\in(0,1)$, suppose that the $\tau$th conditional quantile of $y$ given $\boldsymbol{x}$ follows the linear model
\(
y_i=\boldsymbol{x}_i^\top\boldsymbol{\beta}^*(\tau)+\varepsilon_i,\, \text{with }\xi_\tau^\gamma(\varepsilon_i\mid \boldsymbol{x}_i)=0,
\)
where $\boldsymbol{\beta}^*(\tau)=(\beta_1^*(\tau),\ldots,\beta_p^*(\tau))^\top\in\mathbb{R}^p$ and, for $\varepsilon_i\in\mathbb{R}$, $\xi_\tau^\gamma(\varepsilon_i\mid \boldsymbol{x}_i):=\arg\min_{\theta\in\mathbb{R}}\mathbb{E}\!\left[\mathcal{C}_\tau^\gamma(\varepsilon_i-\theta)-\mathcal{C}_\tau^\gamma(\varepsilon_i)\mid \boldsymbol{x}_i\right]$. The hybrid loss $\mathcal{C}_\tau^\gamma(\cdot)$ proposed in \cite{atanane2025efficient} is defined by
\begin{equation}\label{unsmoothed loss}
\mathcal{C}_\tau^\gamma(s)=(1-\gamma)\rho_\tau(s)+\gamma \ell_\tau(s),\, \tau\in(0,1),\ \gamma\in[0,1],
\end{equation}
where $\rho_\tau(s)=\tfrac{|s|}{2}+(\tau-\tfrac12)s$ is the quantile loss and $\ell_\tau(s)=\tau s_+^2+(1-\tau)s_-^2$ is the asymmetric squared loss, with $s_+=s\vee0$ and $s_-=(-s)\vee 0$. For notational simplicity, we set $x_{i1}\equiv 1$ so that $\beta_1^*$ is the intercept, and we henceforth suppress the dependence on $\tau$, writing simply $\boldsymbol{\beta}^*$. Given an i.i.d. sample $\{(y_i,\boldsymbol{x}_i)\}_{i=1}^n$, the HQER estimator \citep{atanane2025efficient} is defined as
\begin{equation}\label{hybrid model}
\min_{\boldsymbol{\beta}\in\mathbb{R}^p}\frac{1}{n}\sum_{i=1}^n\left\{\mathcal{C}_\tau^\gamma(y_i-\boldsymbol{x}_i^\top\boldsymbol{\beta})-\mathcal{C}_\tau^\gamma(y_i)\right\}.
\end{equation}
Although $\mathcal{C}_\tau^\gamma$ is convex, its lack of differentiability at $0$ through $\rho_\tau$ limits the efficiency of gradient-based optimization.

\begin{figure}[htbp]
\centering
\includegraphics[width=\textwidth]{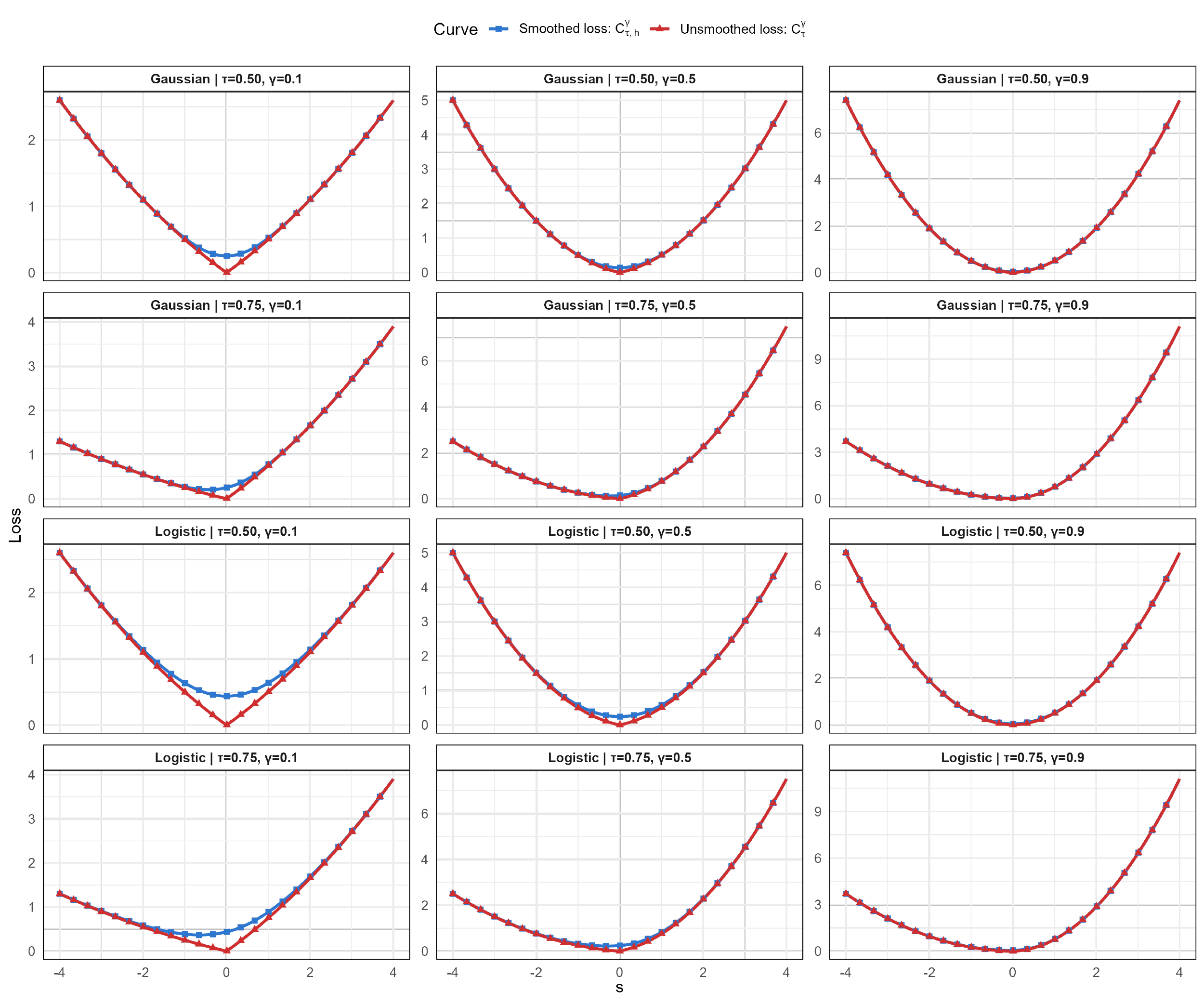}
\caption{Visualization of the HQER loss $\mathcal{C}_\tau^\gamma$ in \eqref{unsmoothed loss} and the smoothed HQER loss $\mathcal{C}_{\tau,h}^\gamma$ in \eqref{smoothed loss} for $\tau\in\{0.50,0.75\}$ and $\gamma\in\{0.1,0.5,0.9\}$ under Gaussian and logistic kernels, with bandwidth $h=0.7$.}
\label{fig:hybrid-loss-grid-wide}
\end{figure}

\noindent Subgradient methods may converge slowly \citep{man2024unified}, whereas linear programming and interior-point methods for nonsmooth problems often entail roughly cubic complexity in $p$ \citep{PortnoyKoenker1997}. In high-dimensional settings, sparsity is usually imposed. For instance, \citet{BelloniChernozhukov2011} studied the Lasso-penalized QR estimator \citep{Tibshirani1996}. Its hybrid counterpart is the $L_1$-penalized HQER estimator
\(
\min_{\boldsymbol{\beta}\in\mathbb{R}^p}\frac{1}{n}\sum_{i=1}^n\left\{\mathcal{C}_\tau^\gamma(y_i-\boldsymbol{x}_i^\top\boldsymbol{\beta})-\mathcal{C}_\tau^\gamma(y_i)\right\}+\lambda\|\boldsymbol{\beta}\|_1,
\)
with tuning parameter $\lambda>0$. More generally, adaptive and folded-concave penalties \citep{Wang2012,Zheng2015} lead to the weighted formulation
\begin{equation}\label{eq:genlasso}
\min_{\boldsymbol{\beta}\in\mathbb{R}^p}\frac{1}{n}\sum_{i=1}^n\left\{\mathcal{C}_\tau^\gamma(y_i-\boldsymbol{x}_i^\top\boldsymbol{\beta})-\mathcal{C}_\tau^\gamma(y_i)\right\}+\sum_{l=1}^p\lambda_l|\beta_l|,
\end{equation}
where $\lambda_l\ge0$ are data-dependent weights. To address the nonsmoothness of $\rho_\tau$, \citet{fernandes2021smoothing} proposed convolution smoothing, which yields a twice-differentiable, convex, and locally strongly convex surrogate. Extending this idea to the hybrid loss gives
\begin{equation}\label{smoothed loss}
\mathcal{C}_{\tau,h}^\gamma(s)=(1-\gamma)\rho_{\tau,h}(s)+\gamma \ell_\tau(s)=\frac{1-\gamma}{h}\int_{-\infty}^{\infty}\rho_\tau(v)\,k\!\left(\frac{v-s}{h}\right)\,dv+\gamma \ell_\tau(s),
\end{equation}
where $\rho_{\tau,h}=\rho_\tau*k_h$, $k_h(s)=h^{-1}k(s/h)$, and $*$ denotes convolution. This smoothing inherits differentiability from the kernel $k$, enabling efficient first-order optimization while preserving the asymmetry indexed by $(\tau,\gamma)$. Replacing $\mathcal{C}_\tau^\gamma$ with $\mathcal{C}_{\tau,h}^\gamma$ in \eqref{eq:genlasso} yields a penalized SHQER estimator. Common choices of $k$ include the uniform, Gaussian, logistic, Epanechnikov, and triangular kernels; see \citet{He2021,fernandes2021smoothing}. In low dimensions, \citet{He2021} showed that smoothed QR is asymptotically equivalent to its nonsmoothed version under weaker assumptions. In high dimensions, \citet{Tan2022} proved that, for $\gamma=0$, the smoothed Lasso estimator achieves the same convergence rate as its nonsmoothed counterpart, with additional improvements under iteratively reweighted regularization. The pure expectile case $\gamma=1$ has also been studied separately; see, for example, \citet{liao2019penalized,retire2024}. The effect of smoothing on HQER and SHQER for several values of $(\tau,\gamma)$ under Gaussian and logistic kernels is illustrated in Figure~\ref{fig:hybrid-loss-grid-wide}.
In what follows, we focus on the genuinely hybrid regime $\gamma\in(0,1)$.

\begin{remark}\label{rem:2.1}
\noindent Let $k$ be a kernel and $h>0$ a bandwidth. Define
\[
\widehat{\mathfrak R}_h^\gamma(\boldsymbol{\beta})=\frac{1}{n}\sum_{i=1}^n\left\{\mathcal{C}_{\tau,h}^\gamma(y_i-\boldsymbol{x}_i^\top\boldsymbol{\beta})-\mathcal{C}_{\tau,h}^\gamma(y_i)\right\}
\]. Two explicit forms of $\rho_{\tau,h}$ are particularly useful. {(i) Gaussian kernel.} If $k(s)=\varphi(s)$ is the $\mathcal N(0,1)$ density, then $\rho_{\tau,h}(s)=\frac12\,\mathbb E|T|+(\tau-\tfrac12)s$, where $T\sim\mathcal N(s,h^2)$. Since $|T|$ is folded normal, $\mathbb E|T|=\sqrt{\frac{2}{\pi}}\,h\,e^{-s^2/(2h^2)}+s\{1-2\Phi(-s/h)\}$, and therefore
\(
\rho_{\tau,h}(s)=\frac{h}{2}\,G\!\left(\frac{s}{h}\right)+\Bigl(\tau-\tfrac12\Bigr)s,
\)
where $G(w)=\sqrt{\frac{2}{\pi}}\,e^{-w^2/2}+w\{1-2\Phi(-w)\}$.
{(ii) Logistic kernel.} If $k(s)=\dfrac{e^{-s}}{(1+e^{-s})^2}$, then $\rho_{\tau,h}(s)=\tau s+h\log(1+e^{-s/h})$.
\end{remark}

\noindent The $L_1$-penalized SHQER estimator is defined by
\begin{equation}\label{eq:sqrlasso}
\min_{\boldsymbol{\beta}\in\mathbb{R}^p}\frac{1}{n}\sum_{i=1}^n\left\{\mathcal{C}_{\tau,h}^\gamma(y_i-\boldsymbol{x}_i^\top\boldsymbol{\beta})-\mathcal{C}_{\tau,h}^\gamma(y_i)\right\}+\lambda\|\boldsymbol{\beta}\|_1.
\end{equation}
For penalized smoothed quantile regression, \citet{Tan2022} proposed coordinate descent and ADMM algorithms tailored to the uniform and Gaussian kernels. Although effective in those settings, such methods remain limited both in kernel choice and in the class of admissible penalties. Motivated by these limitations, we develop in the next subsection a generic optimization procedure for penalized SHQER. Our method accommodates any nonnegative smoothing kernel and a broad family of convex penalties, including the Lasso and Elastic Net. Based on the Local Adaptive Majorize--Minimization (LAMM) framework, it yields a unified proximal optimization strategy in the spirit of \citet{man2024unified}.

\subsection{Local Adaptive Majorize--Minimization Algorithm for Penalized SHQER}\label{section 2.2}
\noindent We now present an optimization framework for penalized SHQER. The general problem is
\(
\min_{\boldsymbol{\beta}\in\mathbb{R}^p}\widehat{\mathcal{H}}_h^\gamma(\boldsymbol{\beta})+p_\lambda(\boldsymbol{\beta}),
\)
where $\widehat{\mathcal{H}}_h^\gamma(\boldsymbol{\beta})=\frac{1}{n}\sum_{i=1}^n\big[\mathcal{C}_{\tau,h}^\gamma(y_i-\boldsymbol{x}_i^\top\boldsymbol{\beta})-\mathcal{C}_{\tau,h}^\gamma(y_i)\big]$ and $p_\lambda(\boldsymbol{\beta})$ is a convex penalty promoting sparsity or shrinkage. Two important examples are the weighted Lasso, $p_\lambda(\boldsymbol{\beta})=\sum_{l=1}^p\lambda_l|\beta_l|$ with $\lambda_l\ge0$, and the Elastic Net, $p_\lambda(\boldsymbol{\beta})=\lambda\alpha\|\boldsymbol{\beta}\|_1+\lambda(1-\alpha)\|\boldsymbol{\beta}\|_2^2$, with $\lambda>0$ and $\alpha\in[0,1]$. Our approach accommodates arbitrary nonnegative kernels and a broad class of convex penalties. It is based on the Local Adaptive Majorize--Minimization (LAMM) framework \citep{Fan2018}, an adaptive version of the classical Majorize--Minimization principle \citep{Lange2000,HunterLange2004}. Let $\nabla\widehat{\mathcal{H}}_h^\gamma(\boldsymbol{\beta})$ denote the gradient of the smoothed loss. At iteration $t$, given $\boldsymbol{\beta}^{(t-1)}$, we approximate $\widehat{\mathcal{H}}_h^\gamma$ locally by the quadratic majorant
\(
\mathcal{G}(\boldsymbol{\beta}\mid\upsilon_t,\boldsymbol{\beta}^{(t-1)})=\widehat{\mathcal{H}}_h^\gamma(\boldsymbol{\beta}^{(t-1)})+\left\langle\nabla\widehat{\mathcal{H}}_h^\gamma(\boldsymbol{\beta}^{(t-1)}),\boldsymbol{\beta}-\boldsymbol{\beta}^{(t-1)}\right\rangle+\frac{\upsilon_t}{2}\|\boldsymbol{\beta}-\boldsymbol{\beta}^{(t-1)}\|_2^2,
\)
where $\upsilon_t>0$ is an adaptive curvature parameter. The update is then obtained from
\begin{equation}\label{eq:lamm_update}
\boldsymbol{\beta}^{(t)}=\arg\min_{\boldsymbol{\beta}\in\mathbb{R}^p}\mathcal{G}(\boldsymbol{\beta}\mid\upsilon_t,\boldsymbol{\beta}^{(t-1)})+p_\lambda(\boldsymbol{\beta}).
\end{equation}
To ensure descent of the objective, $\upsilon_t$ is increased geometrically, for example by $\upsilon_t\leftarrow\delta\upsilon_t$ with $\delta>1$, until $\widehat{\mathcal{H}}_h^\gamma(\boldsymbol{\beta}^{(t)})\le \mathcal{G}(\boldsymbol{\beta}^{(t)}\mid\upsilon_t,\boldsymbol{\beta}^{(t-1)})$. This guarantees monotone convergence:
\(
\widehat{\mathcal{H}}_h^\gamma(\boldsymbol{\beta}^{(t)})+p_\lambda(\boldsymbol{\beta}^{(t)})\le \widehat{\mathcal{H}}_h^\gamma(\boldsymbol{\beta}^{(t-1)})+p_\lambda(\boldsymbol{\beta}^{(t-1)}).
\)
For convex penalties, the minimization in \eqref{eq:lamm_update} admits a closed-form solution through thresholding operators. The first-order optimality condition is $\boldsymbol{0}\in\nabla\widehat{\mathcal{H}}_h^\gamma(\boldsymbol{\beta}^{(t-1)})+\upsilon_t(\boldsymbol{\beta}^{(t)}-\boldsymbol{\beta}^{(t-1)})+\partial p_\lambda(\boldsymbol{\beta})|_{\boldsymbol{\beta}=\boldsymbol{\beta}^{(t)}}$, where $\partial p_\lambda$ denotes the subdifferential. The intercept is usually left unpenalized and updated according to $\beta_1^{(t)}=\beta_1^{(t-1)}-\upsilon_t^{-1}\,\partial\widehat{\mathcal{H}}_h^\gamma(\boldsymbol{\beta}^{(t-1)})/\partial\beta_1$. Algorithm \ref{algorithm} summarizes the procedure.

\begin{algorithm}[H]
\caption{LAMM algorithm for penalized SHQER}\label{algorithm}
\textbf{Input:} kernel $k(\cdot)$, penalty $p_\lambda(\cdot)$, regularization parameters, bandwidth $h$, inflation factor $\delta=1.2$, and tolerance $\epsilon$.\\
\textbf{Initialization:} $\widehat{\boldsymbol{\beta}}^{\,0}\leftarrow\boldsymbol{0}$, $\upsilon_0\leftarrow0.01$.
\begin{algorithmic}[1]
\State Repeat until $\|\widehat{\boldsymbol{\beta}}^{\,t}-\widehat{\boldsymbol{\beta}}^{\,t-1}\|_2\le\epsilon$.
\State Set $\upsilon_t\leftarrow\max\{\upsilon_0,\upsilon_{t-1}/\delta\}$.
\Repeat
\For{$l=1,\dots,p$} \label{step:alg}
\State \textbf{Lasso:} $\widehat{\beta}_l^{\,t}\leftarrow \mathcal{S}\!\left(\widehat{\beta}_l^{\,t-1}-\upsilon_t^{-1}\nabla_{\beta_l}\widehat{\mathcal{H}}_h^\gamma(\widehat{\boldsymbol{\beta}}^{\,t-1}),\,\upsilon_t^{-1}\lambda_l\right)$.
\State \textbf{Elastic Net:} $\widehat{\beta}_l^{\,t}\leftarrow \dfrac{1}{1+2\upsilon_t^{-1}\lambda(1-\alpha)}\mathcal{S}\!\left(\widehat{\beta}_l^{\,t-1}-\upsilon_t^{-1}\nabla_{\beta_l}\widehat{\mathcal{H}}_h^\gamma(\widehat{\boldsymbol{\beta}}^{\,t-1}),\,\upsilon_t^{-1}\lambda\alpha\right)$.
\EndFor
\State If $\mathcal G(\widehat{\boldsymbol{\beta}}^{\,t}\mid\upsilon_t,\widehat{\boldsymbol{\beta}}^{\,t-1})<\widehat{\mathcal{H}}_h^\gamma(\widehat{\boldsymbol{\beta}}^{\,t})$, set $\upsilon_t\leftarrow\delta\upsilon_t$.
\Until{$\mathcal G(\widehat{\boldsymbol{\beta}}^{\,t}\mid\upsilon_t,\widehat{\boldsymbol{\beta}}^{\,t-1})\ge \widehat{\mathcal{H}}_h^\gamma(\widehat{\boldsymbol{\beta}}^{\,t})$}
\State \textbf{Output:} $\widehat{\boldsymbol{\beta}}^{\,t}$.
\end{algorithmic}
\end{algorithm}

\noindent The LAMM algorithm therefore provides a unified optimization strategy for SHQER that is flexible with respect to the kernel and penalty choice. It extends iterative shrinkage--thresholding and majorization--minimization ideas while preserving the monotone-descent property enforced by backtracking. We now turn to the statistical properties of the proposed estimator.
\section{Statistical Theory}\label{section 3}
\noindent We first quantify the deterministic bias induced by convolution smoothing in Section \ref{section 4.1}. Section \ref{section 4.2} then derives high-probability error bounds under the $L_1$- and $L_2$-norms for the $L_1$-penalized smoothed HQER estimator $\boldsymbol{\widehat{\beta}}_h$. With a suitable choice of bandwidth $h$, the resulting rates are near-minimax optimal \citep{WANG2021107262} and coincide with those of the classical $L_1$-penalized quantile regression estimator when $\gamma=0$ \citep{BelloniChernozhukov2011}. Throughout, we write $\lesssim$ for inequalities up to absolute constants independent of $(q,p,n)$. Without loss of generality, we may assume that the covariates are centered, that is, $m_l:=\mathbb{E}(x_l)=0$ for $2\le l\le p$. Otherwise, applying the change of variable $(\beta_1,\beta_2,\ldots,\beta_p)^\top\mapsto(\beta_1+\sum_{l=2}^p m_l\beta_l,\beta_2,\ldots,\beta_p)^\top$ shows that the subsequent results apply to the hybrid conditional quantile model $\xi_{y\mid\boldsymbol{x}}(\tau)=\beta_0^*+\sum_{l=2}^p(x_l-m_l)\beta_l^*$, where $\beta_0^*=\beta_1^*+\sum_{l=2}^p m_l\beta_l^*$.

\subsection{Smoothing Bias}\label{section 4.1}
\noindent Recall the smoothed hybrid loss in \eqref{smoothed loss} and define $e_i(\boldsymbol{\beta})=y_i-\boldsymbol{x}_i^\top\boldsymbol{\beta}$. Let $k$ denote the kernel density, $\bar k(s)=\int_{-\infty}^s k(t)\,dt$ its CDF, and let $k_h(s)=h^{-1}k(s/h)$ and $\bar k_h(s)=\bar k(s/h)$ be the rescaled versions. The empirical smoothed risk is
\(
\widehat{\mathcal{H}}_h^\gamma(\boldsymbol{\beta})=\frac{1}{n}\sum_{i=1}^n\Big\{\mathcal{C}_{\tau,h}^\gamma(e_i(\boldsymbol{\beta}))-\mathcal{C}_{\tau,h}^\gamma(y_i)\Big\}.
\)
Its gradient and generalized Hessian are
\begin{equation}\label{gradient-hessian}
\begin{aligned}
\nabla\widehat{\mathcal{H}}_h^\gamma(\boldsymbol{\beta})
&=\frac{1}{n}\sum_{i=1}^n
\Big[(1-\gamma)\{\bar k(-e_i(\boldsymbol{\beta})/h)-\tau\} \\
&\hspace{4.2cm}-\gamma\psi_\tau(e_i(\boldsymbol{\beta}))\Big]\boldsymbol{x}_i,\\
\nabla^2\widehat{\mathcal{H}}_h^\gamma(\boldsymbol{\beta})
&=\frac{1}{n}\sum_{i=1}^n
\Big[(1-\gamma)k_h(-e_i(\boldsymbol{\beta})) \\
&\hspace{4.2cm}+2\gamma \omega_\tau(e_i(\boldsymbol{\beta}))\Big]
\boldsymbol{x}_i\boldsymbol{x}_i^\top.
\end{aligned}
\end{equation}
where $\psi_\tau(s)=2\omega_\tau(s)s$ and $\omega_\tau(s)=\tau\,\mathds{1}\{s\ge0\}+(1-\tau)\mathds{1}\{s<0\}$. Let
\begin{equation}\label{risk loss}
\mathcal{H}_h^\gamma(\boldsymbol{\beta})=\mathbb{E}\big[\widehat{\mathcal{H}}_h^\gamma(\boldsymbol{\beta})\big],\, \boldsymbol{\beta}\in\mathbb{R}^p,
\end{equation}
and define
\begin{equation}\label{optim smouthed}
\boldsymbol{\beta}_h^*\in\arg\min_{\boldsymbol{\beta}\in\mathbb{R}^p}\mathcal{H}_h^\gamma(\boldsymbol{\beta}).
\end{equation}
In general, $\boldsymbol{\beta}_h^*\neq\boldsymbol{\beta}^*$, and the quantity $\|\boldsymbol{\beta}_h^*-\boldsymbol{\beta}^*\|$ measures the smoothing bias. We assume throughout that $x_{i1}\equiv1$ and that $\boldsymbol{\Sigma}=\mathbb{E}(\boldsymbol{x}\boldsymbol{x}^\top)$ is positive definite, with eigenvalues bounded away from $0$ and $\infty$. Let $\mu_1=\mu_1(\boldsymbol{\Sigma})\ge1$, $\mu_p=\mu_p(\boldsymbol{\Sigma})\in(0,1]$, and $\sigma_x=\max_{1\le l\le p}\sigma_{ll}$.

\begin{assumption}\label{A1}
The conditional density $f_{\varepsilon\mid\boldsymbol{x}}$ exists and satisfies $0<f_{\min}\le f_{\varepsilon\mid\boldsymbol{x}}(0)\le f_{\max}<\infty$ almost surely. Moreover, $f_{\varepsilon\mid\boldsymbol{x}}$ is Lipschitz, that is, $|f_{\varepsilon\mid\boldsymbol{x}}(s)-f_{\varepsilon\mid\boldsymbol{x}}(t)|\le L_f|s-t|$ for all $s,t\in\mathbb{R}$ almost surely.
\end{assumption}

\begin{assumption}\label{A2}
The kernel density $k$ is symmetric, satisfies $\int k(s)\,ds=1$, has finite second moment $\int s^2k(s)\,ds<\infty$, and finite absolute moments $\kappa_l=\int|s|^lk(s)\,ds$ for all $l\in\mathbb{N}^*$.
\end{assumption}

\begin{prop}[Bias bound]\label{proposition 1}
Suppose Assumptions~\ref{A1}--\ref{A2} hold. Let $\tau_{\max}=\max\{\tau,1-\tau\}$, $\tau_{\min}=\min\{\tau,1-\tau\}$, $\boldsymbol{z}=\boldsymbol{\Sigma}^{-1/2}\boldsymbol{x}$, and assume $\mathfrak{m}_3:=\sup_{\boldsymbol{\delta}\in\mathbb{S}^{p-1}}\mathbb{E}|\boldsymbol{z}^\top\boldsymbol{\delta}|^3<\infty$ and $\nu_\varepsilon:=\sup_{\boldsymbol{x}}\mathbb{E}(|\varepsilon|\mid\boldsymbol{x})<\infty$. Define
\begin{equation}\label{eq:prop1-constants}
\begin{aligned}
D_\gamma&=(1-\gamma)f_{\min}+2\gamma\tau_{\min},\qquad
\mathfrak{c}_\gamma=(1-\gamma)L_f,\\
B_\gamma&=2\gamma\tau_{\max}\nu_\varepsilon,\qquad
A_\gamma=\mathfrak{c}_\gamma\left(\frac{\kappa_2+\mathfrak{m}_3}{2}+\kappa_1\right).
\end{aligned}
\end{equation}
Then $\boldsymbol{\beta}_h^*$ is unique whenever $h>0$ satisfies
\begin{equation}\label{eq:prop1-h-range}
\frac{A_\gamma h^2+B_\gamma}{D_\gamma}<h<\frac{D_\gamma}{\mathfrak{c}_\gamma(\kappa_1+\tfrac12\mathfrak{m}_3)}.
\end{equation}
Moreover,
\begin{equation}\label{eq:prop1-M0}
\begin{aligned}
\|\boldsymbol{\beta}_h^*-\boldsymbol{\beta}^*\|_{\boldsymbol{\Sigma}}
&\le \mathfrak{M}_0:=\frac{A_\gamma h^2+B_\gamma}{D_\gamma}\\
&=\frac{\mathfrak{c}_\gamma\left(\frac{\kappa_2+\mathfrak{m}_3}{2}+\kappa_1\right)h^2+2\gamma\tau_{\max}\nu_\varepsilon}{(1-\gamma)f_{\min}+2\gamma\tau_{\min}}.
\end{aligned}
\end{equation}
If, in addition, $\kappa_3<\infty$ and $f'_{\varepsilon\mid\boldsymbol{x}}$ is $L_{f'}$-Lipschitz at $0$, then
\begin{equation}\label{eq:lin_bound_thm}
\Bigl\|\boldsymbol{\Sigma}^{-1}\mathcal{H}(\boldsymbol{\beta}_h^*-\boldsymbol{\beta}^*)+\frac{1-\gamma}{2}\kappa_2 h^2\,\boldsymbol{\Sigma}^{-1}\mathbb{E}[f'_{\varepsilon\mid\boldsymbol{x}}(0)\boldsymbol{x}]+\gamma\,\boldsymbol{\Sigma}^{-1}\mathbb{E}[\psi_\tau(\varepsilon)\boldsymbol{x}]\Bigr\|_{\boldsymbol{\Sigma}}\le \mathfrak{M}_1,
\end{equation}
where
\begin{equation}\label{eq:prop1-M1}
\begin{aligned}
\mathfrak{M}_1&:=\mathfrak{M}+\mathfrak{c}_\gamma\Big(\frac{\mathfrak{m}_3}{2}\mathfrak{M}_0+\kappa_1 h+2\gamma\tau_{\max}\nu_\varepsilon\Big)\mathfrak{M}_0,\\
\mathfrak{M}&:=\frac{(1-\gamma)L_{f'}}{6}\kappa_3 h^3+2\gamma\tau_{\max}\nu_\varepsilon.
\end{aligned}
\end{equation}
\end{prop}

\begin{remark}
If $f_{\varepsilon\mid\boldsymbol{x}}(0)$ and $f'_{\varepsilon\mid\boldsymbol{x}}(0)$ are constant in $\boldsymbol{x}$ and $\mathbb{E}[\boldsymbol{x}]=(1,\boldsymbol{0}_{p-1}^\top)^\top$, then \eqref{eq:lin_bound_thm} reduces to
\(
\Biggl\|f_\varepsilon(0)(\boldsymbol{\beta}_h^*-\boldsymbol{\beta}^*)+\left(\frac{1-\gamma}{2}f'_\varepsilon(0)\kappa_2 h^2+\gamma\,\mathbb{E}[\psi_\tau(\varepsilon)]\right)\begin{bmatrix}1\\ \boldsymbol{0}_{p-1}\end{bmatrix}\Biggr\|_{\boldsymbol{\Sigma}}\le \mathfrak{M}_1.
\)
Hence, under homogeneity, both the kernel-induced and expectile-induced biases affect only the intercept, while the slope components remain unbiased up to second order.
\end{remark}
\noindent Proposition \ref{proposition 1} shows that, under mild smoothness conditions near the origin, the smoothing bias is of order $h^2$ and is intercept-dominant under homogeneity. The next assumption is useful to sharpen this result under independence between $\varepsilon$ and $\boldsymbol{x}$.

\begin{prop}\label{proposition 2}
Assume $\varepsilon\perp\!\!\!\perp\boldsymbol{x}$ and define
\begin{equation}\label{eq c}
\vartheta(\alpha)=\mathbb{E}\Big\{\mathcal{C}_{\tau,h}^\gamma(\varepsilon-\alpha)-\mathcal{C}_{\tau,h}^\gamma(\varepsilon)\Big\},\, \alpha\in\mathbb{R}.
\end{equation}
Then, for $(\tau,\gamma)\in(0,1)\times(0,1)$, the function $\vartheta$ admits a unique minimizer $b_h$.
\end{prop}

\begin{assumption}\label{A3}
The marginal density $f_\varepsilon$ satisfies $f_\varepsilon(0)>0$ and $|f_\varepsilon(s)-f_\varepsilon(t)|\le L_f|s-t|$ for all $s,t\in\mathbb{R}$.
\end{assumption}

\begin{prop}[Smoothed bias bound]\label{proposition 3}
Under Assumptions~\ref{A1}--\ref{A3}, the population objective $\mathcal{H}_h^\gamma(\cdot)$ admits a unique minimizer of the form
\(
\boldsymbol{\beta}_h^*=(\beta_1^*+b_h,\boldsymbol{\beta}_{-}^{*\top})^\top,
\)
where $b_h$ is the unique minimizer of \eqref{eq c} and $\boldsymbol{\beta}_{-}^{*}=(\beta_2^*,\ldots,\beta_p^*)^\top$. Let $\kappa_1=\int|s|k(s)\,ds$ and $\kappa_2=\int s^2k(s)\,ds$. Then, for all sufficiently small $h>0$,
\begin{equation}\label{eq:bh_bound}
|b_h|\le \frac{\mathfrak{c}_\gamma(\kappa_2+\kappa_1\kappa_2^{1/2})h^2+2\gamma\tau_{\max}\nu_\varepsilon}{(1-\gamma)f_\varepsilon(0)+2\gamma\tau_{\min}}.
\end{equation}
Consequently, $|b_h|=O(h^2)+O(\gamma)$ as $h\to0$, and in particular $b_h=O(h^2)$ for  $\gamma\to 0$.
\end{prop}
\begin{remark}
When $\gamma=0$, \eqref{eq:bh_bound} reduces to the pure smoothed quantile result of \citet{Tan2022}: $|b_h|\le L_f\,c_1\,\kappa_2^{1/2}f_\varepsilon(0)^{-1}h^2$, with $c_1=\kappa_1+\kappa_2^{1/2}$. When $\gamma>0$, the additional expectile component modifies the constants but preserves the $h^2$ bias order. As $\gamma\to1$, the kernel term vanishes and the loss approaches the pure expectile case.
\end{remark}
\noindent Proposition \ref{proposition 3} shows that, under independence between $\varepsilon$ and $\boldsymbol{x}$, smoothing affects only the intercept and therefore preserves the sparsity pattern of $\boldsymbol{\beta}^*$. This complements the high-dimensional theory for composite and smoothed quantile procedures \citep{zou2008composite,bradic2011penalized,fernandes2021smoothing}.
\subsection{\texorpdfstring{$L_1$}{L1}-penalized smoothed hybrid regression}\label{section 4.2}
\noindent To incorporate variable selection into SHQER, for given $h>0$ and $\lambda>0$, we consider
\begin{equation}\label{smoothed lasso}
\boldsymbol{\widehat{\beta}}_h\in\arg\min_{\boldsymbol{\beta}\in\mathbb{R}^p}\Big\{\widehat{\mathcal{H}}_h^\gamma(\boldsymbol{\beta})+\lambda\|\boldsymbol{\beta}\|_1\Big\},
\end{equation}
where $\widehat{\mathcal{H}}_h^\gamma(\boldsymbol{\beta})$ is the empirical convolution-smoothed hybrid loss introduced in Section \ref{section 2}. Our goal is to study $\|\boldsymbol{\widehat{\beta}}_h-\boldsymbol{\beta}^*\|_2$ and $\|\boldsymbol{\widehat{\beta}}_h-\boldsymbol{\beta}^*\|_1$. We assume that $x_{i1}\equiv1$ for $i=1,\ldots,n$ and that $m_l=\mathbb{E}[x_l]=0$ for $l\ge2$.

\begin{assumption}\label{A4}
Let $\boldsymbol{\Sigma}=\mathbb{E}(\boldsymbol{x}\boldsymbol{x}^\top)$ be positive definite and let $\boldsymbol{z}=\boldsymbol{\Sigma}^{-1/2}\boldsymbol{x}$ be sub-exponential: there exist constants $a_0,a_1\ge1$ such that
\(
\mathbb{P}\!\left(|\boldsymbol{z}^\top\boldsymbol{\delta}|\ge a_0\|\boldsymbol{\delta}\|_2\,s\right)\le a_1e^{-s},\, \forall\,\boldsymbol{\delta}\in\mathbb{R}^p,\ s\ge0.
\)
\end{assumption}
\noindent Assumption~\ref{A4} is a concentration condition used to obtain the present non-asymptotic high-dimensional bounds. It excludes genuinely heavy-tailed or leverage-prone covariate designs. Accordingly, the robustness studied in this paper concerns asymmetric, contaminated, or heavy-tailed \emph{response errors}; we do not claim robustness to high-leverage covariate outliers. Extending SHQER theory to heavy-tailed designs, for example through covariate truncation or robust covariance/score constructions, is an important direction for future work.
\begin{assumption}[Restricted eigenvalue and finite error variance]\label{A5}
Let $\widehat{\boldsymbol{\Sigma}}=\frac{1}{n}\sum_{i=1}^n\boldsymbol{x}_i\boldsymbol{x}_i^\top$ and fix $\varrho,\eta>0$. There exists $\kappa_{\mathrm{RE}}>0$ such that
\(
\inf_{\boldsymbol{\delta}\in\mathbb{C}_{\boldsymbol{\Sigma}}(\eta)\cap\mathbb{B}_{\boldsymbol{\Sigma}}(\varrho)\setminus\{0\}}\frac{\boldsymbol{\delta}^\top\widehat{\boldsymbol{\Sigma}}\boldsymbol{\delta}}{\|\boldsymbol{\delta}\|_{\boldsymbol{\Sigma}}^2}\ge \kappa_{\mathrm{RE}}.
\)
Moreover, there exists $V_\varepsilon<\infty$ such that $\mathbb{E}(\varepsilon^2\mid\boldsymbol{x})\le V_\varepsilon$ almost surely.
\end{assumption}
\noindent Assume $a_1=1$ and let $\sigma_x^2=\max_{1\le l\le p}\mathbb{E}(x_l^2)$. Define
\begin{equation}\label{eq;boule}
\mathbb{B}_{\boldsymbol{\Sigma}}(\varrho):=\{\boldsymbol{\delta}\in\mathbb{R}^p:\|\boldsymbol{\delta}\|_{\boldsymbol{\Sigma}}\le\varrho\},\, \mathbb{C}_{\boldsymbol{\Sigma}}(\eta):=\{\boldsymbol{\delta}\in\mathbb{R}^p:\|\boldsymbol{\delta}\|_1\le \eta\|\boldsymbol{\delta}\|_{\boldsymbol{\Sigma}}\}.
\end{equation}
Let $k=\bar k'$ denote the kernel density and assume $\kappa_0:=\inf_{|t|\le1}k(t)>0$, which holds for nonnegative compactly supported kernels after rescaling $h$ if necessary.
\begin{definition}[Restricted strong convexity]\label{def:RSC}
For $\varrho,\eta,\kappa>0$, define the event\\
\(
\mathcal{E}_{\mathrm{rsc}}(\varrho,\eta,\kappa,\gamma)=\left\{\frac{\langle\nabla\widehat{\mathcal{H}}_h^\gamma(\boldsymbol{\beta})-\nabla\widehat{\mathcal{H}}_h^\gamma(\boldsymbol{\beta}^*),\,\boldsymbol{\beta}-\boldsymbol{\beta}^*\rangle}{\|\boldsymbol{\beta}-\boldsymbol{\beta}^*\|_{\boldsymbol{\Sigma}}^2}\ge\kappa,\ \forall\,\boldsymbol{\beta}\in\boldsymbol{\beta}^*+\mathbb{B}_{\boldsymbol{\Sigma}}(\varrho)\cap\mathbb{C}_{\boldsymbol{\Sigma}}(\eta)\right\}.
\)
\end{definition}
\noindent The next result states that this local restricted strong convexity event holds with high probability.
\begin{theorem}[Local RSC for the smoothed hybrid loss]\label{theorem 1}
Assume Assumptions~\ref{A1}, \ref{A2}, \ref{A4}, and \ref{A5}, and let \(\kappa_0=\inf_{|t|\le1}k(t)>0\) and \(\tau_{\min}=\min\{\tau,1-\tau\}\). For fixed \(\varrho,\eta>0\), define
\(\Theta(\varrho,\eta)=\{\boldsymbol{\beta}\in\mathbb{R}^p:\boldsymbol{\beta}-\boldsymbol{\beta}^*\in\mathbb{B}_{\boldsymbol{\Sigma}}(\varrho)\cap\mathbb{C}_{\boldsymbol{\Sigma}}(\eta)\}\). If
\(
20a_0^2\varrho\le h\le \frac{f_{\min}}{2L_f},\, n\ge C\left[\sigma_x^2 f_{\max}f_{\min}^{-2}\left(\frac{\eta}{\varrho}\right)^2 h\log(2p)\right],
\) for a universal constant \(C>0\), then with probability at least \(1-(2p)^{-1}\),
\(\inf_{\boldsymbol{\beta}\in\Theta(\varrho,\eta)}\frac{\langle\nabla\widehat{\mathcal{H}}_h^\gamma(\boldsymbol{\beta})-\nabla\widehat{\mathcal{H}}_h^\gamma(\boldsymbol{\beta}^*),\,\boldsymbol{\beta}-\boldsymbol{\beta}^*\rangle}{\|\boldsymbol{\beta}-\boldsymbol{\beta}^*\|_{\boldsymbol{\Sigma}}^2}\ge \kappa\), where \(\kappa=\frac12(1-\gamma)\kappa_0f_{\min}+2\gamma\tau_{\min}\kappa_{\mathrm{RE}}\). Equivalently, the event \(\mathcal{E}_{\mathrm{rsc}}(\varrho,\eta,\kappa,\gamma)\) holds on \(\Theta(\varrho,\eta)\).
\end{theorem}
\noindent Let $\mathcal{Q}:=\{l\in[p]:\beta_l^*\neq0\}$ be the active set and $q:=|\mathcal{Q}|$ its cardinality. Under the standard sparsity regime $q\ll n$, the next theorem gives non-asymptotic bounds for the estimation errors of $\boldsymbol{\widehat{\beta}}_h$.
\begin{theorem}[Estimation error bounds]\label{theorem 2-conv}
Assume model \eqref{hybrid model} with a $q$-sparse $\boldsymbol{\beta}^*$ and Assumptions~\ref{A1}--\ref{A5}. Suppose the bandwidth satisfies
\(
\max\left\{\frac{\sigma_x}{f_{\min}}\sqrt{\frac{q\log(2p)}{n}},\ \frac{\sigma_x^2f_{\max}}{f_{\min}^2}\frac{q\log(2p)}{n}\right\}\lesssim h\le \min\left\{\frac{f_{\min}}{2L_f},\,\sqrt{\sqrt{q}\lambda}\right\}.
\)
Then, with probability at least $1-p^{-1}$,
\(
\|\boldsymbol{\widehat{\beta}}_h-\boldsymbol{\beta}^*\|_2\le C'\sqrt{q}\lambda,\, \|\boldsymbol{\widehat{\beta}}_h-\boldsymbol{\beta}^*\|_1\le C''q\lambda,
\)
where $C',C''>0$ depend only on $(L_f,a_0,\mu_p,\kappa_0,\kappa_2,\gamma)$.
\end{theorem}
\noindent Theorem \ref{theorem 2-conv} shows that, with an appropriate bandwidth choice, the $L_1$-penalized SHQER estimator attains the same order of convergence as the classical $L_1$-penalized estimator under both $L_1$- and $L_2$-losses. The sub-exponential design condition is conventional in high-dimensional concentration arguments \citep{Wainwright2019}, but it is a genuine scope restriction rather than a robustness property. These rates are directly comparable with those obtained for smoothed and composite quantile procedures in \citet{Tan2022}.
\section{Numerical Studies}\label{section 4}
\noindent In this section, we evaluate the performance of SHQER in finite samples through simulations and an illustration using real data. We emphasize the trade-off between robustness and efficiency, governed by $(\tau,\gamma)$, as well as the role of the kernel and bandwidth in statistical accuracy.
\subsection{Simulation results under quantile-heteroscedasticity}
\noindent
We investigate the finite-sample performance of the proposed LAMM solver, presented in Algorithm~\ref{algorithm}, in the context of penalized smoothed hybrid quantile--expectile regression (SHQER). Our experiments focus on a quantile-heteroscedastic, high-dimensional linear model with a sparse ground-truth signal. This setting is most relevant to robust inference under conditional scale shifts.
\noindent
SHQER uses the Gaussian kernel. Logistic, Laplacian, and uniform kernels produced similar results and are omitted for brevity. Guided by our bias and RSC theory, we use the hybrid bandwidth
\(
h_{\mathrm{SHQER}}
\,=\,
\max\Bigl\{0.05,\ \sqrt{(1-\gamma)\,\tau(1-\tau)}\,(\tfrac{\log p}{n})^{0.25}\Bigr\}.
\)
This recovers the quantile rule of \citet{Tan2022} at $\gamma=0$ and automatically decreases smoothing as $\gamma\to 1$, reflecting the intrinsic smoothness of the expectile component. For every fixed $\gamma<1$, the dominant asymptotic order is $h\asymp(\log p/n)^{1/4}$. If $\lambda\asymp\sqrt{\log p/n}$ and $q=o\{(n/\log p)^{1/2}\}$, this order is compatible with the lower and upper bandwidth orders in Theorem~\ref{theorem 2-conv}; when $\gamma=1$, the quantile component disappears and the smoothing bandwidth is immaterial. The finite-sample floor $0.05$ is used only to avoid numerical degeneracy and is not part of the asymptotic prescription. Because the constants in Theorem~\ref{theorem 2-conv} are unknown, we do not claim that this rule exactly solves the finite-sample theoretical inequalities. A scale-adaptive multiplier is not used in the reported experiments.
We generate $\tilde{\boldsymbol{x}}_i \sim \mathcal{N}(\boldsymbol{0}, \boldsymbol{\Sigma})$ with
$(\boldsymbol{\Sigma})_{il} = 0.7^{|i-l|}$, and define
$\boldsymbol{x}_i = (1, \tilde{\boldsymbol{x}}_i^\top)^\top \in \mathbb{R}^{p+1}$.
We consider the following quantile-heteroscedastic data generating process (DGP):
\(
y_i
=
\boldsymbol{x}_i^\top\boldsymbol{\beta}^*
+
\bigl(\frac{1}{2}\,|x_{i,p+1}|+\frac{1}{2}\bigr)\,\bigl\{\varepsilon_i-Q_\varepsilon(\tau)\bigr\},
\)
where centering by $Q_\varepsilon(\tau)$ ensures that the conditional $\tau$-quantile of the noise term is zero. The noise term is either $\varepsilon_i\sim\mathcal{N}(0,3)$ or $\varepsilon_i\sim t(1.9)$.
\noindent
We adopt the following sparse coefficient pattern: $\beta_1^*=4$; nonzeros at even indices $l=2,\dots,20$, with magnitudes decreasing from $1.8$ to $-1.8$; and $\beta_l^*=0$ otherwise.
We fit SHQER with Lasso and Elastic Net (EN; with the mixing parameter fixed at $\alpha\in\{0.3,0.5,0.7\}$). We compare SHQER to:
\begin{enumerate}
\item[(i)] $L_{1}$-penalized asymmetric least squares (SALES; \citealt{GuZou2016});
\item[(ii)] $L_{1}$-penalized quantile regression (QR) via \texttt{rqPen} \citep{SherwoodMaidman2020};
\item[(iii)] CONQUER (convolution-smoothed quantile regression; \citealt{He2021});
\item[(iv)] $L^k$-quantile regression with $k=1.5$ ($L^k\mathrm{QR}$) \citep{jiang2021k}.
\end{enumerate}
\noindent
To avoid an expensive search over a continuous mixing parameter, SHQER is evaluated on the fixed five-point grid $\Gamma=\{0,0.1,0.5,0.9,1\}$. Thus, $\gamma$ is treated as a low-cardinality model index rather than as a continuously optimized parameter, and the same CV folds are reused across the five candidates. Based on the simulation patterns below, $\gamma=0.1$ is a simple robustness-oriented starting value when heavy-tailed response errors are a primary concern, whereas $\gamma=0.5$ provides a neutral hybrid starting point when no such preference is available. These are practical recommendations, not an optimality theorem; a fully data-adaptive theoretical selector for $\gamma$ remains an open problem.
\noindent
We select the sparsity parameter $\lambda$ by 10-fold cross-validation (CV) on each method’s training loss. For EN, we select $\lambda$ by CV for each fixed $\alpha\in\{0.3,0.5,0.7\}$ and report the best-performing choice across $\alpha$ (see the discussion below). For CONQUER, we use the same kernel family and the quantile bandwidth rule as in \citet{Tan2022} (i.e., the rule with $\gamma=0$). For SALES and \(L^k\mathrm{QR}\,(k=1.5)\), we follow their standard CV protocols.  Over 100 replications, we report: (i) estimation error, defined as $\|\boldsymbol{\widehat{\beta}}-\boldsymbol{\beta}^*\|_2$, and (ii) support recovery through the true positive rate (TPR) and false positive rate (FPR). The reported values are the corresponding means and standard deviations over the 100 Monte Carlo runs. Detailed numerical results are reported in the Supplementary Material for clarity of presentation.  Tables~\ref{table_lasso_tau=0.25}--\ref{table_lasso_tau=0.75} summarize the Lasso results for $(n,p)\in\{(600,300),(300,600)\}$ at $\tau\in\{0.25,0.50,0.75\}$ under the linear heteroscedastic model, while Tables~\ref{table:elasticnet_alpha_tau_025}--\ref{table:elasticnet_alpha_tau_075} report the corresponding Elastic-Net comparisons. 
\noindent
Overall, the results reported in Tables~\ref{table_lasso_tau=0.25}--\ref{table_lasso_tau=0.75} and Tables~\ref{table:elasticnet_alpha_tau_025}--\ref{table:elasticnet_alpha_tau_075} confirm that SHQER provides a flexible and competitive framework across a broad range of configurations. Rather than yielding a single uniformly optimal value of $\gamma$, the method adapts to the quantile level, the regularization scheme, and the noise distribution. As shown across these tables, this flexibility is one of the main strengths of SHQER: by interpolating between quantile-oriented and expectile-oriented behaviors, it is able to achieve favorable compromises between estimation accuracy and support recovery in several regimes where purely quantile-based or more rigid competitors may be less well balanced.
\noindent
Under Lasso regularization, the most favorable SHQER configurations are generally obtained for small or moderate values of $\gamma$, especially at the more extreme quantile levels $\tau=0.25$ and $\tau=0.75$, as can be seen from Tables~\ref{table_lasso_tau=0.25} and~\ref{table_lasso_tau=0.75}. In these settings, the configurations $\gamma=0$ and $\gamma=0.1$ consistently yield the best SHQER results in terms of estimation error, while remaining highly competitive in terms of TPR and FPR. The progressive improvement observed in Tables~\ref{table_lasso_tau=0.25} and~\ref{table_lasso_tau=0.75} when moving from $\gamma=1$ to $\gamma=0.5$, then to $\gamma=0.1$ and $\gamma=0$, clearly indicates that the SHQER family can adapt effectively to tail-oriented inference. This behavior is particularly appealing, since it shows that SHQER does not rely on a single rigid specification, but instead offers a continuum of models that can be tuned according to the difficulty and the nature of the estimation problem.
\noindent
At the median level $\tau=0.50$, the differences among the best-performing methods become less pronounced, and SHQER remains competitive with the main alternatives, as reported in Table~\ref{table_lasso_tau=0.50}. In particular, Table~\ref{table_lasso_tau=0.50} shows that the configurations with small $\gamma$ continue to deliver strong results, indicating that SHQER retains good performance even in more central parts of the conditional distribution. This suggests that, beyond its appeal for tail-robust estimation, the proposed hybrid formulation remains sufficiently stable and effective in less extreme regimes as well.
\noindent
Under heavy-tailed noise, the results in Tables~\ref{table_lasso_tau=0.25}--\ref{table_lasso_tau=0.75} further highlight the practical relevance of SHQER. These tables show that expectile-dominant configurations may deteriorate noticeably, whereas quantile-dominant or lightly hybridized versions of SHQER remain substantially more accurate. In particular, according to Tables~\ref{table_lasso_tau=0.25} and~\ref{table_lasso_tau=0.75}, SHQER with $\gamma=0.1$ or $\gamma=0$ repeatedly achieves a strong compromise, combining low estimation error with satisfactory support recovery. Hence, even when the pure quantile limit performs best in some cases, the SHQER family still delivers near-optimal behavior through appropriately chosen hybrid configurations, which further reinforces its robustness and practical usefulness.
\noindent
The Elastic-Net experiments should be interpreted as a computational and empirical extension of SHQER: the high-probability theory in Section~\ref{section 4.2} is established for the $L_1$-penalized estimator and is not asserted for Elastic-Net regularization. Tables~\ref{table:elasticnet_alpha_tau_025}--\ref{table:elasticnet_alpha_tau_075} nevertheless provide finite-sample evidence that several Elastic-Net SHQER configurations are competitive. In these experiments, the best choice is not always $\gamma=0$: depending on the quantile level and the performance criterion, intermediate configurations such as $\gamma=0.1$, $\gamma=0.5$, or $\gamma=0.9$ can match or outperform more extreme choices. These observations are descriptive rather than theoretical guarantees and motivate a separate extension of the error-bound analysis to Elastic-Net penalties.
\noindent
From a variable-selection perspective, the patterns reported in Tables~\ref{table_lasso_tau=0.25}--\ref{table_lasso_tau=0.75} and Tables~\ref{table:elasticnet_alpha_tau_025}--\ref{table:elasticnet_alpha_tau_075} show that SHQER also exhibits an appealing balance. In many configurations, increasing $\gamma$ may improve TPR, but at the cost of a larger FPR, whereas very small values of $\gamma$ tend to provide better control of false discoveries. This trade-off is visible across both the Lasso and Elastic-Net results. One of the main merits of SHQER is precisely that it allows one to navigate this trade-off through the tuning parameter $\gamma$. Therefore, rather than supporting a one-size-fits-all conclusion, the simulations suggest that SHQER should be viewed as a versatile class of estimators whose best configuration depends on the inferential objective: stronger robustness and tighter FPR control tend to favor small $\gamma$, whereas in some Elastic-Net regimes moderate hybridization may yield competitive overall performance.
\noindent
Taken together, the numerical evidence in Tables~\ref{table_lasso_tau=0.25}--\ref{table_lasso_tau=0.75} and Tables~\ref{table:elasticnet_alpha_tau_025}--\ref{table:elasticnet_alpha_tau_075} supports the practical relevance of SHQER in sparse heteroscedastic settings. Under Lasso regularization, quantile-dominant versions of SHQER emerge as the most reliable choices, especially at extreme quantiles and under heavy-tailed noise. Under Elastic-Net regularization, the broader SHQER family becomes even more attractive, since intermediate hybrid configurations can also perform very well. Overall, the evidence shows that SHQER is not only competitive with existing procedures, but also offers an additional degree of flexibility that can be exploited to adapt the estimator to different noise structures, quantile levels, and regularization regimes.

\subsection{Application to Gene Expression Data}

\noindent
We illustrate SHQER on an expression quantitative trait locus (eQTL) gene expression dataset from \citet{Scheetz2006}, which was previously analyzed by \citet{Kim2008} and \citet{Wang2012}. The original study reports expression levels for 31{,}042 probe sets measured in 120 laboratory rats. Using the preprocessing pipeline described in \citet{Wang2012}, we first reduced the number of probes to 18,958. We then retained the 500 probes exhibiting the largest absolute marginal correlation with the target gene \texttt{TRIM32}, which is associated with retinal degeneration. We fit all SHQER models using a Gaussian kernel combined with either $L_{1}$ (lasso) or elastic-net regularization. We consider the following benchmark methods: $L_{1}$- and elastic-net penalized quantile regression implemented via the \texttt{rqPen} package, as well as SALES, and the $L^k$-quantile regression (\(L^k\mathrm{QR}\)). Regularization levels are selected by 10-fold cross-validation; when a single SHQER configuration is required, $\gamma$ is selected over the same fixed five-point grid $\Gamma=\{0,0.1,0.5,0.9,1\}$ using the identical folds. For each method and each quantile level $\tau \in \{0.25,\,0.50,\,0.75\}$, we repeat the following procedure 50 times: in each repetition, the 120 observations are randomly divided into training and test sets of sizes 80 and 40, respectively. We fit the models on the training data and assess their predictive performance on the test set using the corresponding loss function. All models are fitted using the LAMM framework with coordinatewise soft-thresholding updates and adaptive backtracking for the quadratic parameter (see Section~\ref{section 3}). Cross-validation folds are stratified according to response deciles to reduce imbalance across training and validation subsets. The bandwidth parameter $h$ is chosen according to the default rule introduced earlier, whose dominant order is compatible with Theorem~\ref{theorem 2-conv} under the sparsity regime stated above. Additional sensitivity analyses based on $h \in \{0.5,\,1,\,2\}\times h_{\mathrm{default}}$ lead to similar qualitative conclusions and are therefore omitted for brevity. The corresponding real-data results are reported in Table~\ref{table:real} of the Supplementary Material section, which summarizes the average model size together with the corresponding prediction error (PE) on the test sample for the different penalized procedures. The results are presented for the Lasso penalty and for several Elastic Net configurations. Overall, the competing methods exhibit markedly different sparsity patterns across quantile levels and penalty choices. First, SALES consistently yields the smallest prediction errors across all configurations, but this performance is accompanied by relatively large model sizes, typically around 32 to 37 selected variables. CONQUER also produces comparatively dense models, with sizes close to those of SALES, while its prediction error remains systematically higher. In contrast, QR (\texttt{rqPen}) tends to select substantially more variables at the median level $\tau=0.50$, where the model size can become very large, whereas its behavior is more moderate at $\tau=0.25$ and $\tau=0.75$. The \(L^k\mathrm{QR}\) procedure provides the sparsest and most stable solutions overall, with relatively small model sizes and low prediction errors across all settings. For SHQER, the role of the hybrid parameter $\gamma$ is clearly visible at all quantile levels. Under the Lasso penalty, large values of $\gamma$ generally lead to smaller or moderate model sizes and very low prediction errors, especially for $\gamma=1$ and $\gamma=0.9$. As $\gamma$ decreases from $1$ toward $0$, the model size often increases, while the prediction error progressively rises. This pattern is particularly clear at $\tau=0.25$ and $\tau=0.50$, where the configurations with larger $\gamma$ are clearly more favorable than the more quantile-oriented configurations. At the lower quantile level $\tau=0.25$, SHQER shows a strong dependence on the penalty structure. Under the Lasso penalty and under Elastic Net with $\alpha=0.7$, the method remains fairly sparse for large values of $\gamma$, while still maintaining very small prediction errors. For smaller values of $\alpha$, however, the Elastic Net penalty tends to produce larger SHQER models, especially for $\gamma=1$ and $\gamma=0.9$, indicating that the ridge component stabilizes estimation but may reduce sparsity. In comparison, \(L^k\mathrm{QR}\) remains more parsimonious and achieves similarly small prediction errors. At the median level $\tau=0.50$, the differences in sparsity become even more pronounced. QR (\texttt{rqPen}) selects by far the largest models in all penalty settings, whereas SHQER keeps the model size much smaller. In particular, for moderate to large values of $\gamma$, SHQER yields compact models with prediction errors that remain low relative to QR and CONQUER. This suggests that the hybrid formulation provides an appealing compromise at the central quantile level, especially when one seeks to avoid the excessive model sizes produced by standard penalized quantile regression. At the upper quantile level $\tau=0.75$, SHQER again benefits from larger values of $\gamma$, especially under the Lasso penalty and under Elastic Net with larger $\alpha$. In these settings, the method can produce very sparse models, sometimes selecting only a handful of variables, while keeping the prediction error small. As before, decreasing $\gamma$ tends to increase the prediction error and does not systematically improve sparsity. This indicates that, for the present data set, the more expectile-oriented SHQER configurations are often preferable to the more quantile-oriented ones. Taken together, these results show that SHQER is flexible enough to adapt to different quantile levels and penalty structures through the choice of $\gamma$. In the present application, large values of $\gamma$ appear to be the most effective, as they often provide the best compromise between prediction accuracy and model parsimony. Although SALES achieves the smallest prediction errors overall, it does so with less sparse models. By contrast, SHQER and \(L^k\mathrm{QR}\) offer a more balanced tradeoff between predictive performance and sparsity, which is particularly desirable in high-dimensional gene expression analysis.

The present application is intended as a predictive and sparsity illustration rather than a biological marker-discovery analysis. Accordingly, Table~\ref{table:real} summarizes model size and out-of-sample prediction error, but does not quantify gene-by-gene overlap or selection stability across methods. Such overlap and stability analyses would be necessary before interpreting method-specific selected probes as distinct biological signals and are left for a dedicated applied study.


\section{Conclusion}\label{section conclusion}
\noindent
We introduced a class of penalized, convolution--smoothed, hybrid quantile--expectile regression (SHQER) estimators for high--dimensional, sparse modeling. By blending quantile and expectile losses and smoothing the hybrid objective via kernel convolution, our framework retains the interpretability of distributional regression and yields an empirical risk that is differentiable and locally strongly convex with high probability. Based on this structure, we proposed a LAMM proximal algorithm that can handle common convex penalties (e.g., Lasso and Elastic Net) and arbitrary nonnegative kernels. Statistically, we characterized the bias induced by smoothing and established high--probability error bounds for the $L_{1}$--penalized SHQER estimator that match classical $L_{1}$--quantile rates under comparable conditions. Although our theoretical development centers on estimation accuracy, expanding the analysis to beta--min conditions and oracle--type support recovery guarantees is a significant area for future research. These developments would provide a more complete theoretical picture by connecting hybrid convolution smoothing to exact sparsity recovery under suitable signal--strength assumptions. From a computational standpoint, the smoothed hybrid loss allows straightforward gradient--proximal updates with monotone descent and backtracking. The present experiments evaluate statistical performance rather than claiming runtime superiority over alternative solvers. Simulations and an eQTL case study show competitive prediction and sparsity relative to penalized quantile baselines across the settings considered, including heavy--tailed noise. To promote reproducibility and practical application, the proposed methodology has been implemented in the open-source \textsf{R} package \texttt{shqer}, which offers Lasso and Elastic Net fitting, cross-validation, hybrid loss smoothing, and optional bootstrap-based inference. The package and complete source code are publicly available at: \url{https://github.com/abdellah-atanane/shqer}. One open direction is to extend SHQER to include matrix--valued parameters with low--rank structure in addition to vector coefficients and sparsity. In underdetermined regimes, one could combine hybrid convolution smoothing, concave spectral penalties, and multi--step reweighting to obtain fast, provably accurate procedures for quantile/expectile matrix regression and completion. Other avenues of research include structured sparsity (groups and hierarchies), heteroskedastic, time--series designs, online and streaming optimization, and post--selection inference tailored to hybrid losses. Two additional limitations are worth emphasizing: the current non-asymptotic theory assumes sub-exponential covariates and therefore does not cover leverage-prone heavy-tailed designs, and the Elastic-Net results are empirical because the high-probability bounds are proved only for the $L_1$ penalty. Developing theory for these two extensions, together with a principled data-adaptive selector for $\gamma$, is left for future work.
\section*{Statements and Declarations}

\subsection*{Funding}
This work was supported by the Fonds de recherche Québec-Santé (Grant No.~267074 to K.O.) and the Natural Sciences and Engineering Research Council of Canada (Grant No.~RGPIN-2019-06727 to K.O.).

\subsection*{Competing Interests}
The authors have no relevant financial or non-financial interests to disclose.

\subsection*{Data Availability}
The simulation results are based on synthetic data generated according to the designs described in the manuscript. The processed TRIM32 gene-expression dataset used in the real-data application (120 samples and 500 predictors) is publicly available as the \texttt{trim32} dataset in the \textsf{R} package \texttt{abess}; its origin and preprocessing are described in  \cite{Scheetz2006} and in the manuscript.

\subsection*{Code Availability}
The proposed methodology is implemented in the open-source \textsf{R} package \texttt{shqer}. The package and source code are publicly available at \url{https://github.com/abdellah-atanane/shqer}.


\appendix
\section{Appendix}
\noindent This appendix first derives the iterative update formulas used in Step \ref{step:alg} of Algorithm \ref{algorithm} for the weighted Lasso \citep{Tibshirani1996} and the Elastic Net \citep{ZouHastie2005}, and then proves the main theoretical results. The arguments are inspired by \citet{zou2008composite}, \citet{bradic2011penalized}, \citet{fernandes2021smoothing}, and \citet{Tan2022}.

\subsection{Generic proximal-gradient updates}
\noindent
Let $\widehat{\mathcal{H}}^{\gamma}_{h}(\boldsymbol{\beta})$ denote the SHQER empirical loss and let $p_{\lambda}$ be a convex penalty. Given a step size $\upsilon_t>0$, consider the isotropic quadratic surrogate \(\mathcal{G}(\boldsymbol{\beta}\mid \upsilon_t,\widehat{\boldsymbol{\beta}}^{\,t-1})
:=\widehat{\mathcal{H}}^{\gamma}_{h}(\widehat{\boldsymbol{\beta}}^{\,t-1})
+\big\langle \nabla \widehat{\mathcal{H}}^{\gamma}_{h}(\widehat{\boldsymbol{\beta}}^{\,t-1}),\,\boldsymbol{\beta}-\widehat{\boldsymbol{\beta}}^{\,t-1}\big\rangle
+\frac{\upsilon_t}{2}\|\boldsymbol{\beta}-\widehat{\boldsymbol{\beta}}^{\,t-1}\|_2^2
+p_{\lambda}(\boldsymbol{\beta})\). According to the first-order optimality condition, the update $\widehat{\boldsymbol{\beta}}^{\,t}$ satisfies \(\boldsymbol{0}\in \nabla \widehat{\mathcal{H}}^\gamma_h(\widehat{\boldsymbol{\beta}}^{\,t-1})
+\upsilon_t(\widehat{\boldsymbol{\beta}}^{\,t}-\widehat{\boldsymbol{\beta}}^{\,t-1})
+\partial p_{\lambda}(\boldsymbol{\beta})\big|_{\boldsymbol{\beta}=\widehat{\boldsymbol{\beta}}^{\,t}}\), where $\partial p_{\lambda}$ denotes the subdifferential of $p_{\lambda}$. For the unpenalized intercept, the update is \(\widehat{\beta}_1^{\,t}=\widehat{\beta}_1^{\,t-1}-\upsilon_t^{-1}\frac{\partial}{\partial\beta_1}\widehat{\mathcal{H}}^\gamma_h(\widehat{\boldsymbol{\beta}}^{\,t-1})\). Let $\mathcal{S}(u,v)=\mathrm{sign}(u)(|u|-v)_+$ denote the soft-thresholding operator, and write \(\nabla_l\widehat{\mathcal{H}}^\gamma_h(\widehat{\boldsymbol{\beta}}^{\,t-1})=\frac{\partial \widehat{\mathcal{H}}^\gamma_h(\boldsymbol{\beta})}{\partial\beta_l}\big|_{\boldsymbol{\beta}=\widehat{\boldsymbol{\beta}}^{\,t-1}}\). Then the coordinate-wise updates for the penalized coefficients are:
\begin{enumerate}
\item \textbf{Weighted Lasso.} For \(p_{\lambda}(\boldsymbol{\beta})=\sum_{l=1}^p \lambda_l |\beta_l|\), we have
\[
\widehat{\beta}_l^{\,t}
=\mathcal{S}\!\left(\widehat{\beta}_l^{\,t-1}-\upsilon_t^{-1}\nabla_l \widehat{\mathcal{H}}^\gamma_h(\widehat{\boldsymbol{\beta}}^{\,t-1}),\,\upsilon_t^{-1}\lambda_l\right)
\]. \item \textbf{Elastic Net.} For \(p_{\lambda}(\boldsymbol{\beta})=\lambda\alpha\|\boldsymbol{\beta}\|_1+\lambda(1-\alpha)\|\boldsymbol{\beta}\|_2^2\), the update becomes\\ \(\widehat{\beta}_l^{\,t}
=\frac{1}{1+2\lambda(1-\alpha)/\upsilon_t}\,
\mathcal{S}\!\left(\widehat{\beta}_l^{\,t-1}-\upsilon_t^{-1}\nabla_l \widehat{\mathcal{H}}^\gamma_h(\widehat{\boldsymbol{\beta}}^{\,t-1}),\,\upsilon_t^{-1}\lambda\alpha\right)\).
\end{enumerate}
\noindent
Therefore, the explicit coordinate-wise update rules above coincide with the iterative steps implemented in Algorithm \ref{algorithm}. For later use, we measure the local curvature of the population smoothed SHQER objective through $\mathbb{D}:\mathbb{R}^p\times\mathbb{R}^p\to[0,\infty)$ associated with the convex function $\mathcal{H}^{\gamma}_{h}(\cdot)$ by \(\mathbb{D}(\boldsymbol{\beta}_1,\boldsymbol{\beta}_2)
=\big\langle \nabla \mathcal{H}^{\gamma}_{h}(\boldsymbol{\beta}_1)-\nabla \mathcal{H}^{\gamma}_{h}(\boldsymbol{\beta}_2),\,\boldsymbol{\beta}_1-\boldsymbol{\beta}_2\big\rangle\). It is the symmetrized Bregman divergence generated by $\mathcal{H}_{h}^{\gamma}$.
\subsection{Proof of Proposition \ref{proposition 1}}
\noindent In the spirit of \citet{Fan2018}, we derive an upper bound for $\|\boldsymbol{\beta}_h^*-\boldsymbol{\beta}^*\|_{\boldsymbol{\Sigma}}$ via a localized argument. Note that the function $\mathcal{H}_h^\gamma(\boldsymbol{\beta})$ is convex and twice differentiable in a neighborhood of the optimal solution $\boldsymbol{\beta}^*$. We recall the symmetrized Bregman divergence, denoted by
\(\mathbb{D}:\mathbb{R}^p\times\mathbb{R}^p\to[0,\infty)\),  associated with the convex
function \(\mathcal{H}^{\gamma}_{h}(\cdot)\), evaluated at
\(\boldsymbol{\beta}_1,\boldsymbol{\beta}_2\), is
\(
\mathbb{D}(\boldsymbol{\beta}_1,\boldsymbol{\beta}_2)
=\big\langle
\nabla \mathcal{H}^{\gamma}_{h}(\boldsymbol{\beta}_1)
-\nabla \mathcal{H}^{\gamma}_{h}(\boldsymbol{\beta}_2),\,
\boldsymbol{\beta}_1-\boldsymbol{\beta}_2
\big\rangle .
\)
\noindent
To begin, we localize the problem around a boundary point by introducing the corresponding local neighborhood.
\(
\Theta_h=\boldsymbol{\beta}^*+\mathbb{B}_{\boldsymbol{\Sigma}}(h)
=\Bigl\{\boldsymbol{\beta}\in\mathbb{R}^p:\ \|\boldsymbol{\beta}-\boldsymbol{\beta}^*\|_{\boldsymbol{\Sigma}}\le h\Bigr\},
\)
with $\mathbb{B}_{\boldsymbol{\Sigma}}(h)$ defined in \eqref{eq;boule}.
A priori, it is not guaranteed that the minimizer
\(
\boldsymbol{\beta}_h^*\in\arg\min_{\boldsymbol{\beta}\in\mathbb{R}^p}\mathcal{H}^\gamma_h(\boldsymbol{\beta})
\)
lies in $\Theta_h$. Introduce the intermediate point
\(
\boldsymbol{\beta}_h^{\dagger}=(1-\zeta)\boldsymbol{\beta}^*+\zeta\,\boldsymbol{\beta}_h^*,
\,
\zeta=\sup\bigl\{s\in[0,1]:\ \boldsymbol{\beta}^*+s(\boldsymbol{\beta}_h^*-\boldsymbol{\beta}^*)\in\Theta_h\bigr\}.
\)
If $\boldsymbol{\beta}_h^*\notin\Theta_h$, then $\zeta\in(0,1)$ and $\|\boldsymbol{\beta}_h^\dagger-\boldsymbol{\beta}^*\|_{\boldsymbol{\Sigma}}=h$ by construction (boundary point);
otherwise $\zeta=1$ and $\boldsymbol{\beta}_h^\dagger=\boldsymbol{\beta}_h^*$.
\noindent
Next, we derive a basic inequality from convexity and the optimality conditions. Set $\boldsymbol{\delta}:=\boldsymbol{\beta}_h^\dagger-\boldsymbol{\beta}^*$.
Since $\mathcal{H}_h^\gamma$ is convex, its gradient is monotone:
\(
\mathbb{D}(\boldsymbol{\beta},\boldsymbol{\beta}^*)\ge 0,
\, \forall\boldsymbol{\beta}\in\mathbb{R}^p.
\)
Applying this with $\boldsymbol{\beta}=\boldsymbol{\beta}_h^\dagger$ yields the left inequality in \eqref{eq:a5-new} below.
Furthermore, since $\boldsymbol{\beta}_h^\dagger$ lies on the segment joining $\boldsymbol{\beta}^*$ and $\boldsymbol{\beta}_h^*$, Lemma F.2 of \citet{Fan2018} (monotonicity along the segment together with optimality of $\boldsymbol{\beta}_h^*$) implies
\(
\mathbb{D}(\boldsymbol{\beta}_h^\dagger,\boldsymbol{\beta}^*)
\le 
\mathbb{D}(\boldsymbol{\beta}_h^*,\boldsymbol{\beta}^*).
\)
Finally, by first-order optimality at the population minimizer $\boldsymbol{\beta}_h^*$ (i.e., $\nabla \mathcal{H}_h^\gamma(\boldsymbol{\beta}_h^*)=\boldsymbol{0}$), we obtain
\begin{align}\label{eq:a5-new}
0\le \mathbb{D}(\boldsymbol{\beta}_h^\dagger,\boldsymbol{\beta}^*) 
\le \mathbb{D}(\boldsymbol{\beta}_h^*,\boldsymbol{\beta}^*)
= \big\langle -\nabla \mathcal{H}^\gamma_h(\boldsymbol{\beta}^*),\ 
         \boldsymbol{\beta}_h^*-\boldsymbol{\beta}^*\big\rangle .
\end{align}
\noindent
We now express the gradient increment through a mean-value representation.
Define, for $s\in[0,1]$,
$
\boldsymbol{\beta}_s := \boldsymbol{\beta}^* + s(\boldsymbol{\beta}_h^\dagger-\boldsymbol{\beta}^*)
=\boldsymbol{\beta}^*+s\boldsymbol{\delta}.
$ 
Since $\nabla \mathcal{H}_h^\gamma$ is differentiable, the mean value theorem for vector maps gives
\(
\nabla \mathcal{H}^\gamma_h(\boldsymbol{\beta}_h^\dagger)
-\nabla \mathcal{H}^\gamma_h(\boldsymbol{\beta}^*)
=\left(\int_0^1\nabla^2\mathcal{H}^\gamma_h\big(\boldsymbol{\beta}_s\big)\,ds\right)
(\boldsymbol{\beta}_h^\dagger-\boldsymbol{\beta}^*).
\)
Taking the inner product with $\boldsymbol{\delta}$ yields
\begin{align}\label{eq:a6-new-int}
\big\langle \nabla \mathcal{H}^\gamma_h(\boldsymbol{\beta}_h^\dagger)
- \nabla \mathcal{H}^\gamma_h(\boldsymbol{\beta}^*),\
\boldsymbol{\delta}\big\rangle
&=
\int_0^1
\big\langle 
\nabla^2\mathcal{H}^\gamma_h(\boldsymbol{\beta}_s)\,\boldsymbol{\delta},\,
\boldsymbol{\delta}\big\rangle\,ds.
\end{align}
\noindent
We next derive a Hessian representation and establish a local lower bound.
Recall that the Hessian representation of the function $\mathcal{H}^\gamma_h(\boldsymbol{\beta})$ is given by 
\(
\nabla^2\mathcal{H}^\gamma_h(\boldsymbol{\beta})
=\mathbb{E}\!\left[\Big\{(1-\gamma)\,k_h\big(\boldsymbol{x}^\top(\boldsymbol{\beta}-\boldsymbol{\beta}^*)-\varepsilon\big)
+2\gamma\,\omega_\tau\big(\varepsilon-\boldsymbol{x}^\top(\boldsymbol{\beta}-\boldsymbol{\beta}^*)\big)\Big\}\,
\boldsymbol{x}\boldsymbol{x}^\top\right],
\)
where $\varepsilon=y-\boldsymbol{x}^\top\boldsymbol{\beta}^*$ and $k_h(s)=\frac{1}{h}k\!\left(\frac{s}{h}\right)$. For the expectile weight, 
$
\omega_\tau(t)=\tau\,\mathds{1}\{t\ge 0\} + (1-\tau)\,\mathds{1}\{t<0\}$ for $\tau\in[\tau_{\min},\tau_{\max}],$
we have $\omega_\tau(t)\ge \tau_{\min}$ for all $t\in\mathbb{R}$,  and hence
\begin{equation}\label{eq:w-lower}
\mathbb{E}\big[\omega_\tau(\varepsilon-\boldsymbol{x}^\top\boldsymbol{\delta})\mid\boldsymbol{x}\big]\ge \tau_{\min}, 
\end{equation}
where $\boldsymbol{\delta}=\boldsymbol{\beta}-\boldsymbol{\beta}^*$.
\noindent
We now turn to the expansion of the smoothed quantile kernel. Conditioning on $\boldsymbol{x}$ and setting
\(
t=\boldsymbol{x}^\top\boldsymbol{\delta},
\)
we obtain
\(
\mathbb{E}\big[k_h(t-\varepsilon)\mid\boldsymbol{x}\big]
=\int_{-\infty}^{\infty} k_h(t-u)\,f_{\varepsilon|\boldsymbol{x}}(u)\,du
=\int_{-\infty}^{\infty} k(s)\,f_{\varepsilon|\boldsymbol{x}}(t-hs)\,ds.
\)
Then, it follows from Assumption~\ref{A1} that
\(
\Big|\mathbb{E}[k_h(t-\varepsilon)\mid\boldsymbol{x}]-f_{\varepsilon|\boldsymbol{x}}(0)\Big|
\le 
\int |k(s)|\,\big|f_{\varepsilon|\boldsymbol{x}}(t-hs)-f_{\varepsilon|\boldsymbol{x}}(0)\big|\,ds
\le
L_f\int |k(s)|\,\big(|t|+h|s|\big)\,ds.
\)
So that, with $\kappa_1=\int |s|\,|k(s)|\,ds$ (and $\int |k(s)|\,ds=1$ for a nonnegative kernel),
\begin{equation}\label{eq:Delta-bound}
\mathbb{E}\big[k_h(\boldsymbol{x}^\top\boldsymbol{\delta}-\varepsilon)\mid\boldsymbol{x}\big]
= f_{\varepsilon|\boldsymbol{x}}(0) + \Delta_h(\boldsymbol{\delta}),
\end{equation}
\mbox{with} 
$|\Delta_h(\boldsymbol{\delta})|
\le L_f\big(|\boldsymbol{x}^\top\boldsymbol{\delta}|+\kappa_1 h\big)$.
\noindent
Combining \eqref{eq:w-lower} and \eqref{eq:Delta-bound}, we deduce that, for any $\boldsymbol{\delta}$,
\begin{align}
&\mathbb{E}\Big[(1-\gamma)\,k_h(\boldsymbol{x}^\top\boldsymbol{\delta}-\varepsilon)
+2\gamma\,\omega_\tau(\varepsilon-\boldsymbol{x}^\top \boldsymbol{\delta})
\,\Big|\,\boldsymbol{x}\Big] \notag\\
&\quad\ge (1-\gamma)\big\{f_{\varepsilon|\boldsymbol{x}}(0)
+\Delta_h(\boldsymbol{\delta})\big\}+2\gamma\,\tau_{\min} \notag\\
&\quad= (1-\gamma)f_{\varepsilon|\boldsymbol{x}}(0)
+2\gamma\tau_{\min}+(1-\gamma)\Delta_h(\boldsymbol{\delta}). \notag
\end{align}
\noindent
It remains to derive a lower bound for the left-hand side of \eqref{eq:a5-new}. For $s\in[0,1]$, define
\(
\boldsymbol{\delta}_s:=\boldsymbol{\beta}_s-\boldsymbol{\beta}^*:=s\boldsymbol{\delta}.
\)
Using \eqref{eq:a6-new-int} and the Hessian representation, we obtain
\begin{align*}
&\big\langle 
\nabla^2\mathcal{H}^\gamma_h(\boldsymbol{\beta}_s)\,\boldsymbol{\delta}_s,
\boldsymbol{\delta}_s\big\rangle\\
&\quad=
\mathbb{E}\Big[
\mathbb{E}\big[(1-\gamma)k_h(\boldsymbol{x}^\top\boldsymbol{\delta}_s-\varepsilon)
+2\gamma \omega_\tau(\varepsilon-\boldsymbol{x}^\top\boldsymbol{\delta}_s)
\,\big|\,\boldsymbol{x}\big]\,
(\boldsymbol{x}^\top\boldsymbol{\delta}_s)^{2}\Big] \\
&\quad\ge
\mathbb{E}\Big[\big\{(1-\gamma)f_{\varepsilon|\boldsymbol{x}}(0)
+2\gamma\tau_{\min}\big\}(\boldsymbol{x}^\top\boldsymbol{\delta}_s)^{2}\Big]\\
&\qquad +(1-\gamma)\,\mathbb{E}\big[
\Delta_h(\boldsymbol{\delta}_s)\,(\boldsymbol{x}^\top\boldsymbol{\delta}_s)^2\big].
\end{align*}
Define
$
\mathcal{H}:=\mathbb{E}\!\left[\big((1-\gamma)f_{\varepsilon|\boldsymbol{x}}(0)
+2\gamma\,\tau_{\min}\big)\,\boldsymbol{x}\boldsymbol{x}^\top\right],
$ and 
$\boldsymbol{\Sigma}:=\mathbb{E}[\boldsymbol{x}\boldsymbol{x}^\top].
$ 
Then 
\(\|\boldsymbol{u}\|_{\mathcal{H}}^2=\boldsymbol{u}^\top\mathcal{H}\boldsymbol{u}
=\mathbb{E}\big[((1-\gamma)f_{\varepsilon|\boldsymbol{x}}(0)+2\gamma\tau_{\min})(\boldsymbol{x}^\top\boldsymbol{u})^2\big]
\, 
\mbox{and} \, 
\|\boldsymbol{u}\|_{\boldsymbol{\Sigma}}^2=\mathbb{E}[(\boldsymbol{x}^\top\boldsymbol{u})^2].
\)
Using equation \eqref{eq:Delta-bound}, we have
\begin{align*}
(1-\gamma)\,\mathbb{E}\big[\Delta_h(\boldsymbol{\delta}_s)\,(\boldsymbol{x}^\top\boldsymbol{\delta}_s)^2\big]
&\ge
-(1-\gamma)L_f\,\mathbb{E}\Big[\big(|\boldsymbol{x}^\top\boldsymbol{\delta}_s|+\kappa_1 h\big)
|\boldsymbol{x}^\top\boldsymbol{\delta}_s|^2\Big] \\
&= -(1-\gamma)L_f\,\mathbb{E}\big[|\boldsymbol{x}^\top\boldsymbol{\delta}_s|^3\big]
   -(1-\gamma)L_f\,\kappa_1 h\,\|\boldsymbol{\delta}_s\|_{\boldsymbol{\Sigma}}^2.
\end{align*}
Therefore, for each $s\in[0,1]$,
\(
\big\langle 
\nabla^2\mathcal{H}^\gamma_h(\boldsymbol{\beta}_s)\,\boldsymbol{\delta}_s,\,
\boldsymbol{\delta}_s\big\rangle
\ge
\|\boldsymbol{\delta}_s\|_{\mathcal{H}}^2
-(1-\gamma)L_f\,\mathbb{E}\big[|\boldsymbol{x}^\top\boldsymbol{\delta}_s|^3\big]
-(1-\gamma)L_f\,\kappa_1 h\,\|\boldsymbol{\delta}_s\|_{\boldsymbol{\Sigma}}^2.
\)
Since $\boldsymbol{\delta}_s=s\boldsymbol{\delta}$, dividing by $s^2$ gives
\(
\big\langle 
\nabla^2\mathcal{H}^\gamma_h(\boldsymbol{\beta}_s)\,\boldsymbol{\delta},\,
\boldsymbol{\delta}\big\rangle
\ge
\|\boldsymbol{\delta}\|_{\mathcal{H}}^2
-(1-\gamma)L_f\,s\,\mathbb{E}\big[|\boldsymbol{x}^\top\boldsymbol{\delta}|^3\big]
-(1-\gamma)L_f\,\kappa_1 h\,\|\boldsymbol{\delta}\|_{\boldsymbol{\Sigma}}^2.
\)
Substituting this into equation \eqref{eq:a6-new-int} and integrating over $s\in[0,1]$ yields
\begin{align}\label{eq:a8-new-final}
&\big\langle \nabla \mathcal{H}^\gamma_h(\boldsymbol{\beta}_h^\dagger)
- \nabla \mathcal{H}^\gamma_h(\boldsymbol{\beta}^*),
\boldsymbol{\beta}_h^\dagger-\boldsymbol{\beta}^*\big\rangle \notag\\
&\quad\ge \|\boldsymbol{\beta}_h^\dagger-\boldsymbol{\beta}^*\|_{\mathcal{H}}^2
-\frac{(1-\gamma)L_f}{2}\,
\mathbb{E}\!\big[|\boldsymbol{x}^\top(\boldsymbol{\beta}_h^\dagger-\boldsymbol{\beta}^*)|^3\big] \notag\\
&\qquad -(1-\gamma)L_f\,\kappa_1 h\,
\|\boldsymbol{\beta}_h^\dagger-\boldsymbol{\beta}^*\|_{\boldsymbol{\Sigma}}^2.
\end{align}
\noindent
To control the cubic moment term, let $\boldsymbol{z}=\boldsymbol{\Sigma}^{-1/2}\boldsymbol{x}$ and assume
\(
\mathfrak{m}_3
:=\sup_{\boldsymbol{u}\in\mathbb{S}^{p-1}}
\mathbb{E}\big[|\boldsymbol{z}^\top\boldsymbol{u}|^3\big]
<\infty,
\,
\mathbb{S}^{p-1}
:=\big\{\boldsymbol{u}\in\mathbb{R}^p:\|\boldsymbol{u}\|_2=1\big\}.
\)
Then $\mathbb{E}[|\boldsymbol{x}^\top\boldsymbol{\delta}|^3]\le \mathfrak{m}_3\,\|\boldsymbol{\delta}\|_{\boldsymbol{\Sigma}}^3$,
so \eqref{eq:a8-new-final} implies
\begin{align}\label{eq:a8-new-final-mu}
\big\langle \nabla \mathcal{H}^\gamma_h(\boldsymbol{\beta}_h^\dagger)
- \nabla \mathcal{H}^\gamma_h(\boldsymbol{\beta}^*),\
\boldsymbol{\beta}_h^\dagger-\boldsymbol{\beta}^*\big\rangle
&\ge \|\boldsymbol{\beta}_h^\dagger-\boldsymbol{\beta}^*\|_{\mathcal{H}}^2
-\frac{(1-\gamma)L_f\,\mathfrak{m}_3}{2}\,
\|\boldsymbol{\beta}_h^\dagger-\boldsymbol{\beta}^*\|_{\boldsymbol{\Sigma}}^3 \notag\\
&\hspace{3em}
-(1-\gamma)L_f\,\kappa_1 h\,
\|\boldsymbol{\beta}_h^\dagger-\boldsymbol{\beta}^*\|_{\boldsymbol{\Sigma}}^2.
\end{align}
\noindent
We next derive an upper bound for the right-hand side of \eqref{eq:a5-new}. By the Cauchy--Schwarz inequality, we have
\(
\big\langle -\nabla \mathcal{H}^\gamma_h(\boldsymbol{\beta}^*),\ 
\boldsymbol{\beta}_h^\dagger-\boldsymbol{\beta}^*\big\rangle
\le \big\|\boldsymbol{\Sigma}^{-1/2}\nabla \mathcal{H}^\gamma_h(\boldsymbol{\beta}^*)\big\|_2\,
\|\boldsymbol{\beta}_h^\dagger-\boldsymbol{\beta}^*\|_{\boldsymbol{\Sigma}}.
\)
The correct hybrid gradient at $\boldsymbol{\beta}^*$ is
\(
\nabla \mathcal{H}^\gamma_h(\boldsymbol{\beta}^*)
=\mathbb{E}\!\left[\Big((1-\gamma)\big(\bar k(-\varepsilon/h)-\tau\big)
-\gamma\,\psi_\tau(\varepsilon)\Big)\,\boldsymbol{x}\right],
\)
where $k(v)=\int_{-\infty}^{v}k(s)\,ds$. 
A second-order Taylor expansion of $F_{\varepsilon|\boldsymbol{x}}$ combined with integration by parts yields
\(
\left\|\boldsymbol{\Sigma}^{-1/2}\mathbb{E}\!\left[(\bar k(-\varepsilon/h)-\tau)\,\boldsymbol{x}\right]\right\|_2
\le \frac{L_f}{2}\,\kappa_2\,h^2, 
\)
where $\kappa_2:=\int s^2 k(s)\,ds$. Using $|\psi_\tau(u)|\le 2\tau_{\max}|u|$ with $\tau_{\max}=\max(\tau,1-\tau)$ and
$\nu_\varepsilon := \sup_{\boldsymbol{x}}\mathbb{E}(|\varepsilon|\mid\boldsymbol{x})<\infty$, we obtain
\(
\big\|\boldsymbol{\Sigma}^{-1/2}\mathbb{E}[\psi_\tau(\varepsilon)\boldsymbol{x}]\big\|_2\le 2\,\tau_{\max}\,\nu_\varepsilon .
\)
Hence
\begin{equation}\label{eq:a9-new}
\big\|\boldsymbol{\Sigma}^{-1/2}\nabla \mathcal{H}^\gamma_h(\boldsymbol{\beta}^*)\big\|_2
\le (1-\gamma)\,\frac{L_f}{2}\,\kappa_2\,h^2+\,2\,\gamma\tau_{\max}\,\nu_\varepsilon .
\end{equation}
Combining equations \eqref{eq:a5-new}, \eqref{eq:a8-new-final-mu}, and \eqref{eq:a9-new}, we obtain the lower comparison between the local curvature norm and the design norm,
\(
\|\boldsymbol{\beta}_h^\dagger-\boldsymbol{\beta}^*\|_{\mathcal{H}}^2
\,\ge\,
D_\gamma\,\|\boldsymbol{\beta}_h^\dagger-\boldsymbol{\beta}^*\|_{\boldsymbol{\Sigma}}^2, 
\)
where $D_\gamma:=(1-\gamma)f_{\min}+2\gamma\tau_{\min}>0$ and $f_{\min}:=\inf_{\boldsymbol{x}} f_{\varepsilon\mid\boldsymbol{x}}(0)>0$. 
Let
$
r_h:=\|\boldsymbol{\beta}_h^\dagger-\boldsymbol{\beta}^*\|_{\boldsymbol{\Sigma}}.$ The combined upper bound is given by
\begin{equation}\label{eq:rh-ineq-cubic}
D_\gamma\,r_h^2
\,\le\,
\Big(\tfrac{\mathfrak{c}_\gamma}{2}\kappa_2 h^2+2\gamma\tau_{\max}\nu_\varepsilon\Big)\,r_h
\,+\,
\mathfrak{c}_\gamma\!\left(\frac{\mathfrak{m}_3}{2}\,r_h^3+\kappa_1 h\,r_h^2\right).
\end{equation}
By construction, $\boldsymbol{\beta}_h^\dagger\in\Theta_h$, and therefore $r_h\le h$. Using this in the
remainder terms yields the deterministic domination \(
\frac{\mathfrak{m}_3}{2}\,r_h^3+\kappa_1 h\,r_h^2
\,\le\,
\Big(\frac{\mathfrak{m}_3}{2}+\kappa_1\Big)h^2\,r_h,
\) and therefore, equation \eqref{eq:rh-ineq-cubic} implies the linear inequality
\(
D_\gamma\,r_h
\,\le\,
\tfrac{\mathfrak{c}_\gamma}{2}\kappa_2 h^2
+2\gamma\tau_{\max}\nu_\varepsilon
+\mathfrak{c}_\gamma\!\left(\frac{\mathfrak{m}_3}{2}+\kappa_1\right)h^2 .
\)
Equivalently,
\begin{equation}\label{eq:beta-local-bound-compact}
D_\gamma\,r_h
\,\le\,
A_\gamma\,h^2 + B_\gamma,
\end{equation}
where $A_\gamma:=\mathfrak{c}_\gamma\!\left(\frac{\kappa_2+\mathfrak{m}_3}{2}+\kappa_1\right)$ and $B_\gamma:=2\gamma\tau_{\max}\nu_\varepsilon.$ Dividing \eqref{eq:beta-local-bound-compact} by the positive constant $D_\gamma$ gives the explicit bound for $r_h$.
\begin{equation}\label{eq:rh-isolated-simplified}
r_h
\,\le\,
\frac{A_\gamma h^2+B_\gamma}{D_\gamma}
=
\frac{
\mathfrak{c}_\gamma\!\left(\frac{\kappa_2+\mathfrak{m}_3}{2}+\kappa_1\right)h^2
+2\gamma\tau_{\max}\nu_\varepsilon
}{
(1-\gamma)f_{\min}+2\gamma\tau_{\min}
}:=\mathfrak{M}_0.
\end{equation}
\noindent
A sufficient condition for \(r_h<h\) is obtained by assuming that
\begin{equation}\label{eq:interiority-condition}
\frac{A_\gamma h^2+B_\gamma}{D_\gamma}<h.
\end{equation}
Since $D_\gamma>0$ and $h>0$, \eqref{eq:interiority-condition} is equivalent to the quadratic inequality
\begin{equation}\label{eq:quad-cond}
A_\gamma h^2 - D_\gamma h + B_\gamma < 0.
\end{equation}
\noindent
Assuming \(A_\gamma>0\), the discriminant of the quadratic inequality \eqref{eq:quad-cond} with respect to \(h\) is
\(
\Delta_\gamma:=D_\gamma^2-4A_\gamma B_\gamma.
\)
If $\Delta_\gamma\le 0$, then $A_\gamma h^2 - D_\gamma h + B_\gamma\ge 0$ for all $h\in\mathbb{R}$,
so \eqref{eq:quad-cond} has no solution. If $\Delta_\gamma>0$, then the two real roots are
\(
h_{\gamma,-}:=\frac{D_\gamma-\sqrt{\Delta_\gamma}}{2A_\gamma},
\,
h_{\gamma,+}:=\frac{D_\gamma+\sqrt{\Delta_\gamma}}{2A_\gamma}.
\)
Since the parabola opens upward, we have
\(
A_\gamma h^2 - D_\gamma h + B_\gamma < 0
\,\Longleftrightarrow\,
h\in\bigl(h_{\gamma,-},\,h_{\gamma,+}\bigr).
\)
In particular, any choice of bandwidth $h$ satisfying
\(
0<h<h_{\gamma,+}
=\frac{D_\gamma+\sqrt{D_\gamma^2-4A_\gamma B_\gamma}}{2A_\gamma},
\,\text{with }\Delta_\gamma>0,
\)
ensures \eqref{eq:interiority-condition} and hence $r_h<h$.
If $A_\gamma=0$, then \eqref{eq:quad-cond} reduces to $-D_\gamma h + B_\gamma<0$, i.e.
\(
h>\frac{B_\gamma}{D_\gamma}
=
\frac{2\gamma\tau_{\max}\nu_\varepsilon}{(1-\gamma)f_{\min}+2\gamma\tau_{\min}}.
\)
\noindent
Under \eqref{eq:interiority-condition} (equivalently, \eqref{eq:quad-cond}), we have \(r_h<h\), and therefore
$\boldsymbol{\beta}_h^\dagger$ lies in the interior of $\Theta_h$. By definition, 
$\boldsymbol{\beta}_h^\dagger$ is the minimizer over $\Theta_h$, which forces the global minimizer,
$\boldsymbol{\beta}_h^*$, to belong to $\Theta_h$. Consequently,
\(
\boldsymbol{\beta}_h^\dagger=\boldsymbol{\beta}_h^*,
\)
and the bound \eqref{eq:rh-isolated-simplified} applies to
$\|\boldsymbol{\beta}_h^*-\boldsymbol{\beta}^*\|_{\boldsymbol{\Sigma}}$.
\noindent
We now prove uniqueness. Note that  
\begin{equation}\label{eq:B8-hybrid-uniq}
\mathbb{D}(\boldsymbol{\beta}_h^*,\boldsymbol{\beta}^*)
\,\ge\,
\|\boldsymbol{\beta}_h^*-\boldsymbol{\beta}^*\|_{\mathcal H}^{2}
-\frac{\mathfrak{c}_\gamma\,\mathfrak{m}_3}{2}\,
\|\boldsymbol{\beta}_h^*-\boldsymbol{\beta}^*\|_{\boldsymbol{\Sigma}}^{3}
-\mathfrak{c}_\gamma\,\kappa_1 h\,
\|\boldsymbol{\beta}_h^*-\boldsymbol{\beta}^*\|_{\boldsymbol{\Sigma}}^{2},
\end{equation}
Then, by  definition, we have the pointwise bound for the hybrid local curvature norm $\|\cdot\|_{\mathcal H}$,
\(
\|\boldsymbol{\beta}_h^*-\boldsymbol{\beta}^*\|_{\mathcal H}^{2}
\,\ge\,
D_\gamma\,\|\boldsymbol{\beta}_h^*-\boldsymbol{\beta}^*\|_{\boldsymbol{\Sigma}}^{2}.
\)
Substituting into \eqref{eq:B8-hybrid-uniq} gives
\(
\mathbb{D}(\boldsymbol{\beta}_h^*,\boldsymbol{\beta}^*)
\,\ge\,
\Big(D_\gamma-\mathfrak{c}_\gamma\kappa_1 h\Big)\,r_h^{2}
-\frac{\mathfrak{c}_\gamma\,\mathfrak{m}_3}{2}\,r_h^{3},
\)
where $r_h:=\|\boldsymbol{\beta}_h^*-\boldsymbol{\beta}^*\|_{\boldsymbol{\Sigma}}$.
\noindent
On the localized region $\Theta_h$, we have $r_h\le h$.
Hence,
\(
\mathbb{D}(\boldsymbol{\beta}_h^*,\boldsymbol{\beta}^*)
\,\ge\,
\Big(D_\gamma-\mathfrak{c}_\gamma(\kappa_1+\tfrac12\mathfrak{m}_3)h\Big)\,r_h^{2}.
\)
Therefore, if 
$
0<h<
\frac{D_\gamma}{\mathfrak{c}_\gamma(\kappa_1+\tfrac12\mathfrak{m}_3)},
$
then the right-hand side of the latter inequality is strictly positive for all $\boldsymbol{\beta}_h^*\neq\boldsymbol{\beta}^*$.
This implies that the function $\mathcal{H}_h^\gamma$ is strictly convex on $\Theta_h$.
Since the localization step already guarantees that $\boldsymbol{\beta}_h^*\in\Theta_h$,
strict convexity implies that $\boldsymbol{\beta}_h^*$ is the \emph{unique} minimizer of
$\boldsymbol{\beta}\mapsto\mathcal{H}_h^\gamma(\boldsymbol{\beta})$.
\noindent
We now control the bias remainder through a second-order expansion. Define the remainder vector as
$
\boldsymbol{D}_h
=\boldsymbol{\Sigma}^{-1/2}\Big\{\nabla\mathcal{H}^\gamma_h(\boldsymbol{\beta}_h^*)
-\nabla\mathcal{H}^\gamma_h(\boldsymbol{\beta}^*)
-\mathcal{H}(\boldsymbol{\beta}_h^*-\boldsymbol{\beta}^*)\Big\}.
$ 
By the mean value theorem, we have: 
\(
\boldsymbol{D}_h
=\Big\{\boldsymbol{\Sigma}^{-1/2}\Big(\int_0^1\nabla^2\mathcal{H}^\gamma_h\big((1-t)\boldsymbol{\beta}^*+t\boldsymbol{\beta}_h^*\big)\,dt\Big)\boldsymbol{\Sigma}^{-1/2}
-\mathcal{H}_0\Big\}\,\boldsymbol{\Sigma}^{1/2}(\boldsymbol{\beta}_h^*-\boldsymbol{\beta}^*),
\)
where 
$
\mathcal{H}_0=\boldsymbol{\Sigma}^{-1/2}\mathcal{H}\boldsymbol{\Sigma}^{-1/2}
=\mathbb{E}\!\left[((1-\gamma)f_{\varepsilon|\boldsymbol{x}}(0)+2\gamma\tau_{\min})\,\boldsymbol{z}\boldsymbol{z}^\top\right]$ and
$
\boldsymbol{z}=\boldsymbol{\Sigma}^{-1/2}\boldsymbol{x}.
$
Following the procedure outlined in the Hessian expansion and utilizing equation \eqref{eq:Delta-bound} alongside the definitions of $\kappa_1$ and the moment bound $\mathfrak{m}_3$, we arrive at the following result:
\allowdisplaybreaks
\begin{align}
&\left\|
\boldsymbol{\Sigma}^{-1/2}
\int_{0}^{1}\nabla^{2}\mathcal{H}^{\gamma}_h\!\big((1-t)\boldsymbol{\beta}^{*}+t\boldsymbol{\beta}^{*}_{h}\big)\,dt\,
\boldsymbol{\Sigma}^{-1/2}
-\mathcal{H}_{0}
\right\|_{2} \nonumber\\
&\,=
\left\|
\mathbb{E}\!\int_{0}^{1}\!\int_{-\infty}^{\infty}(1-\gamma)
k(s)\Big\{
f_{\varepsilon\mid \boldsymbol{x}}\!\big(t\langle \boldsymbol{x},\boldsymbol{\beta}^{*}_{h}-\boldsymbol{\beta}^{*}\rangle-hs\big)
- f_{\varepsilon\mid \boldsymbol{x}}(0)
\Big\}\,ds\,dt\,\boldsymbol{z}\boldsymbol{z}^{\top}
\right\|_{2} \nonumber\\
&\,\,
+\left\|
\gamma\,\mathbb{E}\!\int_{0}^{1}
\Big(
2\,w_{\tau}\!\big(\langle \boldsymbol{x},(1-t)\boldsymbol{\beta}^{*}+t\boldsymbol{\beta}^{*}_{h}\rangle-y\big)
-
2\,w_{\tau}\!\big(\langle \boldsymbol{x},\boldsymbol{\beta}^{*}\rangle-y\big)
\Big)\,dt\,\boldsymbol{z}\boldsymbol{z}^{\top}
\right\|_{2} \nonumber\\
&\,\le
(1-\gamma)\,L_{f}\sup_{\boldsymbol{\delta}\in\mathbb{S}^{p-1}}
\mathbb{E}\!\int_{0}^{1}\!\int_{-\infty}^{\infty}
k(s)\Big(t\big|\langle \boldsymbol{x},\boldsymbol{\beta}^{*}_{h}-\boldsymbol{\beta}^{*}\rangle\big|+h|s|\Big)
\,ds\,dt\,(\boldsymbol{z}^{\top}\boldsymbol{\delta})^{2} \nonumber\\
&\qquad+\gamma\,\tau_{\max}\sup_{\boldsymbol{\delta}\in\mathbb{S}^{p-1}}
\mathbb{E}\!\int_{0}^{1}(\boldsymbol{z}^{\top}\boldsymbol{\delta})^{2}\,dt \nonumber\\
&\,\le
\frac{(1-\gamma)L_{f}}{2}\sup_{\boldsymbol{\delta}\in\mathbb{S}^{p-1}}
\mathbb{E}\!\Big[\big|\langle \boldsymbol{x},\boldsymbol{\beta}^{*}_{h}-\boldsymbol{\beta}^{*}\rangle\big|\,
(\boldsymbol{z}^{\top}\boldsymbol{\delta})^{2}\Big] \nonumber\\
&\qquad +(1-\gamma)L_{f}\kappa_{1}h
+2\gamma\tau_{\max}\sup_{\boldsymbol{\delta}\in\mathbb{S}^{p-1}}
\mathbb{E}\!\Big[(\boldsymbol{z}^{\top}\boldsymbol{\delta})^{2}\Big] \nonumber\\
&\,\le
\frac{(1-\gamma)L_{f}}{2}\,\mathfrak{m}_{3}\,
\big\|\boldsymbol{\beta}^{*}_{h}-\boldsymbol{\beta}^{*}\big\|_{\boldsymbol{\Sigma}} \nonumber\\
&\qquad +(1-\gamma)L_{f}\kappa_{1}h
+2\gamma \tau_{\max}\nu_\varepsilon.
\end{align}
So that 
\begin{equation}\label{eq:a13-new}
\begin{aligned}
\|\boldsymbol{D}_h\|_2
&\le \mathfrak{c}_\gamma\Big(
\frac{\mathfrak{m}_3}{2}\|\boldsymbol{\beta}_h^*-\boldsymbol{\beta}^*\|_{\boldsymbol{\Sigma}}
+\kappa_1 h+2\gamma \tau_{\max}\nu_\varepsilon \Big)\\
&\qquad\times \|\boldsymbol{\beta}_h^*-\boldsymbol{\beta}^*\|_{\boldsymbol{\Sigma}}\\
&\le \mathfrak{c}_\gamma\Big(
\frac{\mathfrak{m}_3}{2}\mathfrak{M}_0+\kappa_1 h
+2\gamma \tau_{\max}\nu_\varepsilon \Big)\mathfrak{M}_0.
\end{aligned}
\end{equation}
Assume that the conditional density $f_{\varepsilon\mid\boldsymbol{x}}$ is twice continuously differentiable and that its derivative $f'_{\varepsilon\mid\boldsymbol{x}}$ is Lipschitz continuous.  
Consider the gradient of the hybrid population risk evaluated at $\boldsymbol{\beta}^*$,\\
\(
\nabla \mathcal{H}^\gamma_h(\boldsymbol{\beta}^*)
=
\mathbb{E}\!\left[
(1-\gamma)\{\bar{k}(-\varepsilon/h)-\tau\}
-\gamma\,\psi_\tau(\varepsilon)
\,\middle|\,
\boldsymbol{x}
\right]\boldsymbol{x}.
\)
Applying a second-order Taylor expansion of the conditional distribution function,
$F_{\varepsilon\mid\boldsymbol{x}}$, around zero yields
\(
\mathbb{E}\!\left[\bar{k}(-\varepsilon/h)\mid\boldsymbol{x}\right]-\tau
=
\frac{1}{2}\kappa_2 h^2\,f'_{\varepsilon\mid\boldsymbol{x}}(0)
+\mathrm{Rem}_q,
\)
with the remainder term satisfies
\(
|\mathrm{Rem}_q|
\le [\int_{-\infty}^{\infty}\int_{0}^{-hs}\int_{0}^{t}
k(s)\Bigl(f'_{\varepsilon\mid \boldsymbol{x}}(v)-f'_{\varepsilon\mid \boldsymbol{x}}(0)\Bigr)\,dv\,dt\,ds]             \le
\frac{L_{f'}}{6}\kappa_3 h^3,
\)
where 
$\kappa_3 := \int |s|^3|k(s)|\,ds$ and $L_{f'}$ denotes a Lipschitz constant of $f'_{\varepsilon\mid\boldsymbol{x}}$.
Moreover, since
$
|\psi_\tau(\varepsilon)| \le 2\tau_{\max}|\varepsilon|,
$
we have
\(
\big|\mathbb{E}[\psi_\tau(\varepsilon)\mid\boldsymbol{x}]\big|
\le
2\tau_{\max}\sup_{\boldsymbol{x}}\mathbb{E}(|\varepsilon|\mid\boldsymbol{x}) :=2\tau_{\max}\nu_\varepsilon.
\)
By combining the quantile and expectile contributions, we obtain
\(
\mathbb{E}\!\left[
(1-\gamma)\bar{k}(-\varepsilon/h)
-\tau
-\gamma\psi_\tau(\varepsilon)
\,\middle|\,
\boldsymbol{x}
\right]
=
\frac{1-\gamma}{2}\kappa_2 h^2\,f'_{\varepsilon\mid\boldsymbol{x}}(0)
+\mathrm{Rem},
\)
where the total remainder term satisfies
\(
|\mathrm{Rem}|
\le
\mathfrak{M}
:=
\frac{(1-\gamma)L_{f'}}{6}\kappa_3 h^3
+
2\gamma\tau_{\max}\nu_\varepsilon .
\)
Multiplying by $\boldsymbol{x}$ and taking expectations, then normalizing by
$\boldsymbol{\Sigma}^{-1/2}$ yields
\begin{equation}\label{eq:grad_bias_bound}
\left\|
\boldsymbol{\Sigma}^{-1/2}\nabla \mathcal{H}^\gamma_h(\boldsymbol{\beta}^*)
-
\frac{1}{2}(1-\gamma)\kappa_2 h^2\,
\boldsymbol{\Sigma}^{-1/2}\mathbb{E}\!\big[f'_{\varepsilon\mid\boldsymbol{x}}(0)\boldsymbol{x}\big]
-
\gamma\,\boldsymbol{\Sigma}^{-1/2}\mathbb{E}\!\big[\psi_\tau(\varepsilon)\boldsymbol{x}\big]
\right\|_2
\le
\mathfrak{M}.
\end{equation}
\noindent
We now linearize the argument around \texorpdfstring{$\boldsymbol{\beta}^*$}{beta-star}. The population smoothed minimizer \(\boldsymbol{\beta}_h^*\) satisfies the optimality condition
$\nabla \mathcal{H}^\gamma_h(\boldsymbol{\beta}_h^*)=\boldsymbol{0}$. From the definition of $\boldsymbol{D}_h$, we have 
\(
\boldsymbol{\Sigma}^{-1/2}\mathcal{H}(\boldsymbol{\beta}_h^*-\boldsymbol{\beta}^*)
=
-\boldsymbol{\Sigma}^{-1/2}\nabla\mathcal{H}^\gamma_h(\boldsymbol{\beta}^*)
-
\boldsymbol{D}_h .
\)
Now, we introduce the deterministic bias vector
\(
\boldsymbol{b}_h
:=
\frac{1}{2}(1-\gamma)\kappa_2 h^2\,
\boldsymbol{\Sigma}^{-1/2}\mathbb{E}\!\big[f'_{\varepsilon\mid\boldsymbol{x}}(0)\boldsymbol{x}\big]
+
\gamma\,\boldsymbol{\Sigma}^{-1/2}\mathbb{E}\!\big[\psi_\tau(\varepsilon)\boldsymbol{x}\big].
\)
 Adding $\boldsymbol{b}_h$ to both sides of the latter equation yields
\(
\boldsymbol{\Sigma}^{-1/2}\mathcal{H}(\boldsymbol{\beta}_h^*-\boldsymbol{\beta}^*)
+
\boldsymbol{b}_h
=
-
\Big(
\boldsymbol{\Sigma}^{-1/2}\nabla\mathcal{H}^\gamma_h(\boldsymbol{\beta}^*)
-
\boldsymbol{b}_h
\Big)
-
\boldsymbol{D}_h .
\) Taking Euclidean norms and applying the triangle inequality yields
\(
\left\|
\boldsymbol{\Sigma}^{-1/2}\mathcal{H}(\boldsymbol{\beta}_h^*-\boldsymbol{\beta}^*)
+
\boldsymbol{b}_h
\right\|_2
\le
\left\|
\boldsymbol{\Sigma}^{-1/2}\nabla\mathcal{H}^\gamma_h(\boldsymbol{\beta}^*)
-
\boldsymbol{b}_h
\right\|_2
+
\|\boldsymbol{D}_h\|_2 .
\)
The first term is bounded by $\mathfrak{M}$  from the previous expansion in \eqref{eq:grad_bias_bound}.  
Furthermore, as proven in \eqref{eq:a13-new}, the remainder term satisfies 
\(
\|\boldsymbol{D}_h\|_2
\le
\mathfrak{c}_\gamma
\Big(
\frac{\mathfrak{m}_3}{2}\mathfrak{M}_0
+
\kappa_1 h
+
2\gamma\tau_{\max}\nu_\varepsilon
\Big)\mathfrak{M}_0 .
\)
Combining the two bounds yields:
\begin{equation}\label{eq:lin_bound_Hb}
\left\|
\boldsymbol{\Sigma}^{-1/2}\mathcal{H}(\boldsymbol{\beta}_h^*-\boldsymbol{\beta}^*)
+
\boldsymbol{b}_h
\right\|_2
\le
\mathfrak{M}
+
\mathfrak{c}_\gamma
\Big(
\frac{\mathfrak{m}_3}{2}\mathfrak{M}_0
+
\kappa_1 h
+
2\gamma\tau_{\max}\nu_\varepsilon
\Big)\mathfrak{M}_0 .
\end{equation}
Then \eqref{eq:rh-isolated-simplified} and \eqref{eq:lin_bound_Hb} complete the proof of  Proposition \ref{proposition 1}.
\subsection{Proof of Proposition \ref{proposition 2}}
\noindent 
Assume that $\varepsilon \perp\!\!\!\perp \boldsymbol{x}$. From \eqref{eq c}, write \(\mathcal{C}^{\gamma}_{\tau,h}(s)=(1-\gamma)\rho_{\tau,h}(s)+\gamma\ell_\tau(s)\), where $\rho_{\tau,h}$ is the smoothed quantile loss associated with $\rho_\tau(s)=s\{\tau-\mathds{1}\{s<0\}\}$ and satisfies \(\rho_{\tau,h}'(s)=\bar k(s/h)-\tau\) and \(\rho_{\tau,h}''(s)=h^{-1}k(s/h)\), while $\ell_\tau$ is the asymmetric expectile loss with \(\ell_\tau'(s)=2\omega_\tau(s)s\), \(\ell_\tau''(s)=2\omega_\tau(s)\), and \(\omega_\tau(s)=\tau\,\mathds{1}\{s\ge0\}+(1-\tau)\,\mathds{1}\{s<0\}\). Recall that \(\vartheta(\alpha):=\mathbb{E}\!\left[\mathcal{C}^{\gamma}_{\tau,h}(\varepsilon-\alpha)-\mathcal{C}^{\gamma}_{\tau,h}(\varepsilon)\right]\), \(\alpha\in\mathbb{R}\). Differentiating under the expectation gives
\begin{align}\label{eq:vartheta-prime-split}
\vartheta'(\alpha)
&= -\,\mathbb{E}\!\left[\mathcal{C}^{\gamma\,\prime}_{\tau,h}(\varepsilon-\alpha)\right] \notag\\
&= -(1-\gamma)\,\mathbb{E}\!\left[\rho_{\tau,h}'(\varepsilon-\alpha)\right]
-\gamma\,\mathbb{E}\!\left[\ell_\tau'(\varepsilon-\alpha)\right] \notag\\
&=: (1-\gamma)\,\vartheta_q'(\alpha)+\gamma\,\vartheta_e'(\alpha).
\end{align}
where \(\vartheta_q'(\alpha):=-\mathbb{E}\!\left[\rho_{\tau,h}'(\varepsilon-\alpha)\right]\) and \(\vartheta_e'(\alpha):=-\mathbb{E}\!\left[\ell_\tau'(\varepsilon-\alpha)\right]\). For the quantile part, since \(\rho'_{\tau,h}(s)=\bar k(s/h)-\tau\),
\begin{align*}
\vartheta_q'(\alpha)
= -\,\mathbb{E}\!\left[\bar k\!\left(\frac{\varepsilon-\alpha}{h}\right)-\tau\right]
= -\,\mathbb{E}\!\left[\bar k\!\left(\frac{\varepsilon-\alpha}{h}\right)\right]+\tau 
= -\int_{-\infty}^{\infty}\bar k\!\left(\frac{u-\alpha}{h}\right)f_\varepsilon(u)\,du+\tau\\
= -\int_{-\infty}^{\infty}\bar k(s)\,f_\varepsilon(\alpha+hs)\,h\,ds+\tau,
\end{align*}
after the change of variables \(s=(u-\alpha)/h\). Since \(0\le \bar k(\cdot)\le 1\), \(\bar k(t)\to 0\) as \(t\to-\infty\), and \(\bar k(t)\to 1\) as \(t\to+\infty\), dominated convergence yields
\begin{equation}\label{eq:limit-quantile-part}
\lim_{\alpha\to+\infty}\vartheta_q'(\alpha)=\tau,
\,
\lim_{\alpha\to-\infty}\vartheta_q'(\alpha)=-(1-\tau).
\end{equation}

For the expectile part, since \(\ell_\tau'(s)=2\omega_\tau(s)s\),
\(\vartheta_e'(\alpha)=-\mathbb{E}\!\left[2\omega_\tau(\varepsilon-\alpha)(\varepsilon-\alpha)\right]
=2\,\mathbb{E}\!\left[\omega_\tau(\varepsilon-\alpha)(\alpha-\varepsilon)\right]\), and therefore
\begin{align}\label{eq:expectile-improper}
\vartheta_e'(\alpha)
&=2\int_{-\infty}^{\infty}
\omega_\tau(u-\alpha)\,(\alpha-u)\,f_\varepsilon(u)\,du \notag\\
&=2\left\{\tau\int_{\alpha}^{\infty}(\alpha-u)f_\varepsilon(u)\,du
+(1-\tau)\int_{-\infty}^{\alpha}(\alpha-u)f_\varepsilon(u)\,du\right\}.
\end{align}
Assume $\mathbb{E}|\varepsilon|<\infty$. Then both improper integrals are well defined. As \(\alpha\to+\infty\), we have \(\omega_\tau(u-\alpha)\to 1-\tau\) for each fixed \(u\), while \(0\le \omega_\tau(u-\alpha)(\alpha-u)\le \alpha+|u|\) and \(\int(\alpha+|u|)f_\varepsilon(u)\,du=\alpha+\mathbb{E}|\varepsilon|<\infty\). Hence
\(
\vartheta_e'(\alpha)
\sim 2(1-\tau)\int_{-\infty}^{\infty}(\alpha-u)f_\varepsilon(u)\,du
=2(1-\tau)(\alpha-\mathbb{E}\varepsilon)\xrightarrow[\alpha\to+\infty]{}+\infty.
\)
Similarly, as \(\alpha\to-\infty\), \(\omega_\tau(u-\alpha)\to \tau\) and
\(
\vartheta_e'(\alpha)\sim 2\tau(\alpha-\mathbb{E}\varepsilon)\xrightarrow[\alpha\to-\infty]{}-\infty.
\)
Thus
\begin{equation}\label{eq:limit-expectile-part}
\lim_{\alpha\to+\infty}\vartheta_e'(\alpha)=+\infty,
\,
\lim_{\alpha\to-\infty}\vartheta_e'(\alpha)=-\infty.
\end{equation}

Differentiating once more,
\(
\vartheta''(\alpha)
= (1-\gamma)\,\mathbb{E}\!\left[\frac{1}{h}\,k\!\left(\frac{\varepsilon-\alpha}{h}\right)\right]
+2\gamma\,\mathbb{E}\!\left[\omega_\tau(\varepsilon-\alpha)\right]
\ge 2\gamma\,\tau_{\min}>0,
\)
so $\vartheta$ is strictly convex and $\vartheta'$ is strictly increasing. Combining \eqref{eq:vartheta-prime-split}, \eqref{eq:limit-quantile-part}, and \eqref{eq:limit-expectile-part}, we obtain \(\lim_{\alpha\to-\infty}\vartheta'(\alpha)=-\infty\) and \(\lim_{\alpha\to+\infty}\vartheta'(\alpha)=+\infty\). By continuity, the intermediate value theorem ensures the existence of a unique \(b_h\in\mathbb{R}\) such that \(\vartheta'(b_h)=0\). Since $\vartheta$ is strictly convex, this \(b_h\) is the unique finite minimizer of $\vartheta$, which completes the proof of Proposition \ref{proposition 2}. \qed
\subsection{Proof of Proposition \ref{proposition 3}}

\noindent By Proposition \ref{proposition 2}, the one-dimensional population objective \(\vartheta(\alpha):=\mathbb{E}\!\left[\mathcal{C}^\gamma_{\tau,h}(\varepsilon-\alpha)-\mathcal{C}^\gamma_{\tau,h}(\varepsilon)\right]\), \(\alpha\in\mathbb{R}\), is strictly convex and admits a unique minimizer \(b_h\). We show that the global minimizer of \(\mathcal{H}_h^\gamma\) is obtained by shifting only the intercept. Let \(\beta_1^*\) denote the intercept and write \(\boldsymbol{x}=(1,\boldsymbol{x}_{-}^{\top})^{\top}\), with \(\boldsymbol{x}_{-}\in\mathbb{R}^{p-1}\). For any \(\boldsymbol{\beta}=(\beta_1,\ldots,\beta_p)^{\top}\in\mathbb{R}^p\), define \(\Delta(\boldsymbol{\beta}):=(\beta_1-\beta_1^*)+\boldsymbol{x}_{-}^{\top}(\boldsymbol{\beta}_{-}-\boldsymbol{\beta}_{-}^*)\). Then, by the definition of \(\mathcal{H}_h^\gamma\) and iterated expectation,
\begin{align*}
\mathcal{H}^\gamma_h(\boldsymbol{\beta})
&= \mathbb{E}\!\left[
\mathcal{C}^\gamma_{\tau,h}\!\left(\varepsilon-\Delta(\boldsymbol{\beta})\right)
-\mathcal{C}^\gamma_{\tau,h}(\varepsilon)\right]\\
&= \mathbb{E}_{\boldsymbol{x}}\!\left[
\mathbb{E}\!\left[
\mathcal{C}^\gamma_{\tau,h}\!\left(\varepsilon-\Delta(\boldsymbol{\beta})\right)
-\mathcal{C}^\gamma_{\tau,h}(\varepsilon)\,\middle|\,\boldsymbol{x}\right]\right]\\
&= \mathbb{E}\!\left[\vartheta\!\left(\Delta(\boldsymbol{\beta})\right)\right]
\ge \vartheta(b_h)\\
&= \mathbb{E}\!\left[
\mathcal{C}^\gamma_{\tau,h}(\varepsilon-b_h)
-\mathcal{C}^\gamma_{\tau,h}(\varepsilon)\right]
= \mathcal{H}^\gamma_h(\boldsymbol{\beta}^\star).
\end{align*}
where \(\boldsymbol{\beta}^\star=(\beta_1^*+b_h,\boldsymbol{\beta}_{-}^{*\top})^\top\). Hence \(\mathcal{H}^\gamma_h(\boldsymbol{\beta}^\star)=\min_{\boldsymbol{\beta}\in\mathbb{R}^p}\mathcal{H}^\gamma_h(\boldsymbol{\beta})\). We next show that this minimizer is unique. The Hessian of \(\mathcal{H}^\gamma_h\) is
\[
\begin{aligned}
\nabla^2 \mathcal{H}^\gamma_h(\boldsymbol{\beta})
={}&\mathbb{E}\!\left[\Big\{(1-\gamma)\,k_h\!\big(\boldsymbol{x}^{\top}(\boldsymbol{\beta}-\boldsymbol{\beta}^*)-\varepsilon\big)\right.\\
&\left.\qquad+2\gamma\,\omega_\tau\!\big(\varepsilon-\boldsymbol{x}^{\top}(\boldsymbol{\beta}-\boldsymbol{\beta}^*)\big)\Big\}\,\boldsymbol{x}\boldsymbol{x}^{\top}\right].
\end{aligned}
\]
Evaluating at \(\boldsymbol{\beta}^\star\), so that \(\boldsymbol{x}^\top(\boldsymbol{\beta}^\star-\boldsymbol{\beta}^*)=b_h\), gives
\[
\begin{aligned}
\nabla^2 \mathcal{H}^\gamma_h(\boldsymbol{\beta}^\star)
&= \mathbb{E}\!\left[\Big\{(1-\gamma)\,k_h(b_h-\varepsilon)+2\gamma\,\omega_\tau(\varepsilon-b_h)\Big\}\,\boldsymbol{x}\boldsymbol{x}^{\top}\right]\\
&= \vartheta''(b_h)\,\mathbb{E}[\boldsymbol{x}\boldsymbol{x}^{\top}],
\end{aligned}
\]
where the last identity follows from the representation of \(\vartheta''\). Since \(\vartheta''(b_h)>0\) and \(\mathbb{E}[\boldsymbol{x}\boldsymbol{x}^{\top}]\) is positive definite, the Hessian is positive definite. Therefore \(\boldsymbol{\beta}^\star\) is the unique minimizer of \(\mathcal{H}^\gamma_h\), and thus \(\boldsymbol{\beta}_h^*=\boldsymbol{\beta}^\star\).
\noindent
To complete the argument, we introduce a localized comparison point. Define \(\tilde b\) as follows: if \(|b_h|\le \kappa_2^{1/2}h\), set \(\tilde b=b_h\); otherwise take
\(\tilde b=\zeta b_h\) for some \(\zeta\in(0,1)\) such that \(|\tilde b|=\kappa_2^{1/2}h\).
By Assumption~\ref{A3} and \eqref{eq:a5-new} (monotonicity of \(\vartheta'\) and the fact \(\vartheta'(b_h)=0\)), we get
\begin{equation}
0\le \big[\vartheta'(\tilde b)-\vartheta'(0)\big]\tilde b
\le \big[\vartheta'(b_h)-\vartheta'(0)\big]\tilde b
= -\,\vartheta'(0)\,\tilde b.
\label{eq:B1}
\end{equation}
Using \(\vartheta'(\tilde b)-\vartheta'(0)=\int_0^{\tilde b}\vartheta''(t)\,dt\) and the representation of \(\vartheta''\), we have that
\begin{align*}
\vartheta'(\tilde b)-\vartheta'(0)
&= \int_{0}^{\tilde b}\!\!\int_{-\infty}^{\infty}
\Big\{(1-\gamma)\,k(s)f_\varepsilon(t-hs)\\
&\hspace{4.0cm}+2\gamma\,\omega_\tau(s)f_\varepsilon(s)\Big\}\,ds\,dt \\
&= (1-\gamma)f_\varepsilon(0)\,\tilde b
+ \int_{0}^{\tilde b}\!\!\int_{-\infty}^{\infty}
\Big\{(1-\gamma)k(s)
\big[f_\varepsilon(t-hs)-f_\varepsilon(0)\big]\\
&\hspace{4.0cm}+2\gamma\omega_\tau(s)f_{\varepsilon}(s)\Big\}\,ds\,dt,
\end{align*}
where we used \(\int k(s)\,ds=1\). Assumption \ref{A3} assumes that the function \(f_\varepsilon\) is Lipschitz with a constant \(L_f\), i.e.
\(|f_\varepsilon(u)-f_\varepsilon(v)|\le L_f|u-v|\).
Furthermore, since \(\omega_\tau(\cdot)\in[\tau_{\min},\tau_{\max}]\), we have the crude bound
\(|\omega_\tau(s)|\le \tau_{\max}\).Then for \(t\in[0,|\tilde b|]\) and any \(s\), we have 
$
|f_\varepsilon(t-hs)-f_\varepsilon(0)| \le L_f(|t|+h|s|)$ and $
|\omega_\tau(s)| \le \tau_{\max}.
$ Using \(\kappa_1:=\int |s|k(s)\,ds\) and \(\int k(s)\,ds=1\), we obtain
\(
\left|\int_{0}^{\tilde b}\!\!\int_{-\infty}^{\infty}k(s)
      \Big\{(1-\gamma)\big[f_\varepsilon(t-hs)-f_\varepsilon(0)\big]
          \Big\}ds dt\right|
\le \mathfrak{c}_\gamma\!\left(\frac{|\tilde b|^2}{2}+\kappa_1 h|\tilde b|\right).
\) Therefore,
\begin{equation}
\big[\vartheta'(\tilde b)-\vartheta'(0)\big]\tilde b
\,\ge\, (1-\gamma)f_\varepsilon(0)\,\tilde b^{\,2}
- \mathfrak{c}_\gamma\!\left(\frac{|\tilde b|^{3}}{2}+\kappa_1 h\,\tilde b^{\,2}\right)
+2\gamma \tau_{\min}\,\tilde b^{\,2},
\label{eq:LHS-lb}
\end{equation}
where we set \(\mathfrak{c}_\gamma:=(1-\gamma)L_f\). We next derive an upper bound for the right-hand side. Let
\(
F_\varepsilon(u)=\int_{-\infty}^{u} f_\varepsilon(v)\,dv.
\) By the definition of \(\vartheta'\) at \(0\), we have
\(
|\vartheta'(0)|
=
\left|
\int
\Big\{
(1-\gamma)
\,k(s)\big[F_\varepsilon(-hs)-F_\varepsilon(0)\big]
-2\gamma\, s\, \omega_\tau(s) f_\varepsilon(s)
\Big\}
\,ds
\right|.
\) Using the Lipschitz property of $f_\varepsilon$, namely
\(
|F_\varepsilon(u)-F_\varepsilon(0)|\le \tfrac{L_f}{2}|u|^2
\), together with the bound $\omega_\tau\le\tau_{\max}$ and the definition \(
\kappa_2:=\int s^2 k(s)\,ds
\), we obtain \(
|\vartheta'(0)|
\le
\frac{\mathfrak{c}_\gamma}{2}\,\kappa_2\,h^2
+2\gamma\,\tau_{\max}\,\kappa_2^{1/2}\,\nu_\varepsilon .
\) Therefore, using \eqref{eq:B1}, we can conclude that 
\begin{equation}
0
\le
\big[\vartheta'(\tilde b)-\vartheta'(0)\big]\tilde b
\le
\left(
\frac{\mathfrak{c}_\gamma}{2}\,\kappa_2\,h^2
+2\gamma\,\tau_{\max}\,\nu_\varepsilon
\right)
|\tilde b|.
\label{eq:RHS-linear}
\end{equation}
Combining \eqref{eq:B1} and \eqref{eq:RHS-linear}, we obtain
\begin{equation}\label{eq:RHS-raw}
[(1-\gamma)f_\varepsilon(0)+2\gamma \tau_{\min}]\,\tilde b^{\,2}
\le
\mathfrak{c}_\gamma\!\left(\frac{|\tilde b|^3}{2}+\kappa_1 h\,\tilde b^{\,2}\right)
+\left(\frac{\mathfrak{c}_\gamma}{2}\kappa_2 h^2+2\gamma\tau_{\max}\nu_\varepsilon\right)|\tilde b|.
\end{equation}
On the event \( |\tilde b|\le \kappa_2^{1/2}h \), we have \( |\tilde b|^3 \le \kappa_2^{1/2}h\,\tilde b^{\,2} \) and \( \tilde b^{\,2} \le \kappa_2^{1/2}h\,|\tilde b| \). Substituting these bounds into \eqref{eq:RHS-raw} yields
\(
[(1-\gamma)f_\varepsilon(0)+2\gamma \tau_{\min}]\,|\tilde b|^2
\le
\Big[\mathfrak{c}_\gamma h^2\big(\kappa_2+\kappa_1\kappa_2^{1/2}\big)+2\gamma\tau_{\max}\nu_\varepsilon\Big]|\tilde b|.
\)
Cancelling \( |\tilde b| \) when \( \tilde b\neq0 \) gives
\begin{equation}\label{eq:btilde-bound}
[(1-\gamma)f_\varepsilon(0)+2\gamma \tau_{\min}]\,|\tilde b|
\le
\mathfrak{c}_\gamma h^2\big(\kappa_2+\kappa_1\kappa_2^{1/2}\big)+2\gamma\tau_{\max}\nu_\varepsilon.
\end{equation}
Equivalently, \( |\tilde b|
\le
\big[\mathfrak{c}_\gamma h^2\big(\kappa_2+\kappa_1\kappa_2^{1/2}\big)+2\gamma\tau_{\max}\nu_\varepsilon\big]/\big[(1-\gamma)f_\varepsilon(0)+2\gamma \tau_{\min}\big] \). If
\( \big[\mathfrak{c}_\gamma h^2\big(\kappa_2+\kappa_1\kappa_2^{1/2}\big)+2\gamma\tau_{\max}\nu_\varepsilon\big]/\big[(1-\gamma)f_\varepsilon(0)+2\gamma \tau_{\min}\big] < \kappa_2^{1/2}h \),
then \( |\tilde b|\le \kappa_2^{1/2}h \), so the localization argument is self-consistent and the same bound holds for \(b_h\). Combining \eqref{eq:LHS-lb} with \eqref{eq:btilde-bound}, we conclude that \( |\tilde b|=O(h^2)+O(\gamma h) \). In the regime \(h\to0\) with fixed \(\gamma\), this gives the desired rate for \(b_h\). Finally, for sufficiently small \(h\), we necessarily have \(\tilde b=b_h\), which completes the proof of Proposition \ref{proposition 3}. \qed

\subsection{Proof of Theorem \ref{theorem 1}}\label{sec:proof-thm1}

\noindent In analogy with the population curvature measure defined above, we write its empirical counterpart as
 \(\widehat{\mathcal{H}}^{\gamma}_{h}(\cdot)\): \(\mathbb{D}(\boldsymbol{\beta}_1,\boldsymbol{\beta}_2):=\big\langle \nabla \widehat{\mathcal{H}}^{\gamma}_{h}(\boldsymbol{\beta}_1)-\nabla \widehat{\mathcal{H}}^{\gamma}_{h}(\boldsymbol{\beta}_2),\,\boldsymbol{\beta}_1-\boldsymbol{\beta}_2\big\rangle\). We work under Assumptions~\ref{A1}, \ref{A2}, \ref{A4}, and \ref{A5}. Let \(\varrho,\eta>0\) and define the local cone neighborhood
\begin{equation}\label{eq:theta-cone}
\Theta=\Theta(\varrho,\eta)
=\left\{\boldsymbol{\beta}\in\mathbb{R}^p:\ \boldsymbol{\delta}:=\boldsymbol{\beta}-\boldsymbol{\beta}^*
\in \mathbb{B}_{\boldsymbol{\Sigma}}(\varrho)\cap \mathbb{C}_{\boldsymbol{\Sigma}}(\eta)\right\},
\end{equation}
where \(\mathbb{B}_{\boldsymbol{\Sigma}}(\varrho)=\{\boldsymbol{\delta}\in\mathbb{R}^p:\ \|\boldsymbol{\delta}\|_{\boldsymbol{\Sigma}}\le \varrho\}\) and \(\mathbb{C}_{\boldsymbol{\Sigma}}(\eta)=\{\boldsymbol{\delta}\in\mathbb{R}^p:\ \|\boldsymbol{\delta}\|_1\le \eta\,\|\boldsymbol{\delta}\|_{\boldsymbol{\Sigma}}\}\). Throughout, for \(\boldsymbol{\delta}\neq \boldsymbol{0}\), let \(U_{i,\boldsymbol{\delta}}:=\langle \boldsymbol{x}_i,\boldsymbol{\delta}\rangle/\|\boldsymbol{\delta}\|_{\boldsymbol{\Sigma}}\). We begin with the gradient increment and its hybrid decomposition. Recall that the empirical smoothed HQER objective in~\eqref{eq:sqrlasso} is convex and twice continuously differentiable. Using \eqref{gradient-hessian}, for any \(\boldsymbol{\beta}\in\mathbb{R}^p\),
\(
\nabla \widehat{\mathcal{H}}_h^\gamma(\boldsymbol{\beta})
=
\frac{1}{n}\sum_{i=1}^n
\Big\{(1-\gamma)\big(\bar k((\boldsymbol{x}_i^\top\boldsymbol{\beta}-y_i)/h)-\tau\big)
-\gamma\,\psi_\tau(y_i-\boldsymbol{x}_i^\top\boldsymbol{\beta})\Big\}\boldsymbol{x}_i,
\,
\bar k(t):=\int_{-\infty}^t k(s)\,ds .
\)
Fix \(\boldsymbol{\beta}\in\Theta\) and set \(\boldsymbol{\delta}=\boldsymbol{\beta}-\boldsymbol{\beta}^*\), so \(\|\boldsymbol{\delta}\|_{\boldsymbol{\Sigma}}\le\varrho\) and \(\|\boldsymbol{\delta}\|_1\le \eta\|\boldsymbol{\delta}\|_{\boldsymbol{\Sigma}}\) by \eqref{eq:theta-cone}. Let \(\varepsilon_i:=y_i-\boldsymbol{x}_i^\top\boldsymbol{\beta}^*\), so that \(\boldsymbol{x}_i^\top\boldsymbol{\beta}-y_i=\boldsymbol{x}_i^\top\boldsymbol{\delta}-\varepsilon_i\). Define

\begin{align}\label{eq:def-L}
\mathfrak{L}(\boldsymbol{\beta})
&:=\mathbb{D}(\boldsymbol{\beta},\boldsymbol{\beta}^*) \notag\\
&=\big\langle \nabla \widehat{\mathcal{H}}^\gamma_h(\boldsymbol{\beta})
-\nabla \widehat{\mathcal{H}}^\gamma_h(\boldsymbol{\beta}^*),
\boldsymbol{\delta}\big\rangle \notag\\
&=\frac{1-\gamma}{n}\sum_{i=1}^n
\Big\{\bar k\!\Big(\frac{\boldsymbol{x}_i^\top\boldsymbol{\delta}-\varepsilon_i}{h}\Big)
-\bar k\!\Big(\frac{-\varepsilon_i}{h}\Big)\Big\}
\langle \boldsymbol{x}_i,\boldsymbol{\delta}\rangle \notag\\
&\quad-\frac{\gamma}{n}\sum_{i=1}^n
\big(\psi_\tau(\varepsilon_i-\boldsymbol{x}_i^\top\boldsymbol{\delta})
-\psi_\tau(\varepsilon_i)\big)
\langle \boldsymbol{x}_i,\boldsymbol{\delta}\rangle .
\end{align}
where the last line follows by substituting the gradient formula and regrouping the quantile/expectile parts.
\noindent
We next show that the expectile component induces an empirical quadratic curvature under Assumption~\ref{A5}. Recall that \(\psi_\tau(s)=2\omega_\tau(s)s\), where \(\omega_\tau(s)\ge \tau_{\min}:=\min\{\tau,1-\tau\}>0\) by definition of the expectile score.
For any \(s,t\in\mathbb{R}\),
\begin{equation}\label{eq:psi-strong-mono}
\big(\psi_\tau(s+t)-\psi_\tau(s)\big)\,t \ \ge\ 2\tau_{\min}\,t^2
\,\text{(by monotonicity of \(s\mapsto \psi_\tau(s)\) and \(\omega_\tau(\cdot)\ge\tau_{\min}\)).}
\end{equation}
Taking \(s=\varepsilon_i\) and \(t=-\langle \boldsymbol{x}_i,\boldsymbol{\delta}\rangle\) (substitution) gives
\begin{align}
-\frac{\gamma}{n}\sum_{i=1}^n
\big(\psi_\tau(\varepsilon_i-\langle \boldsymbol{x}_i,\boldsymbol{\delta}\rangle)-\psi_\tau(\varepsilon_i)\big)\,
\langle \boldsymbol{x}_i,\boldsymbol{\delta}\rangle \notag
&\,\ge\ \frac{2\gamma\tau_{\min}}{n}\sum_{i=1}^n \langle \boldsymbol{x}_i,\boldsymbol{\delta}\rangle^2
\,\text{(by \eqref{eq:psi-strong-mono})}\notag\\
&\,=\ 2\gamma\tau_{\min}\,\boldsymbol{\delta}^\top
\widehat{\boldsymbol{\Sigma}}\,\boldsymbol{\delta}
\qquad\text{(definition of \(\widehat{\boldsymbol{\Sigma}}\))}\notag\\
&\,\ge\ 2\gamma\tau_{\min}\,\kappa_{\mathrm{RE}}\,
\|\boldsymbol{\delta}\|_{\boldsymbol{\Sigma}}^2
\qquad\text{(Assumption~\ref{A5}).}
\label{eq:expectile-curvature}
\end{align}
\noindent
We now derive a lower bound for the smoothed quantile contribution in~\eqref{eq:def-L} using a discrete truncation argument, following a line of reasoning similar to that of \citep{Tan2022}.
Following the truncation argument of \citet{Tan2022}, for each \(i=1,\dots,n\), define the (discrete) truncation variable and the event
\(
\zeta_i:=\mathds{1}\!\left\{|\varepsilon_i|\le \frac{h}{2}\right\},
\,
E_i:=\left\{|\varepsilon_i|\le \frac{h}{2}\right\}\cap
\left\{|\xi_{i,\boldsymbol{\delta}}|\le \frac{h}{2\varrho}\right\}.
\)
On \(E_i\), we have \(|\varepsilon_i|/h\le 1/2\) (by definition of \(E_i\)), and since \(\|\boldsymbol{\delta}\|_{\boldsymbol{\Sigma}}\le \varrho\) for \(\boldsymbol{\beta}\in\Theta\) (by \eqref{eq:theta-cone}),
\(
\left|\frac{\boldsymbol{x}_i^\top\boldsymbol{\delta}}{h}\right|
=
\frac{\|\boldsymbol{\delta}\|_{\boldsymbol{\Sigma}}}{h}\,|\xi_{i,\boldsymbol{\delta}}|
\le
\frac{\varrho}{h}\cdot \frac{h}{2\varrho}
=\frac{1}{2}
\) (by definition of \(\xi_{i,\boldsymbol{\delta}}\) and \(E_i\)).
Hence,
\(
\left|\frac{\boldsymbol{x}_i^\top\boldsymbol{\delta}-\varepsilon_i}{h}\right|
\le
\left|\frac{\boldsymbol{x}_i^\top\boldsymbol{\delta}}{h}\right|
+\left|\frac{\varepsilon_i}{h}\right|
\le 1
\) (triangle inequality).
Let
\begin{equation}\label{eq:def-kappa0}
\kappa_0:=\min_{|s|\le 1}k(s)>0
\,\text{(positivity of \(k\) on \([-1,1]\)).}
\end{equation}
By the mean value theorem, for each \(i\) there exists a scalar \(\vartheta_i\) between
\(-\varepsilon_i/h\) and \((\boldsymbol{x}_i^\top\boldsymbol{\delta}-\varepsilon_i)/h\) such that
\(
\bar k\!\Big(\frac{\boldsymbol{x}_i^\top\boldsymbol{\delta}-\varepsilon_i}{h}\Big)
-\bar k\!\Big(\frac{-\varepsilon_i}{h}\Big)
=
k(\vartheta_i)\cdot \frac{\boldsymbol{x}_i^\top\boldsymbol{\delta}}{h}
\) (mean value theorem).
On \(E_i\), \(|\vartheta_i|\le 1\) (since both endpoints lie in \([-1,1]\) by the previous bound), hence \(k(\vartheta_i)\ge \kappa_0\) (by \eqref{eq:def-kappa0}), and therefore
\(
\bar k\!\Big(\frac{\boldsymbol{x}_i^\top\boldsymbol{\delta}-\varepsilon_i}{h}\Big)
-\bar k\!\Big(\frac{-\varepsilon_i}{h}\Big)
\ \ge\
\kappa_0\cdot\frac{\boldsymbol{x}_i^\top\boldsymbol{\delta}}{h}
\) (lower bound on \(k(\vartheta_i)\)).
Multiplying both sides by \(\langle \boldsymbol{x}_i,\boldsymbol{\delta}\rangle\) and summing yields
\begin{equation}\label{eq:quantile-discrete}
\begin{aligned}
&\frac{1-\gamma}{n}\sum_{i=1}^n
\Big\{\bar k\!\Big(\frac{\boldsymbol{x}_i^\top\boldsymbol{\delta}-\varepsilon_i}{h}\Big)
-\bar k\!\Big(\frac{-\varepsilon_i}{h}\Big)\Big\}
\langle \boldsymbol{x}_i,\boldsymbol{\delta}\rangle\\
&\qquad\ge
(1-\gamma)\kappa_0\,\frac{1}{nh}\sum_{i=1}^n
\langle \boldsymbol{x}_i,\boldsymbol{\delta}\rangle^2\,\mathds{1}_{E_i}.
\end{aligned}
\end{equation}
\noindent
We next use the Lipschitz continuity of the smoothing function together with a normalization argument to remove the discontinuity in \eqref{eq:quantile-discrete}, we employ the smoothing technique of \citet{loh2017statistical}
(as in \cite{Tan2022}). For \(R>0\), define
\(
\varphi_R(s)=
\begin{cases}
s^2, & |s|\le R/2,\\
\{s-R\,\mathrm{sign}(s)\}^2, & R/2\le |s|\le R,\\
0, & |s|>R,
\end{cases}
\)
which is \(R\)-Lipschitz (by direct inspection) and satisfies
\begin{equation}\label{eq:phi-prop}
\begin{aligned}
\varphi_{cR}(cs)&=c^2\varphi_R(s), \qquad c>0,\\
s^2\mathds{1}\!\left\{|s|\le \frac{R}{2}\right\}
&\le \varphi_R(s)
\le s^2\mathds{1}\!\left\{|s|\le R\right\}.
\end{aligned}
\end{equation}
Taking \(R=h/(2\varrho)\) and using \eqref{eq:phi-prop} gives
\(
\mathds{1}_{E_i}\,\langle \boldsymbol{x}_i,\boldsymbol{\delta}\rangle^2
=
\|\boldsymbol{\delta}\|_{\boldsymbol{\Sigma}}^2\,\zeta_i\,
U_{i,\boldsymbol{\delta}}^2\,\mathds{1}\!\left\{|U_{i,\boldsymbol{\delta}}|\le \frac{h}{2\varrho}\right\}
\ \ge\
\|\boldsymbol{\delta}\|_{\boldsymbol{\Sigma}}^2\,\zeta_i\,\varphi_{h/(2\varrho)}(U_{i,\boldsymbol{\delta}}).
\)
By \eqref{eq:phi-prop} with \(R=h/(2\varrho)\).  Therefore, combining \eqref{eq:quantile-discrete} with \eqref{eq:expectile-curvature}, we obtain that for every \(\boldsymbol{\beta}\in\Theta\),
\begin{align*}
\mathfrak{L}(\boldsymbol{\beta})
&\ge (1-\gamma)\kappa_0\,
\|\boldsymbol{\delta}\|_{\boldsymbol{\Sigma}}^2
\frac{1}{nh}\sum_{i=1}^n
\varphi_{h/(2\varrho)}(U_{i,\boldsymbol{\delta}})\,\zeta_i\\
&\quad +2\gamma\tau_{\min}\kappa_{\mathrm{RE}}\,
\|\boldsymbol{\delta}\|_{\boldsymbol{\Sigma}}^2.
\end{align*}
Equivalently,
\begin{equation}\label{eq:L-normalized}
\frac{\mathfrak{L}(\boldsymbol{\beta})}{\|\boldsymbol{\beta}-\boldsymbol{\beta}^*\|_{\boldsymbol{\Sigma}}^2}
\ \ge\ 
(1-\gamma)\kappa_0\,\mathcal{L}^Q_0(\boldsymbol{\beta})+\gamma\,\mathcal{L}^E_0,
\, \forall\,\boldsymbol{\beta}\in\Theta,
\end{equation}
where
\(
\mathcal{L}^Q_0(\boldsymbol{\beta})
:=
\frac{1}{nh}\sum_{i=1}^n
\varphi_{h/(2\varrho)}(U_{i,\boldsymbol{\delta}})\,\zeta_i,
\,
\mathcal{L}^E_0:=2\tau_{\min}\kappa_{\mathrm{RE}},
\, \boldsymbol{\delta}=\boldsymbol{\beta}-\boldsymbol{\beta}^*.
\)
\noindent
We now derive a lower bound for \(\inf_{\Theta}\mathbb{E}\{\mathcal{L}^Q_0(\boldsymbol{\beta})\}\). First, since \(\zeta_i=\mathds{1}\!\left\{|\varepsilon_i|\le h/2\right\}\),
\(
\mathbb{E}(\zeta_i\mid \boldsymbol{x}_i)=\mathbb{P}(|\varepsilon_i|\le h/2\mid \boldsymbol{x}_i)
=\int_{-h/2}^{h/2} f_{\varepsilon\mid \boldsymbol{x}_i}(s)\,ds.
\)
Using Assumption~\ref{A1} (Lipschitz at \(0\)), we have
\begin{equation}\label{eq:zeta-cond-bnd}
\Big|\mathbb{E}(\zeta_i\mid \boldsymbol{x}_i)-h f_{\varepsilon\mid \boldsymbol{x}_i}(0)\Big|
\le \int_{-h/2}^{h/2}\Big|f_{\varepsilon\mid \boldsymbol{x}_i}(s)-f_{\varepsilon\mid \boldsymbol{x}_i}(0)\Big|\,ds
\le \frac{L_f h^2}{4}
\,\text{(by Assumption~\ref{A1}).}
\end{equation}
Hence, if \(h\le f_{\min}/(2L_f)\) (a bandwidth condition),
\(
\mathbb{E}(\zeta_i\mid \boldsymbol{x}_i)
\ge h f_{\min}-\frac{L_f h^2}{4}
\ge \frac{7}{8}f_{\min}h
\) (since \(L_f h\le f_{\min}/2\)).
Fix any \(\boldsymbol{\beta}\in\Theta\) and let \(\boldsymbol{\delta}=\boldsymbol{\beta}-\boldsymbol{\beta}^*\neq 0\).
Write \(U_{\boldsymbol{\delta}}=\boldsymbol{x}^\top\boldsymbol{\delta}/\|\boldsymbol{\delta}\|_{\boldsymbol{\Sigma}}\), so that \(\mathbb{E}(U_{\boldsymbol{\delta}}^2)=1\) (by definition of \(\|\cdot\|_{\boldsymbol{\Sigma}}\)).
Using \eqref{eq:phi-prop} and the tower property,
\begin{align}
\mathbb{E}\!\left[\varphi_{h/(2\varrho)}(U_{\boldsymbol{\delta}})\,\zeta\right]
&=\mathbb{E}\!\left[\varphi_{h/(2\varrho)}(U_{\boldsymbol{\delta}})\,
\mathbb{E}(\zeta\mid \boldsymbol{x})\right] \notag\\
&\ge \frac{7}{8}f_{\min}h\,
\mathbb{E}\!\left[U_{\boldsymbol{\delta}}^2\,
\mathds{1}\!\left\{|U_{\boldsymbol{\delta}}|\le \frac{h}{4\varrho}\right\}\right] \notag\\
&=\frac{7}{8}f_{\min}h\left(
1-\mathbb{E}\!\left[U_{\boldsymbol{\delta}}^2\,
\mathds{1}\!\left\{|U_{\boldsymbol{\delta}}|> \frac{h}{4\varrho}\right\}\right]\right).
\label{eq:EZphi}
\end{align}
Here the equality uses the tower property, the inequality follows from \eqref{eq:phi-prop} and the bound on \(\mathbb{E}(\zeta\mid\boldsymbol{x})\), and the final equality uses \(\mathbb{E}(U_{\boldsymbol{\delta}}^2)=1\).

Under Assumption~\ref{A4} (with \(a_1=1\)), for all \(s>0\),
\(
\mathbb{P}\!\left\{|U_{\boldsymbol{\delta}}|\ge a_0 s\right\}\le e^{-s}
\) (sub-exponential tail with \(a_1=1\)).
Hence, for any \(s>0\),
\begin{align}
\mathbb{E}\!\left[U_{\boldsymbol{\delta}}^2
\mathds{1}\!\left\{|U_{\boldsymbol{\delta}}|>s\right\}\right]
&=s^{2}\mathbb{P}\!\left\{|U_{\boldsymbol{\delta}}|>s\right\}
+2\int_{s}^{\infty} v\,
\mathbb{P}\!\left\{|U_{\boldsymbol{\delta}}|>v\right\}\,dv \notag\\
&\le s^{2}e^{-s/a_{0}}
+2\int_{s}^{\infty} v\,e^{-v/a_{0}}\,dv \notag\\
&=\bigl(s^{2}+2a_{0}s+2a_{0}^{2}\bigr)e^{-s/a_{0}}.
\label{eq:tail-second-moment}
\end{align}
The first identity is the standard tail-integral formula, the inequality uses the tail bound, and the last equality follows by direct integration.
Take \(s=h/(4\varrho)\). If \(\varrho \le h/(20a_0^2)\), then \(s/a_0\ge 5\) (by algebra), and thus
\(
\mathbb{E}\!\left[U_{\boldsymbol{\delta}}^2\mathds{1}\!\left\{|U_{\boldsymbol{\delta}}|>\frac{h}{4\varrho}\right\}\right]
\le
(25+10+2)e^{-5}
<\frac{1}{4}
\) (numerical bound using \(s/a_0\ge 5\)).
Consequently,
\(
\inf_{\boldsymbol{\beta}\in\Theta}
\mathbb{E}\!\left[
\frac{1}{h}\varphi_{h/(2\varrho)}(U_{\boldsymbol{\delta}})\,\zeta
\right]
\ge
\frac{7}{8}f_{\min}\cdot\frac{3}{4}
=
\frac{21}{32}f_{\min}
\) (plugging into \eqref{eq:EZphi}).
This holds as long as \(20a_0^2\varrho\le h\le f_{\min}/(2L_f)\) (collecting the bandwidth/radius conditions).
\noindent
It remains to establish a uniform deviation bound for \(\mathcal{L}^Q_0\). Define the fluctuation
\(
\Delta
:=
\sup_{\boldsymbol{\delta}\in\mathbb{C}_{\boldsymbol{\Sigma}}(\eta)}
\Big|
\mathcal{L}^Q_0(\boldsymbol{\delta})
-\mathbb{E}\{\mathcal{L}^Q_0(\boldsymbol{\delta})\}
\Big|.
\)
Write
\(
\omega_{\boldsymbol{\delta}}(\boldsymbol{x}_i,\varepsilon_i)
:=
\frac{1}{h}\,
\varphi_{h/(2\varrho)}\!\Big(U_{i,\boldsymbol{\delta}}\Big)\,\zeta_i,
\,
\mathcal{L}^Q_0(\boldsymbol{\delta})=\frac{1}{n}\sum_{i=1}^n \omega_{\boldsymbol{\delta}}(\boldsymbol{x}_i,\varepsilon_i)
\) (by definition of \(\mathcal{L}^Q_0\)).
\noindent
\noindent\textbf{(i) Envelope and variance proxy.}
Using \(\varphi_R(s)\le (R/2)|s|\) (a standard bound for this truncation function) and \(\omega_{\boldsymbol{\delta}}(\boldsymbol{x}_i,\varepsilon_i)\ge 0\),
\(
0\le \omega_{\boldsymbol{\delta}}(\boldsymbol{x}_i,\varepsilon_i)\le (4\varrho)^{-2}h
\) (since \(R=h/(2\varrho)\) and \(|U_{i,\boldsymbol{\delta}}|\le R\) on the support of \(\varphi_R\)).
Moreover, using \(\varphi_R(s)\le s^2\mathds{1}\!\left\{|s|\le R\right\}\) and \(\mathbb{E}(\zeta_i\mid \boldsymbol{x}_i)\le f_{\max}h+L_fh^2/4\) (by \eqref{eq:zeta-cond-bnd} and \(\sup f_{\varepsilon\mid x}(0)\le f_{\max}\)),
\(
\mathbb{E}\omega_{\boldsymbol{\delta}}^2(\boldsymbol{x}_i,\varepsilon_i)
\le (4\varrho)^{-2}\cdot \frac{9f_{\max}h}{8}
\) (same bounding route as \cite{Tan2022}, using \(h\le f_{\min}/(2L_f)\)).
Applying Theorem~7.3 of \citet{bousquet2003concentration} yields that, for any \(s>0\),
\begin{equation}\label{eq:bousquet}
\Delta
\le
\frac{5}{4}\mathbb{E}[\Delta]
+
\frac{3}{2}f_{\max}^{1/2}(4\varrho)^{-1}\sqrt{\frac{hs}{n}}
+\Big(4+\frac{1}{3}\Big)\frac{hs}{(4\varrho)^2n},
\end{equation}
with probability at least \(1-e^{-s}\) (Bousquet’s inequality for bounded empirical processes).
\noindent
\noindent\textbf{(ii) Bounding \(\mathbb{E}[\Delta]\) via symmetrization and contraction.}
By Rademacher symmetrization,
\(
\mathbb{E}[\Delta]
\le
2\,\mathbb{E}\!\left[
\sup_{\boldsymbol\delta\in\mathbb{C}_{\boldsymbol{\Sigma}}(\eta)}
\left|
\frac{1}{n}\sum_{i=1}^n e_i\,\omega_{\boldsymbol{\delta}}(\boldsymbol{x}_i,\varepsilon_i)
\right|
\right]
\)
where \(e_1,\dots,e_n\) are i.i.d.\ Rademacher signs independent of the data.
Since \(\varphi_{h/(2\varrho)}\) is \(\frac{h}{2\varrho}\)-Lipschitz (by construction), the map
\(t\mapsto h^{-1}\varphi_{h/(2\varrho)}(t)\) is \((2\varrho)^{-1}\)-Lipschitz (scaling).
Moreover, \(\omega_{\boldsymbol{\delta}}(\boldsymbol{x}_i,\varepsilon_i)=0\) whenever \(\zeta_iU_{i,\boldsymbol{\delta}}=0\) (by definition).
Thus, Talagrand's contraction principle yields
\begin{align}
\mathbb{E}[\Delta]
&\le
\frac{1}{\varrho}\,
\mathbb{E}\!\left[
\sup_{\boldsymbol{\delta}\in\mathbb{C}_{\boldsymbol{\Sigma}}(\eta)}
\left|
\frac{1}{n}\sum_{i=1}^n e_i\,\zeta_i\,U_{i,\boldsymbol{\delta}}
\right|
\right]
\,\text{(contraction principle)}\notag\\
&=
\frac{1}{\varrho}\,
\mathbb{E}\!\left[
\sup_{\boldsymbol{\delta}\in\mathbb{C}_{\boldsymbol{\Sigma}}(\eta)}
\left|
\left\langle
\frac{1}{n}\sum_{i=1}^n e_i\,\zeta_i\,\boldsymbol{x}_i,\,
\frac{\boldsymbol{\delta}}{\|\boldsymbol{\delta}\|_{\boldsymbol{\Sigma}}}
\right\rangle
\right|
\right]
\label{eq:EDelta-contraction}
\end{align}
The last equality uses \(U_{i,\boldsymbol{\delta}}=\langle \boldsymbol{x}_i,\boldsymbol{\delta}\rangle/\|\boldsymbol{\delta}\|_{\boldsymbol{\Sigma}}\).
Using \(|\langle \boldsymbol{u},\boldsymbol{v}\rangle|\le \|\boldsymbol{u}\|_\infty\|\boldsymbol{v}\|_1\) (Hölder) and the cone constraint
\(\|\boldsymbol{\delta}\|_1\le \eta\|\boldsymbol{\delta}\|_{\boldsymbol{\Sigma}}\) (by \(\boldsymbol{\delta}\in\mathbb{C}_{\boldsymbol{\Sigma}}(\eta)\)),\\
\(
\sup_{\boldsymbol{\delta}\in\mathbb{C}_{\boldsymbol{\Sigma}}(\eta)}
\left|
\left\langle
\frac{1}{n}\sum_{i=1}^n e_i\,\zeta_i\,\boldsymbol{x}_i,\,
\frac{\boldsymbol{\delta}}{\|\boldsymbol{\delta}\|_{\boldsymbol{\Sigma}}}
\right\rangle
\right|
\le
\eta\left\|
\frac{1}{n}\sum_{i=1}^n e_i\,\zeta_i\,\boldsymbol{x}_i
\right\|_\infty
\) (Hölder + cone constraint).
Hence,\\
\(
\mathbb{E}[\Delta]
\le
\frac{\eta}{\varrho}\,
\mathbb{E}\left\|
\frac{1}{n}\sum_{i=1}^n e_i\,\zeta_i\,\boldsymbol{x}_i
\right\|_\infty .
\)
\noindent
\noindent\textbf{(iii) A maximal inequality.}
For each \(1\le l\le p\), define \(S_l:=\sum_{i=1}^n e_i\,\zeta_i\,x_{il}\).
Then \(\mathbb{E}(e_i\zeta_i x_{il})=0\) (since \(\mathbb{E}(e_i)=0\) and \(e_i\) is independent),
and, using \eqref{eq:zeta-cond-bnd},
\(
\mathbb{E}(e_i\zeta_i x_{il})^2
=
\mathbb{E}(\zeta_i x_{il}^2)
\le
\mathbb{E}\!\left\{\mathbb{E}(\zeta_i\mid \boldsymbol{x}_i)\,x_{il}^2\right\}
\le
\sigma_{ll}\,c_h,
\,
c_h:=f_{\max}h+\frac{L_f h^2}{4}
\) (tower property + \eqref{eq:zeta-cond-bnd}).
Under Assumption~\ref{A4} (with \(a_1=1\)), \(\mathbb{P}\!\left\{|x_{il}|\ge a_0\sigma_{ll}^{1/2}s\right\}\le e^{-s}\), which implies
the Bernstein-type moment bound (for all integers \(j\ge 3\))
\(
\mathbb{E}|e_i\zeta_i x_{il}|^j
\le
j!\,a_0^j\sigma_{ll}^{j/2}c_h
\) (sub-exponential moments from the tail bound).
Standard Bernstein maximal inequalities (e.g., \citealp[Theorems~2.5 and~2.10]{boucheron2013concentration}) therefore yield
\(
\mathbb{E}\max_{1\le l\le p}|S_l|
\le
a_0\sigma_x\left\{
\sqrt{2n\,c_h\,\log(2p)}+2\log(2p)
\right\},
\, \sigma_x:=\max_{1\le l\le p}\sigma_{ll}^{1/2}
\) (maximal Bernstein inequality).
Thus,
\begin{equation}\label{eq:EDelta-final}
\mathbb{E}[\Delta]
\le
\frac{\eta}{\varrho}\,a_0\sigma_x\left\{
\sqrt{\frac{2c_h\log(2p)}{n}}+\frac{2\log(2p)}{n}
\right\}.
\end{equation}
Taking \(s=\log(2p)\) in \eqref{eq:bousquet} (a standard choice) and using \eqref{eq:EDelta-final}, we obtain that with probability at least \(1-(2p)^{-1}\),
\begin{equation}\label{eq:Delta-bound0}
\begin{aligned}
\Delta
&\le \frac{\eta}{\varrho}\,a_0\sigma_x\left\{
\frac{15}{8}f_{\max}^{1/2}\sqrt{\frac{h\log(2p)}{n}}
+\frac{5}{2}\frac{\log(2p)}{n}\right\}\\
&\quad +\frac{3}{2}f_{\max}^{1/2}(4\varrho)^{-1}
\sqrt{\frac{h\log(2p)}{n}}
+\Big(4+\frac{1}{3}\Big)\frac{h\log(2p)}{(4\varrho)^2n}.
\end{aligned}
\end{equation}
where we used \(c_h\le 2f_{\max}h\) when \(h\le f_{\min}/(2L_f)\) (by the definition of \(c_h\)).
\noindent
Combining \eqref{eq:L-normalized} with the expectation lower bound established above and the deviation bound \eqref{eq:Delta-bound0} yields the desired empirical restricted strong convexity bound, with probability at least
\(1-(2p)^{-1}\),
\(
\inf_{\boldsymbol{\beta}\in\Theta}
\frac{
\mathbb{D}(\boldsymbol{\beta},\boldsymbol{\beta}^*)
}{
\|\boldsymbol{\beta}-\boldsymbol{\beta}^*\|_{\boldsymbol{\Sigma}}^2
}
\ge
(1-\gamma)\kappa_0\Big(\frac{21}{32}f_{\min}-\Delta\Big)
+2\gamma\tau_{\min}\kappa_{\mathrm{RE}}
\) (by \eqref{eq:L-normalized}).
This holds provided \(20a_0^2\varrho\le h\le f_{\min}/(2L_f)\) (from Step~5).
In particular, if \(n\) is large enough so that \(\Delta\le \frac{5}{32}f_{\min}\) (a sample size condition), then
\(
\inf_{\boldsymbol{\beta}\in\Theta}
\frac{
\mathbb{D}(\boldsymbol{\beta},\boldsymbol{\beta}^*)
}{
\|\boldsymbol{\beta}-\boldsymbol{\beta}^*\|_{\boldsymbol{\Sigma}}^2
}
\ge \kappa:=
\frac{1}{2}(1-\gamma)\kappa_0 f_{\min}
+2\gamma\tau_{\min}\kappa_{\mathrm{RE}}
\) (since \(\frac{21}{32}-\frac{5}{32}=\frac{1}{2}\)).
This completes the proof.
\qed
\subsection{Proof of   Theorem  \ref{theorem 2-conv}}
\noindent
Let \( \mathcal{Q} \subseteq [p] \) denote the active set of \( \boldsymbol{\beta}^* \) with size \( q = |\mathcal{Q}| \).  
Consider the symmetric Bregman divergence between  
\( \boldsymbol{\widehat{\beta}}=\boldsymbol{\widehat{\beta}}_h \)  
and \( \boldsymbol{\beta}^* \),
\begin{equation}\label{a29}
\mathbb{D}(\boldsymbol{\widehat{\beta}},\boldsymbol{\beta}^*)
=
\Big\langle 
\nabla \widehat{\mathcal{H}}^\gamma_h(\boldsymbol{\widehat{\beta}})
- 
\nabla \widehat{\mathcal{H}}^\gamma_h(\boldsymbol{\beta}^*),\,
\boldsymbol{\widehat{\beta}} - \boldsymbol{\beta}^* 
\Big\rangle \ge 0,
\end{equation}
and note that the proof proceeds by deriving matching upper and lower bounds for \eqref{a29}.
\noindent
We begin by establishing an upper bound. By the first-order optimality of \eqref{smoothed lasso}, there exists 
\( \widehat{\boldsymbol{g}} \in \partial \|\boldsymbol{\widehat{\beta}}\|_1 \) such that
\(
\nabla \widehat{\mathcal{H}}^\gamma_h(\boldsymbol{\widehat{\beta}}) + \lambda \widehat{\boldsymbol{g}} = 0.
\)
Let \( \widehat{\boldsymbol{\delta}} := \boldsymbol{\widehat{\beta}} - \boldsymbol{\beta}^* \).  
Using subgradient calculus,
\(
\big\langle \widehat{\boldsymbol{g}}, \boldsymbol{\beta}^* - \boldsymbol{\widehat{\beta}} \big\rangle
\le \|\boldsymbol{\beta}^*\|_1 - \|\boldsymbol{\widehat{\beta}}\|_1 \notag= \|\boldsymbol{\beta}^*_{\mathcal{Q}}\|_1 - \|\boldsymbol{\beta}^*_{\mathcal{Q}}+\widehat{\boldsymbol{\delta}}_{\mathcal{Q}}\|_1-\|\widehat{\boldsymbol{\delta}}_{\mathcal{Q}^c}\|_1 \le \|\widehat{\boldsymbol{\delta}}_{\mathcal{Q}}\|_1 - \|\widehat{\boldsymbol{\delta}}_{\mathcal{Q}^c}\|_1 ,
\)
where the last inequality uses the reverse triangle inequality
\(\|\boldsymbol{\beta}^*_{\mathcal{Q}}+\widehat{\boldsymbol{\delta}}_{\mathcal{Q}}\|_1\ge
\|\boldsymbol{\beta}^*_{\mathcal{Q}}\|_1-\|\widehat{\boldsymbol{\delta}}_{\mathcal{Q}}\|_1\).
\noindent
Substituting the optimality condition into \eqref{a29} and adding/subtracting the population gradient at \(\boldsymbol{\beta}^*\) yields
\begin{align*}
&\Big\langle \nabla \widehat{\mathcal{H}}_h^\gamma(\boldsymbol{\widehat{\beta}})
- \nabla \widehat{\mathcal{H}}_h^\gamma(\boldsymbol{\beta}^*),
\widehat{\boldsymbol{\delta}} \Big\rangle\\
&= -\lambda \big\langle \widehat{\boldsymbol{g}},\widehat{\boldsymbol{\delta}}\big\rangle
+\Big\langle \nabla \widehat{\mathcal{H}}_h^\gamma(\boldsymbol{\beta}^*)
-\nabla \mathcal{H}_h^\gamma(\boldsymbol{\beta}^*),
-\widehat{\boldsymbol{\delta}}\Big\rangle\\
&\quad +\Big\langle \nabla \mathcal{H}_h^\gamma(\boldsymbol{\beta}^*),
-\widehat{\boldsymbol{\delta}}\Big\rangle \\
&\le \lambda\Big(\|\widehat{\boldsymbol{\delta}}_{\mathcal{Q}}\|_1
-\|\widehat{\boldsymbol{\delta}}_{\mathcal{Q}^c}\|_1\Big)
+ \|\boldsymbol{\omega}_h^*\|_\infty\,\|\widehat{\boldsymbol{\delta}}\|_1
+ b_h^*\,\|\widehat{\boldsymbol{\delta}}\|_{\boldsymbol{\Sigma}}.
\end{align*}
where
\(
\boldsymbol{\omega}_h^*:=\nabla \widehat{\mathcal{H}}_h^\gamma(\boldsymbol{\beta}^*)-\nabla \mathcal{H}_h^\gamma(\boldsymbol{\beta}^*),
\,
b_h^*
:=\big\|\boldsymbol{\Sigma}^{-1/2}\nabla \mathcal{H}_h^\gamma(\boldsymbol{\beta}^*)\big\|_2.
\)
(Here we used \(|\langle \boldsymbol{\omega}_h^*,\widehat{\boldsymbol{\delta}}\rangle|
\le \|\boldsymbol{\omega}_h^*\|_\infty\|\widehat{\boldsymbol{\delta}}\|_1\) and Cauchy--Schwarz for the last term.)
\noindent
Assume (as in Proposition \ref{proposition 3}) that
\(
b_h^*\le (1-\gamma)\frac{L_f}{2}\,\kappa_2 h^2 + 2\gamma \tau_{\max}\nu_\varepsilon,
\, \tau_{\max}:=\max\{\tau,1-\tau\}.
\)
\noindent
On the event \( \{\lambda \ge 2\|\boldsymbol{\omega}_h^*\|_\infty\} \), we have
\(\|\boldsymbol{\omega}_h^*\|_\infty\|\widehat{\boldsymbol{\delta}}\|_1\le (\lambda/2)\|\widehat{\boldsymbol{\delta}}\|_1\),
hence
\begin{align}\label{a31}
&\Big\langle \nabla \widehat{\mathcal{H}}_h^\gamma(\boldsymbol{\widehat{\beta}})
- \nabla \widehat{\mathcal{H}}_h^\gamma(\boldsymbol{\beta}^*),
\widehat{\boldsymbol{\delta}} \Big\rangle \notag\\
&\quad\le \lambda
\Big(\|\widehat{\boldsymbol{\delta}}_{\mathcal{Q}}\|_1
-\|\widehat{\boldsymbol{\delta}}_{\mathcal{Q}^c}\|_1
+\tfrac12\|\widehat{\boldsymbol{\delta}}\|_1\Big)
+ b_h^*\,\|\widehat{\boldsymbol{\delta}}\|_{\boldsymbol{\Sigma}} \notag\\
&\quad\le \frac{\lambda}{2}\Big(
3\|\widehat{\boldsymbol{\delta}}_{\mathcal{Q}}\|_1
-\|\widehat{\boldsymbol{\delta}}_{\mathcal{Q}^c}\|_1\Big)
+ b_h^*\,\|\widehat{\boldsymbol{\delta}}\|_{\boldsymbol{\Sigma}} \notag\\
&\quad\le \frac{3}{2}\lambda\,q^{1/2}\|\widehat{\boldsymbol{\delta}}\|_2
+ b_h^*\,\|\widehat{\boldsymbol{\delta}}\|_{\boldsymbol{\Sigma}}.
\end{align}
\noindent
Since the left-hand side of \eqref{a29} is nonnegative, \eqref{a31} implies the cone condition
\(
\|\widehat{\boldsymbol{\delta}}_{\mathcal{Q}^c}\|_1
\le 3\|\widehat{\boldsymbol{\delta}}_{\mathcal{Q}}\|_1
+2\lambda^{-1}b_h^*\,\|\widehat{\boldsymbol{\delta}}\|_{\boldsymbol{\Sigma}},
\)
and thus
\begin{align}\label{a32}
\|\widehat{\boldsymbol{\delta}}\|_1
&\le 4q^{1/2}\|\widehat{\boldsymbol{\delta}}\|_2
+2\lambda^{-1}b_h^*\,\|\widehat{\boldsymbol{\delta}}\|_{\boldsymbol{\Sigma}} \notag
\le \Big(4\mu_p^{-1/2}q^{1/2}+2\lambda^{-1}b_h^*\Big)\|\widehat{\boldsymbol{\delta}}\|_{\boldsymbol{\Sigma}} \notag\\
&\le \Big(4\mu_p^{-1/2}q^{1/2}+\lambda^{-1}(1-\gamma)L_f\,\kappa_2 h^2
+ 4\lambda^{-1}\gamma \tau_{\max}\nu_\varepsilon\Big)\|\widehat{\boldsymbol{\delta}}\|_{\boldsymbol{\Sigma}} \, .
\end{align}
where we used \(\|\boldsymbol{u}\|_2\le \mu_p^{-1/2}\|\boldsymbol{u}\|_{\boldsymbol{\Sigma}}\) (i.e. \(\mu_{p}(\boldsymbol{\Sigma})= \mu_p\)).
\noindent
If we further take \((1-\gamma)L_f\,\kappa_2 h^2 + 4\gamma \tau_{\max}\nu_\varepsilon \leq q^{1/2}\lambda \), then on \(\{\lambda\ge 2\|\boldsymbol{\omega}_h^*\|_\infty\}\),
\(
\boldsymbol{\widehat{\beta}}
\in \boldsymbol{\beta}^*+\mathbb{C}_{\boldsymbol{\Sigma}}(\eta),
\,
\eta=\big(4\mu_p^{-1/2}+c_0\big)\,q^{1/2},
\)
for a universal constant \(c_0:= (1-\gamma)L_f\,\kappa_2 h^2 + 4\gamma \tau_{\max}\nu_\varepsilon>0\).
\noindent
We next derive a lower bound. Set
\(
\varrho=\frac{h}{20a_0^{2}}.
\)
Recall from  Theorem  \ref{theorem 1} that the RSC property only holds (with high probability)
in a local neighborhood
\(
\boldsymbol{\beta}^{*}+\mathbb{B}_{\boldsymbol{\Sigma}}(\varrho)\cap\mathbb{C}_{\boldsymbol{\Sigma}}(\eta),
\)
to which \(\boldsymbol{\widehat{\beta}}\) does not necessarily belong.
Similarly to the proof of Proposition \ref{proposition 1}, we define
\(
\zeta=\sup\Bigl\{u\in[0,1]:\ \boldsymbol{\beta}^{*}+u(\boldsymbol{\widehat{\beta}}-\boldsymbol{\beta}^{*})
\in\mathbb{B}_{\boldsymbol{\Sigma}}(\varrho)\Bigr\},
\)
and an intermediate vector
\(
\widetilde{\boldsymbol{\beta}}
=(1-\zeta)\boldsymbol{\beta}^{*}+\zeta\,\boldsymbol{\widehat{\beta}}
\in \boldsymbol{\beta}^{*}+\mathbb{B}_{\boldsymbol{\Sigma}}(\varrho).
\)
By this definition, we have \(\zeta=1\) if \(\boldsymbol{\widehat{\beta}}\in \boldsymbol{\beta}^{*}+\mathbb{B}_{\boldsymbol{\Sigma}}(\varrho)\),
and \(\zeta\in(0,1)\) if \(\boldsymbol{\widehat{\beta}}\notin \boldsymbol{\beta}^{*}+\mathbb{B}_{\boldsymbol{\Sigma}}(\varrho)\).
In the latter case, \(\widetilde{\boldsymbol{\beta}}\) lies at the boundary of
\(\boldsymbol{\beta}^{*}+\mathbb{B}_{\boldsymbol{\Sigma}}(\varrho)\).
Since \(\widetilde{\boldsymbol{\beta}}-\boldsymbol{\beta}^{*}=\zeta(\boldsymbol{\widehat{\beta}}-\boldsymbol{\beta}^{*})\),
by \eqref{a32} we also have \(\widetilde{\boldsymbol{\beta}}\in \boldsymbol{\beta}^{*}+\mathbb{C}_{\boldsymbol{\Sigma}}(\eta)\)
conditioned on \(\{\lambda\ge 2\|\boldsymbol{\omega}_h^{*}\|_{\infty}\}\).
Consequently, conditioned on
\(\{\lambda\ge 2\|\boldsymbol{\omega}_h^{*}\|_{\infty}\}\cap \mathcal{E}_{\mathrm{rsc}}(\varrho,\eta,\kappa)\)
with
\(
\kappa=\frac{1}{2}(1-\gamma)\,\kappa_0\,f_{\min}
\,+\,2\gamma\,\tau_{\min}\,\kappa_{\mathrm{RE}},
\)
we have
\begin{equation}\label{F4}
\Big\langle
\nabla \widehat{\mathcal{H}}_{h}^{\gamma}(\widetilde{\boldsymbol{\beta}})
-\nabla \widehat{\mathcal{H}}_{h}^{\gamma}(\boldsymbol{\beta}^{*}),\,
\widetilde{\boldsymbol{\beta}}-\boldsymbol{\beta}^{*}
\Big\rangle
\ge
\kappa\,
\big\|\widetilde{\boldsymbol{\beta}}-\boldsymbol{\beta}^{*}\big\|_{\boldsymbol{\Sigma}}^{2}.
\end{equation}
\noindent
Combining the lower and upper bounds obtained above, we apply \eqref{eq:a5-new} with
$(\mathcal{H}_h^\gamma,\boldsymbol{\beta}^*,\boldsymbol{\beta}_h^*)$
replaced by
$(\widehat{\mathcal{H}}_h^\gamma,\widetilde{\boldsymbol{\beta}},\boldsymbol{\widehat{\beta}})$,
to conclude that
\begin{equation}\label{eq:C18-hybrid}
\Big\langle
\nabla\widehat{\mathcal{H}}_h^\gamma(\widetilde{\boldsymbol{\beta}})
-\nabla\widehat{\mathcal{H}}_h^\gamma(\boldsymbol{\beta}^*),\,
\widetilde{\boldsymbol{\beta}}-\boldsymbol{\beta}^*
\Big\rangle
\le
\zeta\,
\Big\langle
\nabla\widehat{\mathcal{H}}_h^\gamma(\boldsymbol{\widehat{\beta}})
-\nabla\widehat{\mathcal{H}}_h^\gamma(\boldsymbol{\beta}^*),\,
\boldsymbol{\widehat{\beta}}-\boldsymbol{\beta}^*
\Big\rangle .
\end{equation}
\noindent
Combining \eqref{eq:C18-hybrid} with \eqref{a32} and \eqref{F4}, we obtain that,
conditionally on
$\{\lambda \ge 2\|\boldsymbol{\omega}_h^*\|_\infty\}\cap\mathcal{E}_{\mathrm{rsc}}(\varrho,\eta,\kappa)$,
\begin{align}\label{eq:C19-hybrid}
\kappa
\,
\|\widetilde{\boldsymbol{\delta}}\|_{\boldsymbol{\Sigma}}^{2}
\le
\frac{3}{2}q^{1/2}\lambda\|\widetilde{\boldsymbol{\delta}}\|_2
+b_h^*\|\widetilde{\boldsymbol{\delta}}\|_{\boldsymbol{\Sigma}} \notag
\le
\frac{3}{2}\mu_p^{-1/2}q^{1/2}\lambda\|\widetilde{\boldsymbol{\delta}}\|_{\boldsymbol{\Sigma}}
+c_0\|\widetilde{\boldsymbol{\delta}}\|_{\boldsymbol{\Sigma}},
\end{align}
where $\widetilde{\boldsymbol{\delta}}=\widetilde{\boldsymbol{\beta}}-\boldsymbol{\beta}^*$.
\noindent
Canceling $\|\widetilde{\boldsymbol{\delta}}\|_{\boldsymbol{\Sigma}}$ from both sides and rearranging the terms yields
\begin{equation}\label{eq:C20-hybrid}
\|\widetilde{\boldsymbol{\delta}}\|_{\boldsymbol{\Sigma}}
\le
\frac{1}{\kappa}((3/2)\mu_p^{-1/2}
+c_0)q^{1/2}\lambda
\end{equation}
\noindent
It remains to control the probability of the event
$\{\lambda \ge 2\|\boldsymbol{\omega}_h^*\|_\infty\}\cap\mathcal{E}_{\mathrm{rsc}}(\varrho,\eta,\kappa)$,
where
\(
\eta=(4\mu_p^{-1/2}+c_0)q^{1/2},
\,
\kappa=\Big((1-\gamma)\frac{1}{2}\kappa_0 f_{\min}+2\gamma\tau_{\min}\kappa_{\mathrm{RE}}\Big).
\)
\noindent
By  Theorem  \ref{theorem 1} and Lemma \ref{lemma 2}, choose
\begin{equation}\label{eq:C21-hybrid}
\lambda
= 2a_0\,\sigma_x\left[
\sqrt{\Big(2(1-\gamma)^2\,V_h
+8\gamma^2\,\tau_{\max}^2\,V_\varepsilon\Big)\frac{4\log(2p)}{n}}
+ L_{\tau,\varepsilon}\,\frac{4\log(2p)}{n}
\right],
\end{equation}
\noindent
with \(
V_h:=\tau(1-\tau)+C h^2,
\,
C:=(\tau+1)\,L_f\,\kappa_2,\, V_\varepsilon:=\mathbb{E}(\varepsilon^2),\, L_{\tau,\varepsilon}
=(1-\gamma)\,\tau_{\max}
+2\gamma\,\tau_{\max}\,\nu_\varepsilon.
\)
\noindent
Then
$\{\lambda \ge 2\|\boldsymbol{\omega}_h^*\|_\infty\}\cap\mathcal{E}_{\mathrm{rsc}}(\varrho,\eta,\kappa,\gamma)$
occurs with probability at least $1-p^{-1}$ provided that
\(
\frac{\sigma_x^2 f_{\max}\, q\log(p)}{
\big((1-\gamma)\kappa_0 f_{\min}+2\gamma\tau_{\min}\kappa_{\mathrm{RE}}\big)^2\,n
}
\,\lesssim\,
h
\,\le\,
\frac{f_{\min}}{2L_f}.
\)
\noindent
This establishes the error bound \eqref{eq:C20-hybrid} for $\widetilde{\boldsymbol{\beta}}$.
Assume further that
\(
\frac{
((3/2)\mu_p^{-1/2}+c_0)q^{1/2}\lambda
}{
(1-\gamma)\kappa_0 f_{\min}+2\gamma\tau_{\min}\kappa_{\mathrm{RE}}
}
< \varrho := \frac{h}{20a_0^2}.
\)
\noindent
Then $\widetilde{\boldsymbol{\beta}}$ lies in the interior of $\mathbb{B}_{\boldsymbol{\Sigma}}(\varrho)$
with high probability. By a contradiction argument, this implies $\zeta=1$, hence
$\widetilde{\boldsymbol{\beta}}=\boldsymbol{\widehat{\beta}}$,
and $\boldsymbol{\widehat{\beta}}$ also satisfies \eqref{eq:C20-hybrid}.
\hfill$\square$
\subsection{Auxiliary lemmas}
\noindent
Define the asymmetric weight function \(\omega_\tau(s):=\tau\,\mathds{1}\{s\ge0\}+(1-\tau)\,\mathds{1}\{s<0\}\) for \(s\in\mathbb{R}\), and set \(\tau_{\max}:=\max\{\tau,1-\tau\}\) and \(\tau_{\min}:=\min\{\tau,1-\tau\}\). Recall that the expectile score is \(\psi_\tau(s)=2\,\omega_\tau(s)\,s\), \(s\in\mathbb{R}\). Since \(\psi_\tau\) is piecewise linear with slopes \(2\tau\) on \([0,\infty)\) and \(2(1-\tau)\) on \((-\infty,0)\), it satisfies \(|\psi_\tau(s)-\psi_\tau(t)|\le 2\tau_{\max}|s-t|\) for all \(s,t\in\mathbb{R}\). Its global Lipschitz constant is therefore \(L_\tau:=\sup_{s\neq t}\frac{|\psi_\tau(s)-\psi_\tau(t)|}{|s-t|}=2\tau_{\max}\in[1,2]\). For the kernel \(k\), recall \(\kappa_1=\int |s|\,k(s)\,ds\) and \(\kappa_2=\int s^2\,k(s)\,ds\). We also define the empirical gradient fluctuation by \(\boldsymbol{\omega}_h^*:=\boldsymbol{\omega}_h(\boldsymbol{\beta}^*)\), where \(\boldsymbol{\omega}_h(\boldsymbol{\beta}):=\nabla \widehat{\mathcal{H}}^{\,\gamma}_h(\boldsymbol{\beta})-\nabla \mathcal{H}^{\,\gamma}_h(\boldsymbol{\beta})\) for \(\boldsymbol{\beta}\in\mathbb{R}^p\). We begin with moment bounds for the standardized covariates and then state the curvature and stability lemmas used in the proofs of Proposition \ref{proposition 3} and Theorem \ref{theorem 1}. This appendix first derives the stochastic ingredients needed for the proofs and then presents several auxiliary lemmas. The arguments are based on standard concentration and empirical-process tools.
\begin{lemma}\label{lemma 1}
Let \( \boldsymbol{z}=\boldsymbol{\Sigma}^{-1/2}\boldsymbol{x}\in\mathbb{R}^p \) denote the standardized feature vector, which is isotropic: \( \mathbb{E}(\boldsymbol{z}\boldsymbol{z}^\top)=\mathbb{I}_p \). Under Assumption~\ref{A4}, the \(l\)-th (\(l\ge3\)) absolute moments of all one-dimensional marginals of \( \boldsymbol{z} \) are uniformly bounded: \( \mathfrak{m}_l:=\sup_{\boldsymbol{\delta}\in\mathbb{S}^{p-1}}\mathbb{E}\big|\boldsymbol{z}^\top\boldsymbol{\delta}\big|^l\le l!\,a_0^l \). In particular, \( \mathfrak{m}_1\le \mathfrak{m}_2^{1/2}=1 \).
\end{lemma}
\subsection*{Proof of Lemma \ref{lemma 1}.}
\noindent By Assumption~\ref{A4}, for every \(\boldsymbol{\delta}\in\mathbb{S}^{p-1}\) and \(s\ge0\), \( \mathbb{P}\big(|\boldsymbol{z}^\top \boldsymbol{\delta}|\ge a_0 s\big)\le e^{-s} \). For any \(l\ge1\), \( \mathbb{E}\big|\boldsymbol{z}^\top\boldsymbol{\delta}\big|^l=\int_0^\infty \mathbb{P}\big(|\boldsymbol{z}^\top\boldsymbol{\delta}|\ge t^{1/l}\big)\,dt \). With the change of variable \(t=a_0^l s^l\), \( \mathbb{E}\big|\boldsymbol{z}^\top\boldsymbol{\delta}\big|^l=l a_0^l\int_0^\infty s^{l-1}e^{-s}\,ds=l!\,a_0^l \). Taking the supremum over \(\boldsymbol{\delta}\) yields the bound for \(\mathfrak{m}_l\). Finally, isotropy gives \( \mathfrak{m}_2=\mathbb{E}|\boldsymbol{z}^\top\boldsymbol{\delta}|^2=\boldsymbol{\delta}^\top\mathbb{I}_p\boldsymbol{\delta}=1 \), hence \( \mathfrak{m}_1\le \mathfrak{m}_2^{1/2}=1 \). \qed

\begin{lemma}\label{lemma 2}
Suppose Assumptions~\ref{A1}--\ref{A5} hold. Let \( \sigma_x=\max_{1\le l\le p}\sigma_{ll} \), where \( \sigma_{ll}:=\mathbb{E}(x_{il}^2) \), and define \( V_h:=\tau(1-\tau)+C h^2 \), with \( C:=(\tau+1)L_f\kappa_2 \), and \( V_\varepsilon:=\mathbb{E}(\varepsilon^2) \). Then, for any \(s>0\), with probability at least \(1-2p e^{-s}\),
\begin{equation}\label{eq:C1-corrected}
\big\| \boldsymbol{\omega}_h^* \big\|_{\infty} 
\le a_0\,\sigma_x\left[
\sqrt{\Big(2(1-\gamma)^2\,V_h
+8\gamma^2\,\tau_{\max}^2\,V_\varepsilon\Big)\frac{2s}{n}}
+ L_{\tau,\varepsilon}\,\frac{2s}{n}
\right],
\end{equation}
where \( L_{\tau,\varepsilon}=(1-\gamma)\tau_{\max}+2\gamma\tau_{\max}\nu_\varepsilon \) and \( \tau_{\max}:=\max\{\tau,1-\tau\} \).
\end{lemma}

\subsection*{Proof of Lemma~\ref{lemma 2}}
\noindent The proof follows a standard empirical-process argument: we control the stochastic gradient deviation by bounding second moments and applying Bernstein's inequality coordinatewise, followed by a union bound.

\noindent
We begin by introducing the centered kernel score \( \delta_i:=\bar k(-\varepsilon_i/h)-\tau \) and the hybrid score \( \zeta_i:=(1-\gamma)\delta_i-\gamma\psi_\tau(\varepsilon_i) \), where \( \psi_\tau(s)=2\bigl(\tau-\mathds{1}(s\le 0)\bigr)s \) is the expectile score function. Evaluating the gradient of the smoothed hybrid risk at the true parameter \(\boldsymbol{\beta}^*\) yields
\(
\nabla \mathcal{H}^{\gamma}_{h}(\boldsymbol{\beta}^*)
=\frac{1}{n}\sum_{i=1}^n \zeta_i \boldsymbol{x}_i .
\)
Hence,
\begin{equation}\label{eq:target-bound}
\bigl\|\nabla \mathcal{H}^{\gamma}_{h}(\boldsymbol{\beta}^*)
-\nabla \widehat{\mathcal{H}}^{\gamma}_{h}(\boldsymbol{\beta}^*)\bigr\|_\infty
=
\Bigl\|\frac{1}{n}\sum_{i=1}^n
\{\zeta_i\boldsymbol{x}_i-\mathbb{E}(\zeta_i\boldsymbol{x}_i)\}\Bigr\|_\infty .
\end{equation}

\noindent
We first derive the second moment of the kernel score. A change of variables together with integration by parts yields \( \mathbb{E}\!\left[\bar k^2(-\varepsilon/h)\mid\boldsymbol{x}\right]=2\int k(v)\bar k(v)\,F_{\varepsilon\mid \boldsymbol{x}}(-vh)\,dv \). Using the Lipschitz continuity of \(f_{\varepsilon\mid\boldsymbol{x}}\) and the kernel moment condition, we obtain
\begin{equation}\label{eq:kernel-second}
\mathbb{E}\!\left[\bar k^2(-\varepsilon/h)\mid\boldsymbol{x}\right]
\le \tau+L_f\kappa_2 h^2 .
\end{equation}
Similarly,
\begin{equation}\label{eq:kernel-bias}
\Big|
\mathbb{E}\!\left[\bar k(-\varepsilon/h)\mid\boldsymbol{x}\right]-\tau
\Big|
\le \frac{L_f}{2}\,\kappa_2\,h^2 .
\end{equation}

\noindent
To justify this, start from \( \mathbb{E}\!\left[\bar k^2(-\varepsilon/h)\mid\boldsymbol{x}\right]=\int_{-\infty}^{\infty}\bar k^2(-u/h)\,f_{\varepsilon\mid \boldsymbol{x}}(u)\,du \). With the change of variables \(u=-vh\), we get
\begin{equation}\label{eq:step1-cv}
\mathbb{E}\!\left[\bar k^2(-\varepsilon/h)\mid\boldsymbol{x}\right]
=h\int_{-\infty}^{\infty}\bar k^2(v)\,f_{\varepsilon\mid \boldsymbol{x}}(-vh)\,dv .
\end{equation}
Let \(g(v)=\bar k^2(v)\) and \(H(v)=F_{\varepsilon\mid \boldsymbol{x}}(-vh)\). Then \(H'(v)=-h f_{\varepsilon\mid \boldsymbol{x}}(-vh)\) and \(g'(v)=2\bar k(v)k(v)\), so \(h f_{\varepsilon\mid \boldsymbol{x}}(-vh)=-H'(v)\), and \eqref{eq:step1-cv} becomes \( \mathbb{E}\!\left[\bar k^2(-\varepsilon/h)\mid\boldsymbol{x}\right]=\int_{-\infty}^{\infty} g(v)(-H'(v))\,dv \). Integrating by parts and using the boundary behavior of \(\bar k\) and \(H\), the boundary term vanishes, giving
\begin{equation}\label{eq:step1-ibp}
\mathbb{E}\!\left[\bar k^2(-\varepsilon/h)\mid\boldsymbol{x}\right]
=\int_{-\infty}^{\infty} 2k(v)\bar k(v)\,F_{\varepsilon\mid \boldsymbol{x}}(-vh)\,dv .
\end{equation}
Next, by the fundamental theorem of calculus and \(F_{\varepsilon\mid \boldsymbol{x}}(0)=\tau\),
\begin{equation}\label{eq:step1-ftc}
F_{\varepsilon\mid \boldsymbol{x}}(-vh)
=\tau+\int_{0}^{-vh} f_{\varepsilon\mid \boldsymbol{x}}(t)\,dt
=\tau-hv\,f_{\varepsilon\mid \boldsymbol{x}}(0)
+\int_{0}^{-vh}\!\bigl\{f_{\varepsilon\mid \boldsymbol{x}}(t)-f_{\varepsilon\mid \boldsymbol{x}}(0)\bigr\}\,dt .
\end{equation}
Substituting \eqref{eq:step1-ftc} into \eqref{eq:step1-ibp} yields
\begin{align}
\mathbb{E}\!\left[\bar k^2(-\varepsilon/h)\mid\boldsymbol{x}\right]
&=2\tau\int k(v)\bar k(v)\,dv \notag\\
&\quad-2h f_{\varepsilon\mid \boldsymbol{x}}(0)
\int v\,k(v)\bar k(v)\,dv \notag\\
&\quad+2\int_{-\infty}^{\infty}\int_{0}^{-vh}
\bigl\{f_{\varepsilon\mid \boldsymbol{x}}(t)-f_{\varepsilon\mid \boldsymbol{x}}(0)\bigr\}
 k(v)\bar k(v)\,dt\,dv . \label{eq:step1-decomp}
\end{align}
Since \( \int k(v)\bar k(v)\,dv=\tfrac12 \), and defining
\(
A_k:=\int_{-\infty}^{\infty} v\,k(v)\bar k(v)\,dv
=\int_{0}^{\infty} k(v)\{1-k(v)\}\,dv>0,
\)
with \(A_k\le \kappa_1:=\int |u|k(u)\,du\), the remainder term is controlled by Lipschitz continuity:
\( \left|\int_{0}^{-vh}\!\bigl\{f_{\varepsilon\mid \boldsymbol{x}}(t)-f_{\varepsilon\mid \boldsymbol{x}}(0)\bigr\}\,dt\right|\le \frac{L_f}{2}v^2h^2 \). Therefore,
\( 2\int_{-\infty}^{\infty}\int_{0}^{-vh}\bigl\{f_{\varepsilon\mid \boldsymbol{x}}(t)-f_{\varepsilon\mid \boldsymbol{x}}(0)\bigr\}k(v)\bar k(v)\,dt\,dv\le L_f h^2\int_{-\infty}^{\infty} v^2 k(v)\bar k(v)\,dv \le L_f\kappa_2 h^2 \), and so
\( \mathbb{E}\!\left[\bar k^2(-\varepsilon/h)\mid\boldsymbol{x}\right]\le \tau-2A_k h f_{\varepsilon\mid \boldsymbol{x}}(0)+L_f\kappa_2 h^2\le \tau+L_f\kappa_2 h^2 \), proving \eqref{eq:kernel-second}. Similarly, \( \mathbb{E}\!\left[\bar k(-\varepsilon/h)\mid\boldsymbol{x}\right]=\int_{-\infty}^{\infty}\bar k(-u/h)\,f_{\varepsilon\mid \boldsymbol{x}}(u)\,du=h\int_{-\infty}^{\infty}\bar k(v)\,f_{\varepsilon\mid \boldsymbol{x}}(-vh)\,dv \). Using again \(H(v)=F_{\varepsilon\mid \boldsymbol{x}}(-vh)\) and integration by parts, the boundary term vanishes and \( \mathbb{E}\!\left[\bar k(-\varepsilon/h)\mid\boldsymbol{x}\right]=\int_{-\infty}^{\infty} k(v)\,F_{\varepsilon\mid \boldsymbol{x}}(-vh)\,dv \). Substituting \eqref{eq:step1-ftc} and using \( \int k(v)\,dv=1 \), we get
\( \mathbb{E}\!\left[\bar k(-\varepsilon/h)\mid\boldsymbol{x}\right]-\tau=-h f_{\varepsilon\mid \boldsymbol{x}}(0)\int v\,k(v)\,dv+\int_{-\infty}^{\infty} k(v)\int_{0}^{-vh}\!\bigl\{f_{\varepsilon\mid \boldsymbol{x}}(t)-f_{\varepsilon\mid \boldsymbol{x}}(0)\bigr\}\,dt\,dv \).
Under symmetry, \( \int v\,k(v)\,dv=0 \), so only the remainder remains. Using the same Lipschitz bound, we obtain \( \Big|\mathbb{E}\!\left[\bar k(-\varepsilon/h)\mid\boldsymbol{x}\right]-\tau\Big|\le \frac{L_f}{2}h^2\int_{-\infty}^{\infty} v^2 k(v)\,dv=\frac{L_f}{2}\kappa_2 h^2 \), proving \eqref{eq:kernel-bias}. Combining \eqref{eq:kernel-second} and \eqref{eq:kernel-bias}, \( \mathbb{E}(\delta_i^2\mid\boldsymbol{x})\le \tau(1-\tau)+(\tau+1)L_f\kappa_2 h^2=:V_h \). Recall that \( \zeta_i=(1-\gamma)\delta_i-\gamma\psi_\tau(\varepsilon_i) \). Expanding the square, \( \zeta_i^2=(1-\gamma)^2 \delta_i^2+\gamma^2 \psi_\tau^2(\varepsilon_i)-2\gamma(1-\gamma)\delta_i\psi_\tau(\varepsilon_i) \).
Using \(2|ab|\le a^2+b^2\), we get \( \zeta_i^2\le 2(1-\gamma)^2\delta_i^2+2\gamma^2\psi_\tau^2(\varepsilon_i) \). Since \( \psi_\tau^2(\varepsilon)\le 4\tau_{\max}^2\varepsilon^2 \), taking conditional expectations yields
\(
\mathbb{E}(\zeta_i^2\mid \boldsymbol{x})
\le 2(1-\gamma)^2\,V_h
+8\gamma^2\,\tau_{\max}^2\,V_\varepsilon .
\)

\noindent
We now apply Bernstein's inequality. Fix \(l\in\{1,\dots,d\}\) and define \(T_{il}:=\zeta_i x_{il}-\mathbb{E}(\zeta_i x_{il})\). Then \((T_{il})_{i=1}^n\) are independent, mean-zero random variables, and
\( \mathbb{E}((\zeta_i x_{il})^2)=\mathbb{E}_{\boldsymbol{x}}[ x_{il}^2\mathbb{E}(\zeta_i ^2|\boldsymbol{x})]\le \sigma_{ll}\big\{2(1-\gamma)^2V_h+8\gamma^2\tau_{\max}^2V_\varepsilon\big\} \).
Since \(|\zeta_i|\le L_{\tau,\varepsilon}\) and \(x_{il}\) has sub-exponential tails, for any integer \(j\ge2\),
\begin{align*}
\mathbb{E}\big(|\zeta_i x_{il}|^j\big)
&\le L_{\tau,\varepsilon}^{\,j-2}\,
\mathbb{E}_x\!\left\{|x_{il}|^j\,\mathbb{E}(\zeta_i^2\mid\boldsymbol{x})\right\} \\
&\le L_{\tau,\varepsilon}^{\,j-2}
\Big\{2(1-\gamma)^2V_h+8\gamma^2\tau_{\max}^2V_\varepsilon\Big\}
a_0^j\sigma_{ll}^{j/2}
\int_0^\infty
\mathbb{P}\!\left(|\sigma_{ll}^{-1/2}x_{il}|\ge a_0 s\right)
j s^{j-1}\,ds \\
&\le j!\,L_{\tau,\varepsilon}^{\,j-2}
\Big\{2(1-\gamma)^2V_h+8\gamma^2\tau_{\max}^2V_\varepsilon\Big\}
a_0^j\sigma_{ll}^{j/2} \\
&\le \frac{j!}{2}\,
\Big\{2(1-\gamma)^2V_h+8\gamma^2\tau_{\max}^2V_\varepsilon\Big\}
a_0^2\sigma_{ll}
\Big\{2L_{\tau,\varepsilon}a_0\sigma_{ll}^{1/2}\Big\}^{j-2}.
\end{align*}
Hence, \(T_{il}\) satisfies Bernstein’s moment condition. Consequently, for any \(s>0\),
\begin{equation}\label{eq:bernstein}
\left|
\frac{1}{n}\sum_{i=1}^n T_{il}
\right|
\le a_0\,\sigma_{ll}^{1/2}\left[
\sqrt{\Big\{2(1-\gamma)^2V_h
+8\gamma^2\tau_{\max}^2V_\varepsilon\Big\}\frac{2s}{n}}
+ L_{\tau,\varepsilon}\,\frac{2s}{n}
\right]
\end{equation}
with probability at least \(1-2e^{-s}\). Applying \eqref{eq:bernstein} coordinatewise and using a union bound, we obtain, with probability at least \(1-2p e^{-s}\),
\( \Bigl\|
\frac{1}{n}\sum_{i=1}^n
\{\zeta_i\boldsymbol{x}_i-\mathbb{E}(\zeta_i\boldsymbol{x}_i)\}
\Bigr\|_\infty
\le a_0\,\sigma_x\left[
\sqrt{\Big\{2(1-\gamma)^2V_h
+8\gamma^2\tau_{\max}^2V_\varepsilon\Big\}\frac{2s}{n}}
+ L_{\tau,\varepsilon}\,\frac{2s}{n}
\right] \).
Combining this with \eqref{eq:target-bound} completes the proof of \eqref{eq:C1-corrected}. \qed

\begin{lemma}[Lemma F.2 of \cite{fan2018single}]\label{lemma 3}
Let \( D_{\mathcal{L}}(\boldsymbol{\beta}_1,\boldsymbol{\beta}_2)=\mathcal{L}(\boldsymbol{\beta}_1)-\mathcal{L}(\boldsymbol{\beta}_2)-\big\langle \nabla \mathcal{L}(\boldsymbol{\beta}_2),\,\boldsymbol{\beta}_1-\boldsymbol{\beta}_2\big\rangle \), and \( D_{\mathcal{L}}^{s}(\boldsymbol{\beta}_1,\boldsymbol{\beta}_2)=D_{\mathcal{L}}(\boldsymbol{\beta}_1,\boldsymbol{\beta}_2)+D_{\mathcal{L}}(\boldsymbol{\beta}_2,\boldsymbol{\beta}_1) \). For \( \boldsymbol{\beta}(t)=\boldsymbol{\beta}^*+t(\boldsymbol{\beta}-\boldsymbol{\beta}^*) \) with \(t\in(0,1]\), we have \( D_{\mathcal{L}}^{s}\bigl(\boldsymbol{\beta}(t),\boldsymbol{\beta}^*\bigr)\le t\,D_{\mathcal{L}}^{s}\bigl(\boldsymbol{\beta},\boldsymbol{\beta}^*\bigr) \).
\end{lemma}

\subsection*{Proof of Lemma \ref{lemma 3}}
\noindent This result is established in \cite{fan2018single}.
\begin{theorem}[\cite{boucheron2013concentration}, Theorem 7.3]
Let $\mathcal{F}$ be a countable set of functions from $\mathcal{X}$ to $\mathbb{R}$ and assume that all functions $f\in\mathcal{F}$ are measurable, square-integrable and satisfy $\mathbb{E}[f(X_k)]=0$. Assume $\sup_{f\in\mathcal{F}}\operatorname*{ess\,sup} f \le 1$ and denote \( Z=\sup_{f\in\mathcal{F}}\sum_{k=1}^n f(X_k) \). Let $\sigma$ be a positive real number such that \( n\sigma^2 \ge \sum_{k=1}^n \sup_{f\in\mathcal{F}} \mathbb{E}\bigl[f^2(X_k)\bigr] \). Then for all $x\ge 0$, we have \( \mathbb{P}\!\left[ Z \ge \mathbb{E}[Z] + x \right]\le \exp\!\left(-v\,h_1\!\left(\frac{x}{v}\right)\right) \), with \(v=n\sigma^2+2\mathbb{E}[Z]\) and \(h_1\) defined in \eqref{eq:h1h2}. Moreover, \( \mathbb{P}\!\left[ Z \ge \mathbb{E}[Z] + \sqrt{2vx} + \frac{x}{3} \right]\le e^{-x} \).
\end{theorem}

\begin{equation}\label{eq:h1h2}
h_1(x) = (1+x)\log(1+x) - x,
\,
h_2(x) = 1 + x - \sqrt{1 + 2x}.
\end{equation}

\subsection*{Proof of Theorem 3}
\noindent We refer the reader to \cite{bousquet2003concentration} for the proof.

\bibliographystyle{sn-basic}
\bibliography{sn-bibliography}

\section{Supplementary Material}
\noindent
This section contains the full numerical results corresponding to the simulation study and the real data analysis presented in the main manuscript. It reports the complete set of tables for all considered configurations together with the corresponding model comparisons.
\subsection{Simulation Results}

\begingroup
\scriptsize
\renewcommand{\arraystretch}{1.15}

\begin{landscape}
\begin{longtable}{llccc|ccc}
\caption{Numerical comparisons under the linear heteroscedastic model with sparse regression coefficients for various SHQER configurations and competing methods (Lasso at $\tau=0.25$).}
\label{table_lasso_tau=0.25}\\

\toprule
\multirow{2}{*}{Noise} & \multirow{2}{*}{Methods}
  & \multicolumn{3}{c|}{\(\left(n = 600, p= 300\right)\)}
  & \multicolumn{3}{c}{\(\left(n = 300, p= 600\right)\)} \\
& & Error & TPR & FPR & Error & TPR & FPR \\
\midrule
\endfirsthead

\caption[]{Numerical comparisons under the linear heteroscedastic model with sparse regression coefficients for various SHQER configurations and competing methods (continued).}\\
\toprule
\multirow{2}{*}{Noise} & \multirow{2}{*}{Methods}
  & \multicolumn{3}{c|}{\(\left(n = 600, p= 300\right)\)}
  & \multicolumn{3}{c}{\(\left(n = 300, p= 600\right)\)} \\
& & Error & TPR & FPR & Error & TPR & FPR \\
\midrule
\endhead

\midrule
\multicolumn{8}{r}{\emph{Continued on next page}}\\
\endfoot

\bottomrule
\endlastfoot

\multirow{9}{*}{\(\mathcal{N}(0,3)\)}

& SALES
  & 0.6511 (0.070) & 0.8900 (0.057) & 0.0652 (0.036)
  & 0.8257 (0.117) & 0.8200 (0.042) & 0.0363 (0.019) \\

& CONQUER
  & 0.5866 (0.138) & 0.8900 (0.057) & 0.0679 (0.037)
  & 0.7699 (0.107) & 0.8200 (0.042) & 0.0288 (0.014) \\

& QR (rqPen)
  & 0.6128 (0.161) & 0.8800 (0.042) & 0.0786 (0.058)
  & 0.7923 (0.133) & 0.8400 (0.052) & 0.0500 (0.027) \\

& \(L^k\mathrm{QR}\,(k=1.5)\)
  & 1.9446 (0.087) & 0.9000 (0.082) & 0.1159 (0.089)
  & 2.1252 (0.170) & 0.8500 (0.053) & 0.1192 (0.085) \\

& SHQER ($\gamma=1$)
  & 1.8717 (0.100) & 0.9400 (0.070) & 0.1721 (0.087)
  & 1.9951 (0.167) & 0.8500 (0.053) & 0.1008 (0.038) \\

& SHQER ($\gamma=0.9$)
  & 1.8208 (0.106) & 0.9200 (0.079) & 0.1548 (0.090)
  & 1.9869 (0.182) & 0.8600 (0.070) & 0.1080 (0.051) \\

& SHQER ($\gamma=0.5$)
  & 1.4768 (0.091) & 0.9200 (0.079) & 0.1286 (0.064)
  & 1.6491 (0.206) & 0.8600 (0.052) & 0.0769 (0.029) \\

& SHQER ($\gamma=0.1$)
  & 0.7180 (0.069) & 0.9000 (0.082) & 0.0707 (0.048)
  & 0.9267 (0.170) & 0.8400 (0.052) & 0.0429 (0.020) \\

& SHQER ($\gamma=0$)
  & 0.5531 (0.097) & 0.8900 (0.057) & 0.0634 (0.030)
  & 0.7386 (0.108) & 0.8400 (0.052) & 0.0339 (0.020) \\

\midrule

\multirow{9}{*}{\(t(1.9)\)}

& SALES
  & 1.0756 (0.305) & 0.8300 (0.095) & 0.0597 (0.038)
  & 1.1624 (0.273) & 0.7800 (0.079) & 0.0308 (0.022) \\

& CONQUER
  & 0.5045 (0.081) & 0.9100 (0.074) & 0.0631 (0.035)
  & 0.7672 (0.164) & 0.8200 (0.042) & 0.0276 (0.010) \\

& QR (rqPen)
  & 0.5171 (0.071) & 0.9100 (0.088) & 0.1083 (0.080)
  & 0.8012 (0.178) & 0.9000 (0.067) & 0.1768 (0.172) \\

& \(L^k\mathrm{QR}\,(k=1.5)\)
  & 1.6322 (0.087) & 0.9400 (0.070) & 0.1114 (0.039)
  & 1.8265 (0.221) & 0.8500 (0.053) & 0.0758 (0.043) \\

& SHQER ($\gamma=1$)
  & 1.9503 (0.518) & 0.8500 (0.097) & 0.1031 (0.046)
  & 2.1774 (0.463) & 0.7800 (0.079) & 0.0717 (0.043) \\

& SHQER ($\gamma=0.9$)
  & 1.8615 (0.422) & 0.8600 (0.097) & 0.1086 (0.045)
  & 2.1499 (0.465) & 0.8100 (0.088) & 0.0859 (0.055) \\

& SHQER ($\gamma=0.5$)
  & 1.4968 (0.363) & 0.8800 (0.079) & 0.1038 (0.050)
  & 1.6709 (0.303) & 0.8100 (0.088) & 0.0563 (0.031) \\

& SHQER ($\gamma=0.1$)
  & 0.6716 (0.150) & 0.9600 (0.070) & 0.0986 (0.040)
  & 0.9690 (0.144) & 0.8400 (0.052) & 0.0386 (0.019) \\

& SHQER ($\gamma=0$)
  & 0.5320 (0.076) & 0.9000 (0.082) & 0.0631 (0.021)
  & 0.8571 (0.146) & 0.8300 (0.048) & 0.0341 (0.021) \\

\end{longtable}
\end{landscape}
\endgroup


\begingroup
\scriptsize
\renewcommand{\arraystretch}{1.15}

\begin{landscape}
\begin{longtable}{llccc|ccc}
\caption{Numerical comparisons under the linear heteroscedastic model with sparse regression coefficients for various SHQER configurations and competing methods (Lasso at $\tau=0.50$).}
\label{table_lasso_tau=0.50}\\

\toprule
\multirow{2}{*}{Noise} & \multirow{2}{*}{Methods}
  & \multicolumn{3}{c|}{\(\left(n = 600, p= 300\right)\)}
  & \multicolumn{3}{c}{\(\left(n = 300, p= 600\right)\)} \\
& & Error & TPR & FPR & Error & TPR & FPR \\
\midrule
\endfirsthead

\caption[]{Numerical comparisons under the linear heteroscedastic model with sparse regression coefficients for various SHQER configurations and competing methods (continued).}\\
\toprule
\multirow{2}{*}{Noise} & \multirow{2}{*}{Methods}
  & \multicolumn{3}{c|}{\(\left(n = 600, p= 300\right)\)}
  & \multicolumn{3}{c}{\(\left(n = 300, p= 600\right)\)} \\
& & Error & TPR & FPR & Error & TPR & FPR \\
\midrule
\endhead

\midrule
\multicolumn{8}{r}{\emph{Continued on next page}}\\
\endfoot

\bottomrule
\endlastfoot

\multirow{9}{*}{\(\mathcal{N}(0,3)\)}

& SALES
  & 0.4855 (0.071) & 0.8870 (0.071) & 0.0588 (0.034)
  & 0.6819 (0.121) & 0.8320 (0.049) & 0.0335 (0.020) \\

& CONQUER
  & 0.5092 (0.075) & 0.8880 (0.069) & 0.0638 (0.040)
  & 0.7061 (0.127) & 0.8360 (0.048) & 0.0423 (0.030) \\

& QR (rqPen)
  & 0.5314 (0.086) & 0.9010 (0.073) & 0.1029 (0.083)
  & 0.7373 (0.138) & 0.8580 (0.059) & 0.1302 (0.140) \\

& \(L^k\mathrm{QR}\,(k=1.5)\)
  & 0.4741 (0.064) & 0.8980 (0.071) & 0.0705 (0.035)
  & 0.6637 (0.122) & 0.8340 (0.052) & 0.0408 (0.024) \\

& SHQER ($\gamma=1$)
  & 0.4875 (0.072) & 0.8920 (0.071) & 0.0702 (0.040)
  & 0.6875 (0.123) & 0.8370 (0.051) & 0.0407 (0.024) \\

& SHQER ($\gamma=0.9$)
  & 0.4870 (0.070) & 0.8960 (0.074) & 0.0737 (0.043)
  & 0.6879 (0.126) & 0.8350 (0.048) & 0.0408 (0.026) \\

& SHQER ($\gamma=0.5$)
  & 0.4790 (0.069) & 0.8960 (0.075) & 0.0721 (0.042)
  & 0.6727 (0.125) & 0.8340 (0.048) & 0.0396 (0.024) \\

& SHQER ($\gamma=0.1$)
  & 0.4701 (0.065) & 0.8980 (0.072) & 0.0741 (0.036)
  & 0.6583 (0.118) & 0.8350 (0.050) & 0.0383 (0.022) \\

& SHQER ($\gamma=0$)
  & 0.4761 (0.067) & 0.8970 (0.073) & 0.0737 (0.040)
  & 0.6604 (0.117) & 0.8350 (0.048) & 0.0395 (0.026) \\

\midrule

\multirow{9}{*}{\(t(1.9)\)}

& SALES
  & 0.8391 (0.308) & 0.8390 (0.072) & 0.0624 (0.036)
  & 1.1986 (0.582) & 0.7790 (0.117) & 0.0344 (0.019) \\

& CONQUER
  & 0.3878 (0.063) & 0.9560 (0.062) & 0.0719 (0.037)
  & 0.5787 (0.107) & 0.8550 (0.064) & 0.0364 (0.020) \\

& QR (rqPen)
  & 0.4003 (0.068) & 0.9660 (0.057) & 0.1312 (0.087)
  & 0.5970 (0.107) & 0.8950 (0.072) & 0.1181 (0.107) \\

& \(L^k\mathrm{QR}\,(k=1.5)\)
  & 0.4493 (0.075) & 0.9290 (0.074) & 0.0706 (0.029)
  & 0.6863 (0.154) & 0.8440 (0.061) & 0.0382 (0.024) \\

& SHQER ($\gamma=1$)
  & 0.8416 (0.314) & 0.8480 (0.080) & 0.0673 (0.033)
  & 1.2111 (0.587) & 0.7900 (0.118) & 0.0404 (0.024) \\

& SHQER ($\gamma=0.9$)
  & 0.8221 (0.301) & 0.8480 (0.077) & 0.0682 (0.032)
  & 1.1911 (0.588) & 0.7890 (0.118) & 0.0379 (0.020) \\

& SHQER ($\gamma=0.5$)
  & 0.7294 (0.265) & 0.8650 (0.069) & 0.0687 (0.031)
  & 1.0904 (0.554) & 0.7990 (0.115) & 0.0379 (0.019) \\

& SHQER ($\gamma=0.1$)
  & 0.4880 (0.100) & 0.9220 (0.073) & 0.0740 (0.040)
  & 0.7855 (0.344) & 0.8370 (0.079) & 0.0362 (0.022) \\

& SHQER ($\gamma=0$)
  & 0.3969 (0.068) & 0.9520 (0.064) & 0.0777 (0.035)
  & 0.5940 (0.109) & 0.8550 (0.064) & 0.0361 (0.018) \\

\end{longtable}
\end{landscape}
\endgroup

\begingroup
\scriptsize
\renewcommand{\arraystretch}{1.15}

\begin{landscape}
\begin{longtable}{llccc|ccc}
\caption{Numerical comparisons under the linear heteroscedastic model with sparse regression coefficients for various SHQER configurations and competing methods (Lasso at $\tau=0.75$).}
\label{table_lasso_tau=0.75}\\

\toprule
\multirow{2}{*}{Noise} & \multirow{2}{*}{Methods}
  & \multicolumn{3}{c|}{\(\left(n = 600, p= 300\right)\)}
  & \multicolumn{3}{c}{\(\left(n = 300, p= 600\right)\)} \\
& & Error & TPR & FPR & Error & TPR & FPR \\
\midrule
\endfirsthead

\caption[]{Numerical comparisons under the linear heteroscedastic model with sparse regression coefficients for various SHQER configurations and competing methods (continued).}\\
\toprule
\multirow{2}{*}{Noise} & \multirow{2}{*}{Methods}
  & \multicolumn{3}{c|}{\(\left(n = 600, p= 300\right)\)}
  & \multicolumn{3}{c}{\(\left(n = 300, p= 600\right)\)} \\
& & Error & TPR & FPR & Error & TPR & FPR \\
\midrule
\endhead

\midrule
\multicolumn{8}{r}{\emph{Continued on next page}}\\
\endfoot

\bottomrule
\endlastfoot

\multirow{9}{*}{\(\mathcal{N}(0,3)\)}

& SALES
  & 0.5822 (0.095) & 0.8800 (0.079) & 0.0693 (0.040)
  & 0.8054 (0.129) & 0.8400 (0.052) & 0.0375 (0.013) \\

& CONQUER
  & 0.5156 (0.125) & 0.8700 (0.082) & 0.0559 (0.033)
  & 0.6996 (0.133) & 0.8200 (0.042) & 0.0303 (0.017) \\

& QR (rqPen)
  & 0.5331 (0.130) & 0.9000 (0.082) & 0.0907 (0.061)
  & 0.7300 (0.146) & 0.8500 (0.071) & 0.1259 (0.094) \\

& \(L^k\mathrm{QR}\,(k=1.5)\)
  & 1.8781 (0.105) & 0.9400 (0.070) & 0.1762 (0.069)
  & 2.1141 (0.200) & 0.8600 (0.070) & 0.1081 (0.068) \\

& SHQER ($\gamma=1$)
  & 1.7969 (0.100) & 0.9300 (0.067) & 0.1679 (0.088)
  & 1.9626 (0.186) & 0.8500 (0.071) & 0.1007 (0.022) \\

& SHQER ($\gamma=0.9$)
  & 1.7378 (0.090) & 0.9300 (0.067) & 0.1555 (0.095)
  & 1.9041 (0.146) & 0.8600 (0.070) & 0.1056 (0.030) \\

& SHQER ($\gamma=0.5$)
  & 1.4155 (0.086) & 0.9300 (0.067) & 0.1366 (0.043)
  & 1.5748 (0.186) & 0.8500 (0.071) & 0.0614 (0.029) \\

& SHQER ($\gamma=0.1$)
  & 0.7007 (0.080) & 0.9100 (0.088) & 0.0828 (0.055)
  & 0.8966 (0.153) & 0.8500 (0.071) & 0.0412 (0.020) \\

& SHQER ($\gamma=0$)
  & 0.5014 (0.103) & 0.8900 (0.088) & 0.0893 (0.048)
  & 0.7108 (0.133) & 0.8300 (0.048) & 0.0429 (0.032) \\

\midrule

\multirow{9}{*}{\(t(1.9)\)}

& SALES
  & 0.9085 (0.271) & 0.8500 (0.085) & 0.0762 (0.033)
  & 1.4529 (0.581) & 0.7600 (0.143) & 0.0380 (0.018) \\

& CONQUER
  & 0.4919 (0.073) & 0.9300 (0.067) & 0.0576 (0.022)
  & 0.7108 (0.098) & 0.8300 (0.048) & 0.0317 (0.020) \\

& QR (rqPen)
  & 0.4895 (0.070) & 0.9300 (0.067) & 0.0693 (0.014)
  & 0.7327 (0.113) & 0.8600 (0.070) & 0.0878 (0.091) \\

& \(L^k\mathrm{QR}\,(k=1.5)\)
  & 1.6334 (0.142) & 0.9400 (0.052) & 0.1576 (0.055)
  & 1.7272 (0.111) & 0.8400 (0.070) & 0.0703 (0.019) \\

& SHQER ($\gamma=1$)
  & 1.8561 (0.274) & 0.8800 (0.063) & 0.1414 (0.066)
  & 1.9897 (0.211) & 0.8600 (0.097) & 0.0778 (0.034) \\

& SHQER ($\gamma=0.9$)
  & 1.7867 (0.266) & 0.8700 (0.082) & 0.1193 (0.065)
  & 1.9035 (0.172) & 0.8300 (0.067) & 0.0681 (0.037) \\

& SHQER ($\gamma=0.5$)
  & 1.4015 (0.218) & 0.8700 (0.082) & 0.1003 (0.047)
  & 1.5829 (0.282) & 0.8300 (0.067) & 0.0617 (0.022) \\

& SHQER ($\gamma=0.1$)
  & 0.6458 (0.138) & 0.9300 (0.067) & 0.0862 (0.061)
  & 0.8806 (0.155) & 0.8500 (0.071) & 0.0634 (0.031) \\

& SHQER ($\gamma=0$)
  & 0.5186 (0.090) & 0.9300 (0.082) & 0.0617 (0.015)
  & 0.8286 (0.112) & 0.8400 (0.052) & 0.0336 (0.018) \\

\end{longtable}
\end{landscape}
\endgroup

\newpage

\begingroup
\scriptsize
\renewcommand{\arraystretch}{1.15}

\begin{landscape}
\begin{longtable}{lllccc|ccc}
\caption{Numerical comparisons under the linear heteroscedastic model with sparse regression coefficients
for various SHQER configurations and competing methods (elastic net at $\tau=0.25$ with
$\alpha \in \{0.3,0.5,0.7\}$).}
\label{table:elasticnet_alpha_tau_025}\\

\toprule
\multirow{2}{*}{Noise} & \multirow{2}{*}{$\alpha$} & \multirow{2}{*}{Methods}
  & \multicolumn{3}{c|}{\(\left(n = 600, p= 300\right)\)}
  & \multicolumn{3}{c}{\(\left(n = 300, p= 600\right)\)} \\
& & & Error & TPR & FPR & Error & TPR & FPR \\
\midrule
\endfirsthead

\caption[]{Numerical comparisons under the linear heteroscedastic model with sparse regression coefficients
for various SHQER configurations and competing methods (continued).}\\
\toprule
\multirow{2}{*}{Noise} & \multirow{2}{*}{$\alpha$} & \multirow{2}{*}{Methods}
  & \multicolumn{3}{c|}{\(\left(n = 600, p= 300\right)\)}
  & \multicolumn{3}{c}{\(\left(n = 300, p= 600\right)\)} \\
& & & Error & TPR & FPR & Error & TPR & FPR \\
\midrule
\endhead

\midrule
\multicolumn{9}{r}{\emph{Continued on next page}}\\
\endfoot

\bottomrule
\endlastfoot

\multirow{1}{*}{\(\mathcal{N}(0,3)\)}

& \multirow{9}{*}{0.3}
& SALES (EN)
  & 2.7620 (0.015) & 1.0000 (0.000) & 0.5828 (0.029)
  & 2.8609 (0.061) & 0.9500 (0.071) & 0.2347 (0.064) \\
& & CONQUER (EN)
  & 1.5272 (0.005) & 0.9500 (0.071) & 0.5000 (0.132)
  & 2.2007 (0.031) & 0.9500 (0.071) & 0.2729 (0.017) \\
& & QR (rqPen, EN)
  & 1.4628 (0.013) & 1.0000 (0.000) & 0.3603 (0.110)
  & 2.0229 (0.003) & 1.0000 (0.000) & 0.6008 (0.011) \\
& & \(L^k\mathrm{QR}\,(\mathrm{EN}, k=1.5)\)
  & 1.6513 (0.030) & 1.0000 (0.000) & 0.3345 (0.039)
  & 2.2483 (0.034) & 0.9500 (0.071) & 0.2322 (0.017) \\
& & SHQER ($\gamma=1$)
  & 1.6933 (0.002) & 0.9500 (0.071) & 0.3379 (0.054)
  & 2.2864 (0.037) & 0.9500 (0.071) & 0.2229 (0.011) \\
& & SHQER ($\gamma=0.9$)
  & 1.5914 (0.030) & 1.0000 (0.000) & 0.3897 (0.068)
  & 2.2262 (0.033) & 0.9500 (0.071) & 0.2746 (0.005) \\
& & SHQER ($\gamma=0.5$)
  & 1.6663 (0.037) & 1.0000 (0.000) & 0.3500 (0.032)
  & 2.2537 (0.009) & 0.9000 (0.141) & 0.2356 (0.002) \\
& & SHQER ($\gamma=0.1$)
  & 1.7358 (0.179) & 0.9500 (0.071) & 0.3845 (0.237)
  & 2.3448 (0.207) & 0.9000 (0.141) & 0.2314 (0.140) \\
& & SHQER ($\gamma=0$)
  & 1.7225 (0.139) & 0.9500 (0.071) & 0.4172 (0.229)
  & 2.3171 (0.174) & 0.9000 (0.141) & 0.2593 (0.141) \\
\\

& \multirow{9}{*}{0.5}
& SALES (EN)
  & 2.4181 (0.005) & 1.0000 (0.000) & 0.3362 (0.051)
  & 2.5425 (0.047) & 0.8500 (0.071) & 0.1381 (0.044) \\
& & CONQUER (EN)
  & 1.3892 (0.001) & 0.9000 (0.141) & 0.1879 (0.027)
  & 1.8226 (0.009) & 0.8500 (0.071) & 0.1797 (0.010) \\
& & QR (rqPen, EN)
  & 1.1746 (0.033) & 0.9500 (0.071) & 0.1845 (0.056)
  & 1.6717 (0.022) & 0.9000 (0.141) & 0.2831 (0.292) \\
& & \(L^k\mathrm{QR}\,(\mathrm{EN}, k=1.5)\)
  & 1.3775 (0.010) & 0.9500 (0.071) & 0.1707 (0.022)
  & 1.8748 (0.054) & 0.9000 (0.141) & 0.1602 (0.116) \\
& & SHQER ($\gamma=1$)
  & 1.4150 (0.027) & 0.9000 (0.000) & 0.1672 (0.037)
  & 1.9874 (0.072) & 0.8000 (0.000) & 0.0958 (0.004) \\
& & SHQER ($\gamma=0.9$)
  & 1.3005 (0.008) & 0.9500 (0.071) & 0.2034 (0.024)
  & 1.8887 (0.053) & 0.8500 (0.071) & 0.1237 (0.002) \\
& & SHQER ($\gamma=0.5$)
  & 1.3895 (0.011) & 1.0000 (0.000) & 0.1879 (0.017)
  & 1.9860 (0.029) & 0.8500 (0.071) & 0.0941 (0.020) \\
& & SHQER ($\gamma=0.1$)
  & 1.4561 (0.248) & 0.9000 (0.141) & 0.2034 (0.146)
  & 1.8664 (0.001) & 0.8500 (0.071) & 0.1771 (0.008) \\
& & SHQER ($\gamma=0$)
  & 1.4278 (0.208) & 0.9000 (0.141) & 0.2345 (0.176)
  & 1.8514 (0.002) & 0.8500 (0.071) & 0.1992 (0.020) \\
\\

& \multirow{9}{*}{0.7}
& SALES (EN)
  & 2.0249 (0.017) & 0.9500 (0.071) & 0.2034 (0.068)
  & 2.1753 (0.032) & 0.8000 (0.000) & 0.0966 (0.031) \\
& & CONQUER (EN)
  & 1.0930 (0.017) & 0.9000 (0.141) & 0.1034 (0.005)
  & 1.4644 (0.016) & 0.8500 (0.071) & 0.0831 (0.007) \\
& & QR (rqPen, EN)
  & 0.9381 (0.040) & 0.9000 (0.141) & 0.1103 (0.024)
  & 1.2954 (0.120) & 0.9500 (0.071) & 0.3703 (0.040) \\
& & \(L^k\mathrm{QR}\,(\mathrm{EN}, k=1.5)\)
  & 1.0704 (0.002) & 0.8500 (0.071) & 0.1000 (0.010)
  & 1.4148 (0.036) & 0.9000 (0.000) & 0.1542 (0.000) \\
& & SHQER ($\gamma=1$)
  & 1.0992 (0.052) & 0.8500 (0.071) & 0.0948 (0.017)
  & 1.6602 (0.131) & 0.8000 (0.000) & 0.0407 (0.002) \\
& & SHQER ($\gamma=0.9$)
  & 1.0165 (0.023) & 0.8500 (0.071) & 0.1207 (0.024)
  & 1.5159 (0.095) & 0.8000 (0.000) & 0.0551 (0.004) \\
& & SHQER ($\gamma=0.5$)
  & 1.0988 (0.007) & 0.8500 (0.071) & 0.0897 (0.005)
  & 1.5914 (0.244) & 0.8000 (0.000) & 0.1042 (0.095) \\
& & SHQER ($\gamma=0.1$)
  & 1.1289 (0.195) & 0.8500 (0.071) & 0.1172 (0.093)
  & 1.4856 (0.019) & 0.8500 (0.071) & 0.0873 (0.008) \\
& & SHQER ($\gamma=0$)
  & 1.2547 (0.039) & 0.8500 (0.071) & 0.0603 (0.012)
  & 1.4736 (0.004) & 0.8500 (0.071) & 0.1068 (0.010) \\
\midrule

\multirow{1}{*}{\(t(1.9)\)}

& \multirow{9}{*}{0.3}
& SALES (EN)
  & 2.8498 (0.139) & 1.0000 (0.000) & 0.7017 (0.066)
  & 2.9189 (0.059) & 0.9000 (0.000) & 0.2169 (0.084) \\
& & CONQUER (EN)
  & 1.4703 (0.254) & 1.0000 (0.000) & 0.4603 (0.027)
  & 2.2946 (0.088) & 0.9000 (0.000) & 0.2373 (0.000) \\
& & QR (rqPen, EN)
  & 1.2267 (0.232) & 1.0000 (0.000) & 0.6534 (0.129)
  & 2.0820 (0.028) & 1.0000 (0.000) & 0.5788 (0.099) \\
& & \(L^k\mathrm{QR}\,(\mathrm{EN}, k=1.5)\)
  & 1.6456 (0.257) & 0.9500 (0.071) & 0.4017 (0.046)
  & 2.2883 (0.006) & 0.9000 (0.000) & 0.2483 (0.030) \\
& & SHQER ($\gamma=1$)
  & 1.9672 (0.512) & 0.9500 (0.071) & 0.4086 (0.051)
  & 2.3629 (0.086) & 0.9000 (0.000) & 0.2822 (0.016) \\
& & SHQER ($\gamma=0.9$)
  & 2.0119 (0.281) & 0.9500 (0.071) & 0.3362 (0.076)
  & 2.4609 (0.055) & 0.9000 (0.000) & 0.1542 (0.012) \\
& & SHQER ($\gamma=0.5$)
  & 1.6371 (0.259) & 0.9500 (0.071) & 0.4172 (0.000)
  & 2.3011 (0.024) & 0.9000 (0.000) & 0.2475 (0.019) \\
& & SHQER ($\gamma=0.1$)
  & 1.4555 (0.282) & 1.0000 (0.000) & 0.5155 (0.046)
  & 2.5293 (0.048) & 0.9000 (0.000) & 0.1458 (0.026) \\
& & SHQER ($\gamma=0$)
  & 1.4333 (0.284) & 1.0000 (0.000) & 0.5483 (0.049)
  & 2.4988 (0.065) & 0.9000 (0.000) & 0.1610 (0.022) \\
\\

& \multirow{9}{*}{0.5}
& SALES (EN)
  & 2.4989 (0.178) & 0.9500 (0.071) & 0.3155 (0.105)
  & 2.5973 (0.063) & 0.9000 (0.000) & 0.1483 (0.035) \\
& & CONQUER (EN)
  & 1.0927 (0.316) & 0.9500 (0.071) & 0.3172 (0.044)
  & 1.8533 (0.008) & 0.9000 (0.000) & 0.1492 (0.031) \\
& & QR (rqPen, EN)
  & 0.8935 (0.207) & 0.9500 (0.071) & 0.3586 (0.059)
  & 1.4811 (0.041) & 0.9000 (0.000) & 0.4068 (0.113) \\
& & \(L^k\mathrm{QR}\,(\mathrm{EN}, k=1.5)\)
  & 1.3252 (0.241) & 0.9500 (0.071) & 0.2276 (0.024)
  & 1.9859 (0.057) & 0.9000 (0.000) & 0.1068 (0.014) \\
& & SHQER ($\gamma=1$)
  & 1.7169 (0.574) & 0.9000 (0.141) & 0.2362 (0.056)
  & 2.0833 (0.143) & 0.8000 (0.000) & 0.1407 (0.019) \\
& & SHQER ($\gamma=0.9$)
  & 1.7532 (0.300) & 0.8500 (0.071) & 0.1569 (0.071)
  & 1.9799 (0.096) & 0.9000 (0.000) & 0.1576 (0.014) \\
& & SHQER ($\gamma=0.5$)
  & 1.3004 (0.259) & 0.9500 (0.071) & 0.2293 (0.027)
  & 1.9884 (0.015) & 0.9000 (0.000) & 0.1017 (0.005) \\
& & SHQER ($\gamma=0.1$)
  & 1.1097 (0.265) & 0.9500 (0.071) & 0.3448 (0.015)
  & 1.8298 (0.059) & 0.9000 (0.000) & 0.1712 (0.017) \\
& & SHQER ($\gamma=0$)
  & 1.0886 (0.264) & 0.9500 (0.071) & 0.3638 (0.012)
  & 1.8052 (0.076) & 0.9000 (0.000) & 0.1856 (0.013) \\
\\

& \multirow{9}{*}{0.7}
& SALES (EN)
  & 2.1075 (0.250) & 0.8500 (0.071) & 0.1690 (0.024)
  & 2.2315 (0.022) & 0.7500 (0.071) & 0.0856 (0.004) \\
& & CONQUER (EN)
  & 0.8166 (0.171) & 0.9500 (0.071) & 0.1569 (0.037)
  & 1.3213 (0.023) & 0.9000 (0.000) & 0.0788 (0.011) \\
& & QR (rqPen, EN)
  & 0.6731 (0.162) & 0.9000 (0.000) & 0.1569 (0.051)
  & 1.1076 (0.114) & 0.9000 (0.000) & 0.1322 (0.089) \\
& & \(L^k\mathrm{QR}\,(\mathrm{EN}, k=1.5)\)
  & 0.9745 (0.230) & 0.9000 (0.000) & 0.1207 (0.039)
  & 1.6214 (0.110) & 0.8000 (0.000) & 0.0492 (0.010) \\
& & SHQER ($\gamma=1$)
  & 1.4175 (0.611) & 0.9000 (0.141) & 0.1345 (0.054)
  & 1.7613 (0.220) & 0.7500 (0.071) & 0.0729 (0.022) \\
& & SHQER ($\gamma=0.9$)
  & 1.2291 (0.609) & 0.9000 (0.141) & 0.1448 (0.049)
  & 1.6009 (0.159) & 0.8000 (0.000) & 0.0831 (0.014) \\
& & SHQER ($\gamma=0.5$)
  & 0.9647 (0.244) & 0.8500 (0.071) & 0.1241 (0.039)
  & 1.6053 (0.082) & 0.8000 (0.000) & 0.0466 (0.008) \\
& & SHQER ($\gamma=0.1$)
  & 0.7828 (0.173) & 0.9500 (0.071) & 0.2069 (0.005)
  & 1.3281 (0.036) & 0.9000 (0.000) & 0.0864 (0.014) \\
& & SHQER ($\gamma=0$)
  & 0.7660 (0.171) & 0.9500 (0.071) & 0.2379 (0.015)
  & 1.2973 (0.014) & 0.9000 (0.000) & 0.0941 (0.008) \\

\end{longtable}
\end{landscape}
\endgroup

\begingroup
\scriptsize
\renewcommand{\arraystretch}{1.15}

\begin{landscape}
\begin{longtable}{lllccc|ccc}
\caption{Numerical comparisons under the linear heteroscedastic model with sparse regression coefficients
for various SHQER configurations and competing methods (elastic net at $\tau=0.50$ with
$\alpha \in \{0.3,0.5,0.7\}$).}
\label{table:elasticnet_alpha_tau_05}\\

\toprule
\multirow{2}{*}{Noise} & \multirow{2}{*}{$\alpha$} & \multirow{2}{*}{Methods}
  & \multicolumn{3}{c|}{\(\left(n = 600, p= 300\right)\)}
  & \multicolumn{3}{c}{\(\left(n = 300, p= 600\right)\)} \\
& & & Error & TPR & FPR & Error & TPR & FPR \\
\midrule
\endfirsthead

\caption[]{Numerical comparisons under the linear heteroscedastic model with sparse regression coefficients
for various SHQER configurations and competing methods (continued).}\\
\toprule
\multirow{2}{*}{Noise} & \multirow{2}{*}{$\alpha$} & \multirow{2}{*}{Methods}
  & \multicolumn{3}{c|}{\(\left(n = 600, p= 300\right)\)}
  & \multicolumn{3}{c}{\(\left(n = 300, p= 600\right)\)} \\
& & & Error & TPR & FPR & Error & TPR & FPR \\
\midrule
\endhead

\midrule
\multicolumn{9}{r}{\emph{Continued on next page}}\\
\endfoot

\bottomrule
\endlastfoot

\multirow{1}{*}{\(\mathcal{N}(0,3)\)}

& \multirow{9}{*}{0.3}
& SALES (EN)
  & 2.6919 (0.021) & 1.0000 (0.000) & 0.5397 (0.066)
  & 2.7873 (0.041) & 0.9000 (0.000) & 0.1907 (0.030) \\
& & CONQUER (EN)
  & 1.4250 (0.015) & 1.0000 (0.000) & 0.5034 (0.083)
  & 2.1516 (0.057) & 0.9000 (0.000) & 0.3093 (0.006) \\
& & QR (rqPen, EN)
  & 1.3026 (0.134) & 1.0000 (0.000) & 0.7638 (0.207)
  & 1.9341 (0.131) & 0.9000 (0.000) & 0.6814 (0.029) \\
& & \(L^k\mathrm{QR}\,(\mathrm{EN}, k=1.5)\)
  & 1.5294 (0.108) & 1.0000 (0.000) & 0.4603 (0.198)
  & 2.2433 (0.065) & 0.9000 (0.000) & 0.2212 (0.001) \\
& & SHQER ($\gamma=1$)
  & 1.6270 (0.036) & 1.0000 (0.000) & 0.3397 (0.056)
  & 2.2648 (0.029) & 0.9500 (0.071) & 0.2229 (0.018) \\
& & SHQER ($\gamma=0.9$)
  & 1.5361 (0.006) & 0.9500 (0.071) & 0.3983 (0.046)
  & 2.2111 (0.041) & 0.9500 (0.071) & 0.2644 (0.014) \\
& & SHQER ($\gamma=0.5$)
  & 1.6339 (0.039) & 1.0000 (0.000) & 0.3569 (0.022)
  & 2.2438 (0.076) & 0.9000 (0.000) & 0.2390 (0.007) \\
& & SHQER ($\gamma=0.1$)
  & 1.5055 (0.084) & 1.0000 (0.000) & 0.4776 (0.046)
  & 2.1936 (0.038) & 0.9000 (0.000) & 0.3297 (0.011) \\
& & SHQER ($\gamma=0$)
  & 1.4884 (0.085) & 1.0000 (0.000) & 0.5086 (0.046)
  & 2.1937 (0.023) & 0.9000 (0.000) & 0.3449 (0.008) \\
\\

& \multirow{9}{*}{0.5}
& SALES (EN)
  & 2.3403 (0.018) & 1.0000 (0.000) & 0.3276 (0.059)
  & 2.4724 (0.041) & 0.8500 (0.071) & 0.1314 (0.040) \\
& & CONQUER (EN)
  & 1.1677 (0.030) & 1.0000 (0.000) & 0.3190 (0.090)
  & 1.7990 (0.124) & 0.9000 (0.000) & 0.1568 (0.006) \\
& & QR (rqPen, EN)
  & 1.0999 (0.040) & 1.0000 (0.000) & 0.3362 (0.188)
  & 1.5088 (0.064) & 0.9500 (0.071) & 0.5161 (0.071) \\
& & \(L^k\mathrm{QR}\,(\mathrm{EN}, k=1.5)\)
  & 1.2307 (0.148) & 0.9500 (0.071) & 0.2879 (0.178)
  & 1.8619 (0.025) & 0.9000 (0.000) & 0.1568 (0.099) \\
& & SHQER ($\gamma=1$)
  & 1.3367 (0.055) & 0.8500 (0.071) & 0.1793 (0.034)
  & 1.9834 (0.061) & 0.8500 (0.071) & 0.0915 (0.002) \\
& & SHQER ($\gamma=0.9$)
  & 1.2450 (0.013) & 0.9500 (0.071) & 0.2207 (0.034)
  & 1.8845 (0.067) & 0.9000 (0.000) & 0.1178 (0.004) \\
& & SHQER ($\gamma=0.5$)
  & 1.3570 (0.047) & 1.0000 (0.000) & 0.2069 (0.039)
  & 1.8599 (0.009) & 0.9000 (0.000) & 0.1678 (0.115) \\
& & SHQER ($\gamma=0.1$)
  & 1.2291 (0.097) & 1.0000 (0.000) & 0.3138 (0.039)
  & 1.8168 (0.110) & 0.9000 (0.000) & 0.1619 (0.004) \\
& & SHQER ($\gamma=0$)
  & 1.2104 (0.101) & 1.0000 (0.000) & 0.3328 (0.046)
  & 1.8100 (0.101) & 0.9000 (0.000) & 0.1695 (0.002) \\
\\

& \multirow{9}{*}{0.7}
& SALES (EN)
  & 1.9165 (0.022) & 0.9000 (0.000) & 0.2224 (0.051)
  & 2.0724 (0.001) & 0.8500 (0.071) & 0.0881 (0.010) \\
& & CONQUER (EN)
  & 0.9079 (0.008) & 0.9500 (0.071) & 0.2276 (0.083)
  & 1.4136 (0.043) & 0.9000 (0.000) & 0.0958 (0.013) \\
& & QR (rqPen, EN)
  & 0.8000 (0.045) & 0.9500 (0.071) & 0.2569 (0.144)
  & 1.2232 (0.030) & 0.9000 (0.000) & 0.2475 (0.170) \\
& & \(L^k\mathrm{QR}\,(\mathrm{EN}, k=1.5)\)
  & 0.9329 (0.133) & 0.9000 (0.000) & 0.1948 (0.163)
  & 1.3835 (0.103) & 0.9000 (0.000) & 0.1449 (0.004) \\
& & SHQER ($\gamma=1$)
  & 0.9996 (0.081) & 0.8500 (0.071) & 0.1034 (0.020)
  & 1.6572 (0.095) & 0.8000 (0.000) & 0.0415 (0.001) \\
& & SHQER ($\gamma=0.9$)
  & 0.9332 (0.052) & 0.8500 (0.071) & 0.1397 (0.046)
  & 1.5377 (0.078) & 0.8500 (0.071) & 0.0585 (0.001) \\
& & SHQER ($\gamma=0.5$)
  & 1.0587 (0.027) & 0.9000 (0.141) & 0.1207 (0.024)
  & 1.5021 (0.017) & 0.8500 (0.071) & 0.1008 (0.097) \\
& & SHQER ($\gamma=0.1$)
  & 0.9527 (0.058) & 0.9500 (0.071) & 0.2172 (0.044)
  & 1.4426 (0.051) & 0.9000 (0.000) & 0.0873 (0.006) \\
& & SHQER ($\gamma=0$)
  & 0.9332 (0.067) & 0.9500 (0.071) & 0.2414 (0.039)
  & 1.4316 (0.057) & 0.9000 (0.000) & 0.0924 (0.001) \\
\midrule

\multirow{1}{*}{\(t(1.9)\)}

& \multirow{9}{*}{0.3}
& SALES (EN)
  & 2.7314 (0.053) & 1.0000 (0.000) & 0.6052 (0.251)
  & 2.7925 (0.099) & 0.9000 (0.000) & 0.1780 (0.117) \\
& & CONQUER (EN)
  & 1.2374 (0.256) & 1.0000 (0.000) & 0.4897 (0.078)
  & 2.0704 (0.092) & 0.8500 (0.071) & 0.2686 (0.020) \\
& & QR (rqPen, EN)
  & 1.0008 (0.160) & 1.0000 (0.000) & 0.7517 (0.190)
  & 1.8097 (0.017) & 0.9000 (0.000) & 0.4797 (0.249) \\
& & \(L^k\mathrm{QR}\,(\mathrm{EN}, k=1.5)\)
  & 1.4305 (0.358) & 1.0000 (0.000) & 0.5103 (0.151)
  & 2.1944 (0.002) & 0.8500 (0.071) & 0.2119 (0.038) \\
& & SHQER ($\gamma=1$)
  & 2.0486 (0.149) & 0.9500 (0.071) & 0.2966 (0.029)
  & 2.2912 (0.036) & 0.9000 (0.000) & 0.2619 (0.011) \\
& & SHQER ($\gamma=0.9$)
  & 1.9206 (0.161) & 0.9500 (0.071) & 0.3034 (0.024)
  & 2.2025 (0.023) & 0.9000 (0.000) & 0.2805 (0.016) \\
& & SHQER ($\gamma=0.5$)
  & 1.3931 (0.328) & 1.0000 (0.000) & 0.5362 (0.158)
  & 2.1742 (0.014) & 0.8500 (0.071) & 0.2186 (0.024) \\
& & SHQER ($\gamma=0.1$)
  & 1.2712 (0.185) & 1.0000 (0.000) & 0.5086 (0.002)
  & 2.0637 (0.030) & 0.8500 (0.071) & 0.3085 (0.026) \\
& & SHQER ($\gamma=0$)
  & 1.2379 (0.187) & 1.0000 (0.000) & 0.5293 (0.012)
  & 2.2025 (0.246) & 0.8500 (0.071) & 0.2220 (0.113) \\
\\

& \multirow{9}{*}{0.5}
& SALES (EN)
  & 2.3976 (0.117) & 0.9500 (0.071) & 0.4086 (0.202)
  & 2.4742 (0.109) & 0.8000 (0.000) & 0.1229 (0.085) \\
& & CONQUER (EN)
  & 0.8668 (0.193) & 1.0000 (0.000) & 0.3448 (0.000)
  & 1.5998 (0.054) & 0.8500 (0.071) & 0.1661 (0.012) \\
& & QR (rqPen, EN)
  & 0.6910 (0.182) & 1.0000 (0.000) & 0.3397 (0.051)
  & 1.3669 (0.008) & 0.8500 (0.071) & 0.2627 (0.141) \\
& & \(L^k\mathrm{QR}\,(\mathrm{EN}, k=1.5)\)
  & 1.1068 (0.344) & 1.0000 (0.000) & 0.3259 (0.124)
  & 1.8827 (0.071) & 0.8000 (0.000) & 0.0746 (0.029) \\
& & SHQER ($\gamma=1$)
  & 1.6435 (0.413) & 0.9000 (0.141) & 0.2138 (0.107)
  & 1.9900 (0.081) & 0.8000 (0.000) & 0.1119 (0.002) \\
& & SHQER ($\gamma=0.9$)
  & 1.6543 (0.187) & 0.9000 (0.141) & 0.1466 (0.007)
  & 1.8563 (0.049) & 0.8500 (0.071) & 0.1347 (0.016) \\
& & SHQER ($\gamma=0.5$)
  & 1.0704 (0.277) & 1.0000 (0.000) & 0.3448 (0.156)
  & 1.8588 (0.045) & 0.8000 (0.000) & 0.0814 (0.034) \\
& & SHQER ($\gamma=0.1$)
  & 0.9184 (0.190) & 1.0000 (0.000) & 0.3086 (0.007)
  & 1.6766 (0.009) & 0.8500 (0.071) & 0.1415 (0.040) \\
& & SHQER ($\gamma=0$)
  & 0.8947 (0.181) & 1.0000 (0.000) & 0.3345 (0.010)
  & 1.6460 (0.024) & 0.8500 (0.071) & 0.1542 (0.034) \\
\\

& \multirow{9}{*}{0.7}
& SALES (EN)
  & 1.9309 (0.165) & 0.8500 (0.071) & 0.1379 (0.010)
  & 2.0497 (0.104) & 0.7500 (0.071) & 0.0653 (0.035) \\
& & CONQUER (EN)
  & 0.5979 (0.156) & 1.0000 (0.000) & 0.2017 (0.051)
  & 1.1870 (0.020) & 0.8500 (0.071) & 0.0847 (0.046) \\
& & QR (rqPen, EN)
  & 0.4854 (0.144) & 1.0000 (0.000) & 0.1931 (0.034)
  & 1.0400 (0.046) & 0.9500 (0.071) & 0.1839 (0.119) \\
& & \(L^k\mathrm{QR}\,(\mathrm{EN}, k=1.5)\)
  & 0.7902 (0.279) & 0.9500 (0.071) & 0.2086 (0.124)
  & 1.2402 (0.065) & 0.8500 (0.071) & 0.1322 (0.029) \\
& & SHQER ($\gamma=1$)
  & 1.3255 (0.444) & 0.9000 (0.141) & 0.1224 (0.105)
  & 1.6339 (0.104) & 0.7500 (0.071) & 0.0500 (0.001) \\
& & SHQER ($\gamma=0.9$)
  & 1.3009 (0.220) & 0.8500 (0.071) & 0.0552 (0.005)
  & 1.4614 (0.080) & 0.8000 (0.000) & 0.0568 (0.004) \\
& & SHQER ($\gamma=0.5$)
  & 0.7398 (0.228) & 1.0000 (0.000) & 0.1328 (0.017)
  & 1.5011 (0.095) & 0.8000 (0.000) & 0.0271 (0.002) \\
& & SHQER ($\gamma=0.1$)
  & 0.6008 (0.187) & 1.0000 (0.000) & 0.2052 (0.012)
  & 1.2806 (0.009) & 0.8000 (0.000) & 0.0585 (0.030) \\
& & SHQER ($\gamma=0$)
  & 0.5792 (0.174) & 1.0000 (0.000) & 0.2276 (0.010)
  & 1.2504 (0.015) & 0.8500 (0.071) & 0.0669 (0.035) \\

\end{longtable}
\end{landscape}
\endgroup

\begingroup
\scriptsize
\renewcommand{\arraystretch}{1.15}

\begin{landscape}
\begin{longtable}{lllccc|ccc}
\caption{Numerical comparisons under the linear heteroscedastic model with sparse regression coefficients
for various SHQER configurations and competing methods (elastic net at $\tau=0.75$ with
$\alpha \in \{0.3,0.5,0.7\}$).}
\label{table:elasticnet_alpha_tau_075}\\

\toprule
\multirow{2}{*}{Noise} & \multirow{2}{*}{$\alpha$} & \multirow{2}{*}{Methods}
  & \multicolumn{3}{c|}{\(\left(n = 600, p= 300\right)\)}
  & \multicolumn{3}{c}{\(\left(n = 300, p= 600\right)\)} \\
& & & Error & TPR & FPR & Error & TPR & FPR \\
\midrule
\endfirsthead

\caption[]{Numerical comparisons (continued).}\\
\toprule
\multirow{2}{*}{Noise} & \multirow{2}{*}{$\alpha$} & \multirow{2}{*}{Methods}
  & \multicolumn{3}{c|}{\(\left(n = 600, p= 300\right)\)}
  & \multicolumn{3}{c}{\(\left(n = 300, p= 600\right)\)} \\
& & & Error & TPR & FPR & Error & TPR & FPR \\
\midrule
\endhead

\midrule
\multicolumn{9}{r}{\emph{Continued on next page}}\\
\endfoot

\bottomrule
\endlastfoot

\multirow{1}{*}{\(\mathcal{N}(0,3)\)}

& \multirow{9}{*}{0.3}
& SALES (EN)
  & 2.7741 (0.033) & 1.0000 (0.000) & 0.5103 (0.059)
  & 2.8644 (0.034) & 0.8500 (0.071) & 0.2119 (0.022) \\
& & CONQUER (EN)
  & 1.4834 (0.024) & 0.9500 (0.071) & 0.5259 (0.056)
  & 2.3108 (0.115) & 0.9000 (0.000) & 0.2627 (0.038) \\
& & QR (rqPen, EN)
  & 1.3206 (0.002) & 1.0000 (0.000) & 0.8621 (0.122)
  & 2.1597 (0.103) & 0.9000 (0.000) & 0.3924 (0.119) \\
& & \(L^k\mathrm{QR}\,(\mathrm{EN}, k=1.5)\)
  & 1.5608 (0.119) & 1.0000 (0.000) & 0.4724 (0.166)
  & 2.3108 (0.032) & 0.9000 (0.000) & 0.2169 (0.002) \\
& & SHQER ($\gamma=1$)
  & 1.6475 (0.051) & 0.9500 (0.071) & 0.3414 (0.024)
  & 2.3318 (0.011) & 0.9000 (0.000) & 0.2237 (0.007) \\
& & SHQER ($\gamma=0.9$)
  & 1.5556 (0.031) & 1.0000 (0.000) & 0.4224 (0.041)
  & 2.2796 (0.024) & 0.9000 (0.000) & 0.2585 (0.008) \\
& & SHQER ($\gamma=0.5$)
  & 1.6396 (0.050) & 0.9500 (0.071) & 0.3879 (0.012)
  & 2.3238 (0.038) & 0.9000 (0.000) & 0.2263 (0.018) \\
& & SHQER ($\gamma=0.1$)
  & 1.5438 (0.014) & 0.9500 (0.071) & 0.5138 (0.000)
  & 2.5587 (0.018) & 0.8500 (0.071) & 0.1381 (0.030) \\
& & SHQER ($\gamma=0$)
  & 1.5393 (0.008) & 0.9500 (0.071) & 0.5603 (0.002)
  & 2.4050 (0.193) & 0.9000 (0.000) & 0.2585 (0.123) \\
\\

& \multirow{9}{*}{0.5}
& SALES (EN)
  & 2.4158 (0.047) & 1.0000 (0.000) & 0.3155 (0.046)
  & 2.5632 (0.032) & 0.8500 (0.071) & 0.1331 (0.004) \\
& & CONQUER (EN)
  & 1.2260 (0.036) & 0.8500 (0.071) & 0.2655 (0.024)
  & 2.0281 (0.014) & 0.9000 (0.000) & 0.1119 (0.038) \\
& & QR (rqPen, EN)
  & 1.0401 (0.038) & 0.8500 (0.071) & 0.2241 (0.083)
  & 1.7300 (0.214) & 0.9000 (0.000) & 0.2703 (0.174) \\
& & \(L^k\mathrm{QR}\,(\mathrm{EN}, k=1.5)\)
  & 1.3134 (0.015) & 0.8500 (0.071) & 0.1897 (0.020)
  & 1.9673 (0.073) & 0.8500 (0.071) & 0.1542 (0.108) \\
& & SHQER ($\gamma=1$)
  & 1.3562 (0.048) & 0.8500 (0.071) & 0.1845 (0.022)
  & 2.0654 (0.035) & 0.8500 (0.071) & 0.0949 (0.002) \\
& & SHQER ($\gamma=0.9$)
  & 1.2396 (0.005) & 0.8500 (0.071) & 0.2345 (0.024)
  & 1.9746 (0.035) & 0.9000 (0.000) & 0.1178 (0.001) \\
& & SHQER ($\gamma=0.5$)
  & 1.3391 (0.024) & 0.8500 (0.071) & 0.2069 (0.010)
  & 2.0718 (0.071) & 0.8000 (0.000) & 0.0898 (0.012) \\
& & SHQER ($\gamma=0.1$)
  & 1.2190 (0.008) & 0.9000 (0.141) & 0.3172 (0.020)
  & 2.1711 (0.352) & 0.8500 (0.071) & 0.0966 (0.072) \\
& & SHQER ($\gamma=0$)
  & 1.2115 (0.006) & 0.9000 (0.141) & 0.3569 (0.017)
  & 2.1324 (0.337) & 0.8500 (0.071) & 0.1127 (0.078) \\
\\

& \multirow{9}{*}{0.7}
& SALES (EN)
  & 2.0046 (0.057) & 0.9500 (0.071) & 0.1914 (0.041)
  & 2.2219 (0.030) & 0.8500 (0.071) & 0.0966 (0.005) \\
& & CONQUER (EN)
  & 0.9459 (0.015) & 0.8500 (0.071) & 0.1310 (0.049)
  & 1.5425 (0.008) & 0.8500 (0.071) & 0.0898 (0.058) \\
& & QR (rqPen, EN)
  & 0.7761 (0.035) & 0.8500 (0.071) & 0.0810 (0.022)
  & 1.3637 (0.103) & 0.8000 (0.000) & 0.0746 (0.053) \\
& & \(L^k\mathrm{QR}\,(\mathrm{EN}, k=1.5)\)
  & 0.9311 (0.089) & 0.8500 (0.071) & 0.1845 (0.154)
  & 1.4915 (0.037) & 0.9000 (0.000) & 0.1424 (0.010) \\
& & SHQER ($\gamma=1$)
  & 1.0402 (0.054) & 0.8500 (0.071) & 0.0897 (0.020)
  & 1.7255 (0.056) & 0.8000 (0.000) & 0.0449 (0.006) \\
& & SHQER ($\gamma=0.9$)
  & 0.9271 (0.014) & 0.8500 (0.071) & 0.1259 (0.027)
  & 1.6224 (0.052) & 0.8000 (0.000) & 0.0576 (0.010) \\
& & SHQER ($\gamma=0.5$)
  & 0.9738 (0.015) & 0.8500 (0.071) & 0.0966 (0.015)
  & 1.7465 (0.083) & 0.8000 (0.000) & 0.0356 (0.005) \\
& & SHQER ($\gamma=0.1$)
  & 1.1308 (0.279) & 0.8000 (0.000) & 0.1224 (0.105)
  & 1.5757 (0.110) & 0.9000 (0.000) & 0.0814 (0.026) \\
& & SHQER ($\gamma=0$)
  & 1.0954 (0.238) & 0.8000 (0.000) & 0.1431 (0.129)
  & 1.5480 (0.101) & 0.9000 (0.000) & 0.0915 (0.026) \\
\\

\multirow{1}{*}{\(t(1.9)\)}

& \multirow{9}{*}{0.3}
& SALES (EN)
  & 2.9070 (0.073) & 1.0000 (0.000) & 0.6310 (0.361)
  & 2.9155 (0.047) & 0.9000 (0.000) & 0.2305 (0.149) \\
& & CONQUER (EN)
  & 1.4506 (0.205) & 1.0000 (0.000) & 0.4897 (0.098)
  & 2.3551 (0.135) & 0.8500 (0.071) & 0.2280 (0.035) \\
& & QR (rqPen, EN)
  & 1.2865 (0.342) & 1.0000 (0.000) & 0.6845 (0.193)
  & 2.1130 (0.067) & 0.8500 (0.071) & 0.5339 (0.074) \\
& & \(L^k\mathrm{QR}\,(\mathrm{EN}, k=1.5)\)
  & 1.6560 (0.188) & 0.9500 (0.071) & 0.4328 (0.012)
  & 2.2929 (0.029) & 0.8500 (0.071) & 0.2398 (0.035) \\
& & SHQER ($\gamma=1$)
  & 2.1346 (0.034) & 0.9500 (0.071) & 0.3190 (0.051)
  & 2.3269 (0.002) & 0.8500 (0.071) & 0.2678 (0.019) \\
& & SHQER ($\gamma=0.9$)
  & 2.0318 (0.052) & 0.9500 (0.071) & 0.3397 (0.037)
  & 2.3317 (0.091) & 0.8500 (0.071) & 0.2288 (0.084) \\
& & SHQER ($\gamma=0.5$)
  & 1.6620 (0.200) & 1.0000 (0.000) & 0.4328 (0.002)
  & 2.3325 (0.069) & 0.8500 (0.071) & 0.2525 (0.024) \\
& & SHQER ($\gamma=0.1$)
  & 1.4844 (0.300) & 1.0000 (0.000) & 0.5293 (0.017)
  & 2.4675 (0.374) & 0.8500 (0.071) & 0.2331 (0.143) \\
& & SHQER ($\gamma=0$)
  & 1.8203 (0.175) & 1.0000 (0.000) & 0.2897 (0.034)
  & 2.5620 (0.164) & 0.8500 (0.071) & 0.1602 (0.018) \\
\\

& \multirow{9}{*}{0.5}
& SALES (EN)
  & 2.5298 (0.006) & 0.9500 (0.071) & 0.3707 (0.158)
  & 2.5756 (0.081) & 0.8000 (0.000) & 0.1407 (0.079) \\
& & CONQUER (EN)
  & 1.1536 (0.188) & 1.0000 (0.000) & 0.3121 (0.056)
  & 1.9446 (0.084) & 0.8000 (0.000) & 0.1246 (0.020) \\
& & QR (rqPen, EN)
  & 0.9393 (0.229) & 0.9500 (0.071) & 0.3776 (0.037)
  & 1.5805 (0.029) & 0.8000 (0.000) & 0.1780 (0.029) \\
& & \(L^k\mathrm{QR}\,(\mathrm{EN}, k=1.5)\)
  & 1.3279 (0.217) & 0.9500 (0.071) & 0.2431 (0.017)
  & 1.8355 (0.114) & 0.8500 (0.071) & 0.1686 (0.076) \\
& & SHQER ($\gamma=1$)
  & 1.8950 (0.087) & 0.9000 (0.141) & 0.1534 (0.051)
  & 2.0303 (0.049) & 0.8000 (0.000) & 0.1263 (0.018) \\
& & SHQER ($\gamma=0.9$)
  & 1.7750 (0.116) & 0.9000 (0.141) & 0.1707 (0.056)
  & 1.9385 (0.043) & 0.8500 (0.071) & 0.1492 (0.024) \\
& & SHQER ($\gamma=0.5$)
  & 1.3450 (0.185) & 0.9500 (0.071) & 0.2466 (0.002)
  & 2.0319 (0.139) & 0.8000 (0.000) & 0.0890 (0.035) \\
& & SHQER ($\gamma=0.1$)
  & 1.1497 (0.276) & 1.0000 (0.000) & 0.3414 (0.010)
  & 1.8746 (0.079) & 0.8000 (0.000) & 0.1636 (0.025) \\
& & SHQER ($\gamma=0$)
  & 1.5483 (0.200) & 0.9500 (0.071) & 0.1534 (0.007)
  & 1.8279 (0.047) & 0.8000 (0.000) & 0.1856 (0.018) \\
\\

& \multirow{9}{*}{0.7}
& SALES (EN)
  & 2.0994 (0.066) & 0.8500 (0.071) & 0.1621 (0.005)
  & 2.1721 (0.128) & 0.7500 (0.071) & 0.0712 (0.058) \\
& & CONQUER (EN)
  & 0.8662 (0.181) & 0.9500 (0.071) & 0.1793 (0.020)
  & 1.4150 (0.034) & 0.8000 (0.000) & 0.0763 (0.007) \\
& & QR (rqPen, EN)
  & 0.7159 (0.160) & 1.0000 (0.000) & 0.2259 (0.056)
  & 1.2093 (0.015) & 0.8000 (0.000) & 0.0669 (0.052) \\
& & \(L^k\mathrm{QR}\,(\mathrm{EN}, k=1.5)\)
  & 0.9890 (0.223) & 0.9000 (0.141) & 0.1500 (0.007)
  & 1.3833 (0.114) & 0.8500 (0.071) & 0.1407 (0.012) \\
& & SHQER ($\gamma=1$)
  & 1.5834 (0.120) & 0.8500 (0.071) & 0.0897 (0.034)
  & 1.6612 (0.023) & 0.7500 (0.071) & 0.0559 (0.010) \\
& & SHQER ($\gamma=0.9$)
  & 1.4360 (0.144) & 0.8500 (0.071) & 0.0879 (0.032)
  & 1.5523 (0.016) & 0.7500 (0.071) & 0.0644 (0.017) \\
& & SHQER ($\gamma=0.5$)
  & 1.0042 (0.185) & 0.9000 (0.141) & 0.1431 (0.022)
  & 1.6669 (0.212) & 0.7500 (0.071) & 0.0339 (0.017) \\
& & SHQER ($\gamma=0.1$)
  & 0.8147 (0.189) & 0.9500 (0.071) & 0.2172 (0.010)
  & 1.4332 (0.059) & 0.8000 (0.000) & 0.0780 (0.019) \\
& & SHQER ($\gamma=0$)
  & 1.1805 (0.175) & 0.9000 (0.141) & 0.0845 (0.002)
  & 1.3894 (0.026) & 0.8000 (0.000) & 0.0898 (0.007) \\

\end{longtable}
\end{landscape}
\endgroup

\subsection{Results for Real Data Analysis}

\noindent
\begingroup
\scriptsize
\renewcommand{\arraystretch}{1.20}

\begin{landscape}
\begin{longtable}{llcccccc}
\caption{Average model size (Size) and prediction error (PE) with standard deviations in parentheses, computed over random train/test splits, for SHQER, SALES, CONQUER, QR (rqPen), and \(L^k\mathrm{QR}\,(k=1.5)\).}
\label{table:real}\\

\toprule
\multirow{2}{*}{Penalty / $\alpha$} & \multirow{2}{*}{Method}
  & \multicolumn{2}{c}{$\tau=0.25$}
  & \multicolumn{2}{c}{$\tau=0.50$}
  & \multicolumn{2}{c}{$\tau=0.75$} \\
\cmidrule(lr){3-4}\cmidrule(lr){5-6}\cmidrule(lr){7-8}
& & Size & PE & Size & PE & Size & PE \\
\midrule
\endfirsthead

\caption[]{Average model size (Size) and prediction error (PE, hybrid quantile--expectile loss) with standard errors in parentheses, computed over random train/test splits, for SHQER, SALES, CONQUER, QR (rqPen) and \(L^k\mathrm{QR}\,(k=1.5)\) (continued).}\\
\toprule
\multirow{2}{*}{Penalty / $\alpha$} & \multirow{2}{*}{Method}
  & \multicolumn{2}{c}{$\tau=0.25$}
  & \multicolumn{2}{c}{$\tau=0.50$}
  & \multicolumn{2}{c}{$\tau=0.75$} \\
\cmidrule(lr){3-4}\cmidrule(lr){5-6}\cmidrule(lr){7-8}
& & Size & PE & Size & PE & Size & PE \\
\midrule
\endhead

\midrule
\multicolumn{8}{r}{\emph{Continued on next page}}\\
\endfoot

\bottomrule
\endlastfoot

\multirow{9}{*}{Lasso ($\alpha=1$)}
& SALES
  & 36.00 (0.00) & 0.00 (0.00)
  & 35.00 (0.00) & 0.00 (0.00)
  & 35.00 (0.00) & 0.00 (0.00) \\
& CONQUER
  & 35.00 (0.00) & 0.08 (0.00)
  & 35.00 (0.00) & 0.11 (0.00)
  & 34.00 (0.00) & 0.08 (0.00) \\
& QR (rqPen)
  & 27.00 (0.00) & 0.02 (0.00)
  & 71.00 (0.00) & 0.03 (0.00)
  & 26.00 (0.00) & 0.02 (0.00) \\
& \(L^k\mathrm{QR}\,(k=1.5)\)
  & 22.00 (0.00) & 0.01 (0.00)
  & 23.00 (0.00) & 0.01 (0.00)
  & 14.00 (0.00) & 0.01 (0.00) \\
& SHQER ($\gamma=1$)
  & 28.00 (0.00) & 0.00 (0.00)
  & 22.00 (0.00) & 0.00 (0.00)
  & 0.00 (0.00) & 0.01 (0.00) \\
& SHQER ($\gamma=0.9$)
  & 19.00 (0.00) & 0.01 (0.00)
  & 22.00 (0.00) & 0.01 (0.00)
  & 0.00 (0.00) & 0.01 (0.00) \\
& SHQER ($\gamma=0.5$)
  & 25.00 (0.00) & 0.03 (0.00)
  & 23.00 (0.00) & 0.04 (0.00)
  & 10.00 (0.00) & 0.03 (0.00) \\
& SHQER ($\gamma=0.1$)
  & 34.00 (0.00) & 0.07 (0.00)
  & 23.00 (0.00) & 0.09 (0.00)
  & 9.00 (0.00) & 0.07 (0.00) \\
& SHQER ($\gamma=0$)
  & 35.00 (0.00) & 0.08 (0.00)
  & 24.00 (0.00) & 0.11 (0.00)
  & 9.00 (0.00) & 0.08 (0.00) \\
\midrule

\multirow{9}{*}{Elastic Net ($\alpha=0.3$)}
& SALES
  & 37.00 (0.00) & 0.00 (0.00)
  & 32.00 (0.00) & 0.00 (0.00)
  & 37.00 (0.00) & 0.01 (0.00) \\
& CONQUER
  & 34.00 (0.00) & 0.08 (0.00)
  & 35.00 (0.00) & 0.11 (0.00)
  & 37.00 (0.00) & 0.08 (0.00) \\
& QR (rqPen)
  & 31.00 (0.00) & 0.02 (0.00)
  & 63.00 (0.00) & 0.03 (0.00)
  & 23.00 (0.00) & 0.03 (0.00) \\
& \(L^k\mathrm{QR}\,(k=1.5)\)
  & 35.00 (0.00) & 0.01 (0.00)
  & 30.00 (0.00) & 0.01 (0.00)
  & 24.00 (0.00) & 0.01 (0.00) \\
& SHQER ($\gamma=1$)
  & 43.00 (0.00) & 0.00 (0.00)
  & 29.00 (0.00) & 0.00 (0.00)
  & 23.00 (0.00) & 0.00 (0.00) \\
& SHQER ($\gamma=0.9$)
  & 42.00 (0.00) & 0.01 (0.00)
  & 30.00 (0.00) & 0.01 (0.00)
  & 26.00 (0.00) & 0.01 (0.00) \\
& SHQER ($\gamma=0.5$)
  & 37.00 (0.00) & 0.03 (0.00)
  & 33.00 (0.00) & 0.04 (0.00)
  & 25.00 (0.00) & 0.03 (0.00) \\
& SHQER ($\gamma=0.1$)
  & 40.00 (0.00) & 0.07 (0.00)
  & 33.00 (0.00) & 0.09 (0.00)
  & 24.00 (0.00) & 0.07 (0.00) \\
& SHQER ($\gamma=0$)
  & 41.00 (0.00) & 0.08 (0.00)
  & 33.00 (0.00) & 0.11 (0.00)
  & 24.00 (0.00) & 0.08 (0.00) \\
\midrule

\multirow{9}{*}{Elastic Net ($\alpha=0.5$)}
& SALES
  & 33.00 (0.00) & 0.00 (0.00)
  & 36.00 (0.00) & 0.00 (0.00)
  & 35.00 (0.00) & 0.00 (0.00) \\
& CONQUER
  & 35.00 (0.00) & 0.08 (0.00)
  & 31.00 (0.00) & 0.11 (0.00)
  & 35.00 (0.00) & 0.08 (0.00) \\
& QR (rqPen)
  & 27.00 (0.00) & 0.02 (0.00)
  & 70.00 (0.00) & 0.03 (0.00)
  & 25.00 (0.00) & 0.02 (0.00) \\
& \(L^k\mathrm{QR}\,(k=1.5)\)
  & 24.00 (0.00) & 0.01 (0.00)
  & 23.00 (0.00) & 0.01 (0.00)
  & 15.00 (0.00) & 0.01 (0.00) \\
& SHQER ($\gamma=1$)
  & 33.00 (0.00) & 0.00 (0.00)
  & 25.00 (0.00) & 0.00 (0.00)
  & 11.00 (0.00) & 0.00 (0.00) \\
& SHQER ($\gamma=0.9$)
  & 32.00 (0.00) & 0.01 (0.00)
  & 24.00 (0.00) & 0.01 (0.00)
  & 12.00 (0.00) & 0.01 (0.00) \\
& SHQER ($\gamma=0.5$)
  & 34.00 (0.00) & 0.03 (0.00)
  & 24.00 (0.00) & 0.04 (0.00)
  & 11.00 (0.00) & 0.03 (0.00) \\
& SHQER ($\gamma=0.1$)
  & 36.00 (0.00) & 0.07 (0.00)
  & 27.00 (0.00) & 0.09 (0.00)
  & 10.00 (0.00) & 0.07 (0.00) \\
& SHQER ($\gamma=0$)
  & 37.00 (0.00) & 0.08 (0.00)
  & 27.00 (0.00) & 0.11 (0.00)
  & 10.00 (0.00) & 0.08 (0.00) \\
\midrule

\multirow{9}{*}{Elastic Net ($\alpha=0.7$)}
& SALES
  & 34.00 (0.00) & 0.00 (0.00)
  & 35.00 (0.00) & 0.00 (0.00)
  & 35.00 (0.00) & 0.00 (0.00) \\
& CONQUER
  & 35.00 (0.00) & 0.08 (0.00)
  & 34.00 (0.00) & 0.11 (0.00)
  & 35.00 (0.00) & 0.08 (0.00) \\
& QR (rqPen)
  & 26.00 (0.00) & 0.02 (0.00)
  & 65.00 (0.00) & 0.03 (0.00)
  & 27.00 (0.00) & 0.02 (0.00) \\
& \(L^k\mathrm{QR}\,(k=1.5)\)
  & 26.00 (0.00) & 0.01 (0.00)
  & 18.00 (0.00) & 0.01 (0.00)
  & 9.00 (0.00) & 0.01 (0.00) \\
& SHQER ($\gamma=1$)
  & 24.00 (0.00) & 0.00 (0.00)
  & 17.00 (0.00) & 0.00 (0.00)
  & 4.00 (0.00) & 0.00 (0.00) \\
& SHQER ($\gamma=0.9$)
  & 26.00 (0.00) & 0.01 (0.00)
  & 20.00 (0.00) & 0.01 (0.00)
  & 7.00 (0.00) & 0.01 (0.00) \\
& SHQER ($\gamma=0.5$)
  & 30.00 (0.00) & 0.03 (0.00)
  & 20.00 (0.00) & 0.04 (0.00)
  & 18.00 (0.00) & 0.03 (0.00) \\
& SHQER ($\gamma=0.1$)
  & 31.00 (0.00) & 0.07 (0.00)
  & 27.00 (0.00) & 0.09 (0.00)
  & 17.00 (0.00) & 0.07 (0.00) \\
& SHQER ($\gamma=0$)
  & 32.00 (0.00) & 0.08 (0.00)
  & 22.00 (0.00) & 0.11 (0.00)
  & 6.00 (0.00) & 0.08 (0.00) \\

\end{longtable}
\end{landscape}
\endgroup

\end{document}